\documentclass[longauth]{aa}
\makeatletter
\renewcommand*\maketitle{%
  \thispagestyle{firstpage}
\begingroup
    \if@wideboxfn
    \setlength\bibindent{1.4\parindent}
    \else
    \setlength\bibindent{\parindent}
    \fi
    \renewcommand*\thefootnote{\@fnsymbol\c@footnote}%
    \renewcommand\@makefntext[1]{%
    \ifaa@longfn\hsize\textwidth\fi
    \noindent
    \hb@xt@\bibindent{\hss\@makefnmark\enspace}##1}
  \ifaa@twocolumn
  \begingroup
    \begin{aa@strip}
          \aa@maketitle
    \end{aa@strip}
    \@thanks
  \endgroup
  \else
    \begingroup
      \let\thanks\footnote
      \aa@maketitle
    \endgroup
  \fi
\endgroup
  \setcounter{footnote}{0}%
}
\makeatother
\usepackage{natbib}
\bibpunct{(}{)}{;}{a}{}{,}
\usepackage{txfonts}
\usepackage{newtxtext,newtxmath}
\usepackage{flushend}
\usepackage{xcolor}
\usepackage{graphicx}
\usepackage[breaklinks, colorlinks, urlcolor=cyan,
    linkcolor=blue,citecolor=blue,pdfencoding=auto]{hyperref}
\usepackage{longtable}  %long table in appendix
\usepackage{pdflscape}  %landscape
\usepackage{bm}         %for bold ell
\newcommand{\orcit}[1]{\protect\href{https://orcid.org/#1}{\protect\includegraphics[width=8pt]{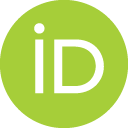}}}
\makeatletter
\renewcommand*\aa@pageof{, page \thepage{} of \pageref*{LastPage}}
\makeatother

\begin{document}

\title{The Atacama Cosmology Telescope}

\subtitle{Extragalactic Point Sources in the Southern Surveys \\at 150, 220 and 280 GHz observed between 2008-2010}

\author{Cristian~Vargas\orcit{0000-0001-5327-1400}\inst{\ref{inst:01}}
\and
Carlos H. L\'opez-Caraballo\orcit{0000-0002-6439-5385}\inst{\ref{inst:02},\ref{inst:03}}
\and
Elia S. Battistelli\orcit{0000-0001-5210-7625}\inst{\ref{inst:04}}
\and
Rolando Dunner\orcit{0000-0003-3892-1860}\inst{\ref{inst:01}}
\and
Gerrit Farren\orcit{0000-0001-5704-1127}\inst{\ref{inst:05},\ref{inst:06}}
\and
Megan Gralla\orcit{0000-0001-9032-1585}\inst{\ref{inst:07}}
\and
Kirsten R. Hall\orcit{0000-0002-4176-845X}\inst{\ref{inst:08}}
\and
Carlos Herv\'ias-Caimapo\orcit{0000-0002-4765-3426}\inst{\ref{inst:01}}
\and
Matt Hilton\orcit{0000-0002-8490-8117}\inst{\ref{inst:09},\ref{inst:10}}
\and
Adam D. Hincks\orcit{0000-0003-1690-6678}\inst{\ref{inst:11},\ref{inst:12}}
\and
Kevin Huffenberger\orcit{0000-0001-7109-0099}\inst{\ref{inst:13}}
\and
Tobias Marriage\orcit{0000-0003-4496-6520}\inst{\ref{inst:14}}
\and
Tony Mroczkowski\orcit{0000-0003-3816-5372}\inst{\ref{inst:15}}
\and
Michael D. Niemack\orcit{0000-0001-7125-3580}\inst{\ref{inst:16},\ref{inst:17}}
\and
Lyman Page\orcit{0000-0002-9828-3525}\inst{\ref{inst:18}}
\and
Bruce Partridge\orcit{0000-0001-6541-9265}\inst{\ref{inst:19}}
\and
Felipe Rojas\orcit{0000-0002-6721-4746}\inst{\ref{inst:01}}
\and
Francesca Rizzo\orcit{0000-0001-9705-2461}\inst{\ref{inst:20},\ref{inst:21}}
\and
Crist\'obal Sif\'on\orcit{0000-0002-8149-1352}\inst{\ref{inst:22}}
\and
Suzanne Staggs\orcit{0000-0002-7020-7301}\inst{\ref{inst:18}}
\and
Edward J. Wollack\orcit{0000-0002-7567-4451}\inst{\ref{inst:23}}
}

\institute{Instituto de Astrof\'isica and Centro de Astro-Ingenier\'ia, Facultad de F\'isica, Pontificia Universidad Cat\'olica de Chile, Av. Vicu\~na Mackenna 4860, 7820436 Macul, Santiago, Chile
\label{inst:01}
\and
Instituto de Astrof\'{\i}sica de Canarias, E-38200 La Laguna, Tenerife, Spain
\label{inst:02}
\and
Departamento de Astrof\'{\i}sica, Universidad de La Laguna, E-38206 La Laguna, Tenerife, Spain
\label{inst:03}
\and
Sapienza - University of Rome - Physics department, Piazzale Aldo Moro 5 I-00185, Rome, Italy
\label{inst:04}
\and 
DAMTP, Centre for Mathematical Sciences, University of Cambridge, Wilberforce Road, Cambridge CB3 OWA, UK
\label{inst:05}
\and
Kavli Institute for Cosmology Cambridge, Madingley Road, Cambridge CB3 0HA, UK
\label{inst:06}
\and 
Department of Astronomy/Steward Observatory, University of Arizona, 933 N. Cherry Ave., Tucson, AZ 85721, USA
\label{inst:07}
\and 
Center for Astrophysics $\vert$ Harvard \& Smithsonian, 60 Garden St., Cambridge, MA 02138, USA
\label{inst:08}
\and
Wits Centre for Astrophysics, School of Physics, University of the Witwatersrand, Private Bag 3, 2050, Johannesburg, South Africa
\label{inst:09}
\and
Astrophysics Research Centre, School of Mathematics, Statistics, and Computer Science, University of KwaZulu-Natal, Westville Campus, Durban 4041, South Africa
\label{inst:10}
\and
David A. Dunlap Department of Astronomy \& Astrophysics, University of Toronto, 50 St. George St., Toronto ON M5S 3H4, Canada
\label{inst:11}
\and
Specola Vaticana (Vatican Observatory), V-00120 Vatican City State
\label{inst:12}
\and
Department of Physics, Florida State University, Tallahassee FL, USA 32306
\label{inst:13}
\and
The William H. Miller III Department of Physics and Astronomy, Johns Hopkins University, 3701 San Martin Drive, Baltimore, MD 21220, USA
\label{inst:14}
\and
European Southern Observatory, Karl-Schwarzschild-Str. 2, D-85748, Garching, Germany
\label{inst:15}
\and
Department of Physics, Cornell University, Ithaca, NY 14853,USA
\label{inst:16}
\and
Department of Astronomy,Cornell University, Ithaca, NY 14853, USA
\label{inst:17}
\and
Joseph Henry Laboratories of Physics, Jadwin Hall, Princeton University, Princeton, NJ, USA 08544
\label{inst:18}
\and
Department of Physics and Astronomy, Haverford College, Haverford, PA, USA 19041
\label{inst:19}
\and
Cosmic Dawn Center, Denmark
\label{inst:20}
\and
Niels Bohr Institute, University of Copenhagen, Jagtvej 128, 2200 Copenhagen N, Denmark
\label{inst:21}
\and
Instituto de F\'isica, Pontificia Universidad Cat\'olica de Valpara\'iso, Casilla 4059, Valpara\'iso, Chile
\label{inst:22}
\and
NASA/Goddard Space Flight Center, Greenbelt, MD, USA 20771
\label{inst:23}}

\date{Received October 25, 2023; accepted Month Day, Year}
\titlerunning{ACT Southern Sources}
\authorrunning{C. Vargas et al}

\date{Received Day Month Year / Accepted Day Month Year}

\abstract{We present a multi-frequency, multi-epoch catalog of extragalactic sources. The catalog is based on 150, 220, and 280\,GHz observations carried out in 2008, 2009, and 2010 using the Millimeter Bolometric Array Camera on the Atacama Cosmology Telescope. We also present and release 280\,GHz maps from 2008 and 2010. {The catalog contains 483} sources found in a sky area of ${\sim}600$ square degrees. It was obtained by cross-matching sources found in 11 sub-catalogs, one for each season and frequency band. We also include co-added data from ${\sim}150$ and ${\sim}160$ square degrees using two and three years of overlapping observations. {We divide the sources into two populations, synchrotron and dusty emitters, based on their spectral behavior in the 150--280\,GHz frequency range. We find 284 synchrotron sources and 183 dusty source candidates. Our cross-matching with catalogs from radio to X-ray results in 251 synchrotron sources (88\%) and 92 dusty sources (51\%) with counterparts and suggests that 91 dusty candidates are not in existing catalogs}. We study the variability and number counts of each population. In the case of synchrotron sources, we find year-to-year variability, with a mean value around 35\%. As expected, we find no evidence of dusty source variability. Our number counts generally agree with previous measurements and models, except for dusty sources at 280\,GHz, where some models overestimate our results.} 


\maketitle

\section{Introduction}

The Atacama Cosmology Telescope (ACT) was a 6-meter millimeter-wavelength telescope installed on Cerro Toco, Chile, that mapped the skies for more than a decade, leading to advances in both cosmology and Galactic science \citep[e.g.,][]{Staggs2018}.
From 2008 to 2010, ACT conducted equatorial and southern sky surveys with the Millimeter Bolometer Array Camera (MBAC; \citealt{Fowler2007}, \citealt{Swetz2011}) in three bands centered at 150, 220, and 280 GHz with arcminute resolution. 
The primary science goal of these surveys was the study of the cosmic microwave background (CMB; \citep{Das2011,Dunkley2011,Das2014,Sievers2013}). In addition, the data were ideal for finding galaxy clusters through their Sunyaev-Zeldovich (SZ) effect \citep{Hincks2010,Menanteau2010,Marriage2011clusters,Menanteau2013,Hasselfield2013c,Hilton2021} and finding extragalactic source emission \citep{Marriage2011,Marsden2014,Gralla2020}.
The surveys' combination of large area, high resolution, and low noise allows for the detection at millimeter wavelengths of sources with flux densities down to a few millijanskys. 
Moreover, the multi-frequency aspect of the surveys offers the opportunity to separate source populations based on their spectral characteristics.
{To unveil the wealth of cosmological information within the CMB, it is essential to accurately identify and separate these foreground emissions.}

{Within the ACT millimeter wavelength range (1.1 -- 2.0 mm), we can broadly describe the emission mechanism from extragalactic sources as either non-thermal or thermal, each of these having their own spectral properties. We modeled the spectra as a power law $ S_\nu \propto \nu^{\alpha}$, relating the flux density $S$ to frequency $\nu$.
Flat or negative spectral indices ($\alpha \lesssim 1$) are usually associated with non-thermal synchrotron radiation. An example of intense synchrotron emission occurs when jets of relativistic electrons emerging from active galactic nuclei (AGNs) interact with gas and magnetic fields in the interstellar media (ISM) in galaxies \citep{Padovani2017}.
On the other hand, a positive spectral index ($\alpha \gtrsim 1$) is generally associated with thermal emission, which can be produced by diffuse interstellar material in certain states of local thermal equilibrium. An important example are dust grains heated by optical and ultraviolet radiation from massive young stars. Dust emits with a modified blackbody spectrum and temperatures ranging from 30 to 60 K. This blackbody spectrum has a positive slope near 300 GHz for a wide range of redshifts ($0<z<10$). However, high-redshift galaxies dominated by cold dust emission could be subject to the Wien displacement law, moving their peak close to this frequency range and thus showing a decreased spectral index at higher frequencies \citep{Casey2014}. 
Another relevant source of emission from galaxies is gas emission. On top of the continuum emission, these frequency bands are also subject to gas emission lines. Molecules such as CO are relevant at millimeter wavelengths, and transitions such as (2-1) occur at 230.538 GHz. Hence, they can significantly contaminate broadband measurements of nearby galaxies.}

According to the currently accepted unified AGN model \citep{Urry1995,Netzer2015}, AGNs correspond to the same type of physical object consisting of supermassive black holes at the center of distant galaxies that strongly accelerate material from their surrounding accretion disks, emitting strong radiation over large portions of the spectrum and for approximately 10\% of the population, expelling relativistic jets of hot plasma. 
The spectral properties depend on the observer's orientation relative to the accretion disk and jets. For side-on observers, the optically thin lobes create a steep spectrum component at radio frequencies, and the dusty accretion torus obscures the central black hole engine. For lines of sight aligned with the jet, the object appears as a compact flat-spectrum source, referred to as a blazar. 
{The synchrotron source population in CMB surveys typically consists of blazars.}
AGNs show variability at all wavelengths \citep[][]{Ulrich1997} and at different timescales from days \citep[][]{Aranzana2018} to years \citep[][]{Chen2013,Soldi2014}. The mechanism causing the variations is not well known, but they are believed to come mostly from accretion-disk instabilities (in radio-quiet sources), shocks from the relativistic jets (in blazars) at short timescales, and geometric effects at long timescales \citep[][]{Bach2005}.  Seasonal surveys such as ACT can provide information about the variability of AGNs, with sampling timescales ranging from weeks to over a decade, depending on the scan strategy and frequency of visits to the same portion of the sky. 

{While dust obscuration often hides dusty sources from visual and UV observations, they are bright in millimeter and submillimeter wavebands, making these wavelengths useful for identifying and observing the high-redshift population. The characterization of dusty sources has progressed significantly as millimeter and submillimeter-wave surveys have grown in size and resolving power in the last decade.}
In 1983, the Infrared Astronomical Satellite (IRAS) mapped the entire sky at infrared wavelengths, leading to the discovery of a population of 18,000 extragalactic sources \citep{Saunders2000}. 
Most of these were at relatively low redshifts, z < 0.3, with emission dominated by dust, and were classified as luminous infrared galaxies (LIRGs) (10$^{11} < L_{\textrm{IR}} < 10^{12}$ L$_\odot$) and ultra-luminous infrared galaxies (ULIRGs) (10$^{12} < L_{\textrm{IR}} < 10^{13}$ L$_\odot$), as they were orders of magnitude more luminous in the infrared compared to typical spiral galaxies with luminosities around 10$^{10}$ L$_\odot$ \citep{Blain2002}. 
Beginning in the late 1990s, observations at 450 and 850~$\mu$m with the Sub-millimetre Common-User Bolometer Array (SCUBA) instrument on the James Clerk Maxwell Telescope (JCMT; \citep{Holland1999}) discovered a high-redshift component of the dusty star-forming galaxy (DSFG) population. 
This population was named submillimeter galaxies (SMGs) due to the wavelength of their discovery. These early surveys of SMGs covered relatively small areas, only a few square degrees at most, and as a result, they traced out populations of rather dim sources~\citep[][]{Smail1997,Hughes1998,Barger1998,Eales2000,Cowie2002,Scott2002,Bertoldi2007,Weiss2009}.  

In the last decade, the sky-coverage, multi-frequency, and multi-epoch nature of submillimeter survey data have allowed for new DSFG candidates \citep{Vieira2010,Mocanu2013, Marsden2014,Gralla2020,Everett2020}. 
Follow-up observations of these sources (and a similarly bright population of sources detected in early \textit{Herschel} surveys, \citealt{2012MNRAS.424.1614O}) using facilities such as the Australia Telescope Compact Array (ATCA), the Atacama Large Millimeter/submillimeter Array (ALMA), and the Submillimeter Array (SMA) have demonstrated that these objects are indeed at high redshift \citep[see][and references therein]{Vieira2013,Weiss2013,2013MNRAS.433..498A,2017MNRAS.464..968S,2017MNRAS.466.2825B,2020ApJ...902...78R}.
Moreover, the high resolution achieved with interferometers {has shown that most bright sources \citep[i.e., fluxes higher than 100 mJy at 500 microns][]{Negrello2017}} undergo magnification via strong gravitational lensing by a massive source along the line of sight \citep[for example,][]{2017ApJ...844..110R,2019ApJ...879...95R} or that they are complex systems, such as multiple interacting pairs or protoclusters \citep{Casey2015,Miller2018}. Thus, the combination of negative K-correction \citep{Blain2002} and gravitational lensing makes large-area millimeter and submillimeter surveys uniquely powerful in studying the nature of star formation at high redshift ($z > 2$).

In this work, we report new measurements of the spectral distribution of AGNs and DSFGs found in the southern survey performed by the MBAC instrument. Here we analyze data from the  2008, 2009, and 2010 seasons and the multi-frequency data at 150 and 220\,GHz as well as new data at 280\,GHz.
The main result of this paper is a catalog including all three bands and seasons. From this catalog, we obtain the following: a) a classification of sources based on their spectral properties, b) a calculation of the number counts of these populations at different frequencies, c) data on the variability of the AGN population at a yearly time scale, and d) the SEDs of selected sources.
This study is complementary to previous MBAC-ACT works that include source catalogs from the southern survey at 150\,GHz from season 2008 \citep{Marriage2011}, at 150 and 220\,GHz from season 2008 \citep{Marsden2014}, and the equatorial survey  at 150 and 220\,GHz (2009 and 2010 seasons) and 280\,GHz (2010 season) presented by \cite{Gralla2020}.
The latter publication only presented data for the equatorial region. Here we extend that work to the southern region, including three seasons for 150--220\,GHz and two for 280\,GHz.
In addition, \cite{Datta2019} {presented} a catalog of intensity and polarization properties of sources located in the  equatorial region at 150\,GHz.

This paper is organized as follows: Section~\ref{sec:data} describes ACT observations, maps, and beam properties. Section~\ref{sec:source_extraction} details our methods of source detection and flux measurement and data validation via simulations and statistical analysis of different regions. Section~\ref{sec:act_catalog} presents the catalogs, source classification, and cross-matches with other experiments. Section~\ref{sec:source_analysis} shows a variability and number counts analysis. We conclude in Sect.~\ref{sec:summary} with a summary of our results.
%Sect.~\ref{sec:characterization_seds} treats analysis of an individual source.

\section{Data}\label{sec:data}

\subsection{Observations}

With MBAC, ACT observed the equatorial and southern sky during 2008 (season~2; s2), 2009 (season~3; s3) and 2010 (season~4; s4) campaigns,\footnote{Season 1 corresponds to commissioning observations done in the calendar
year 2007.} which were performed simultaneously at three different frequencies 150, 220, and 280 GHz with angular resolutions of 1$\farcm$4, 1$\farcm$0 and 0$\farcm$9 respectively \citep{Hasselfield2013}. 

In 2008 the survey was dedicated to a southern stripe area of approximately 440~$\mathrm{deg}^2$ centered at declination $\delta = -52.25^\circ$ with a size of $\Delta  \delta \approx 7^\circ$, and a wide range in right ascension $\mathrm{RA} \approx 0^\circ - 110^\circ$, corresponding to approximately 830 hours of observing time. During 2009 and 2010, roughly 1040 and 1560 hours of observations were split between an equatorial stripe of approximately 480~$\mathrm{deg}^2$ centered around $\delta = 0^\circ$ with a size of $\Delta \delta \approx 4.4^\circ$, and right ascension $\mathrm{RA} \approx 296^\circ$ to   $64^\circ$, as well as two southern areas centered around $\delta = -53^\circ$ with sizes of $\Delta \delta \approx 4^\circ - 6^\circ$, and right ascensions $\mathrm{RA} \approx 63^\circ$ to  $107^\circ$ and $\mathrm{RA} \approx 260^\circ$ to $297^\circ$, respectively. We name these two southern fields left (L) and right (R) respectively, due to their orientation in the maps. In this work we focus on the southern surveys.

\subsection{Maps}\label{sec:maps}

Our data set consists of southern observations from 2008, 2009 and 2010 at 150, 220 and 280\,GHz frequency bands (from here on called f150, f220 and f280 respectively)\footnote{Monochromatic detector arrays were assigned the names AR1, AR2 and AR3, and each contained the frequency bands called 148, 218 and 277\,GHz as described in \citealt{Swetz2011}; here we use the current ACT notation and naming convention.}.
The f280 2009 map was excluded because of its high noise levels (above 200$\mu$K per $\mathrm{deg}^2$ rms in the deepest area, due to technical problems of the camera). Thus, the analysis was done on a total set of 8 maps. Each map was generated following a map-making procedure described in \citealt{Dunner2013}. 

The maps were calibrated in two steps: first the f150 2008 map was cross-calibrated using the CMB power spectrum from WMAP 7-year data \citep{Jarosik2011}, and used as a reference map; after that, the f150 maps from the remaining seasons and f220 maps from all seasons were calibrated against the reference map again using the CMB (procedure explained in \citealt{Hajian2011}, map calibration presented in \citealt{Das2014}). 
{Despite the calibration was referenced to WMAP, \citet{Louis2014} verified for consistency and found an excellent agreement between ACT and \textit{Planck} maps. The calibration factor respect to \textit{Planck} is $0.998 \pm 0.007$ for f150 2008 and $1.001 \pm 0.025$ for f220 2008.}%Hence, we preserve the public map calibrations.

All f150 and f220 maps and weights are publicly available on the Legacy Archive for Microwave Background Data Analysis (LAMBDA).\footnote{\url{https://lambda.gsfc.nasa.gov/product/act/act_prod_table.html} as Data Release Products for the 2008-2010 Season at 148 GHz and 218 GHz.} We use the public versions throughout this work. Additionally, we present 2008 and 2010 maps and weights in the f280 band,\footnote{\url{https://lambda.gsfc.nasa.gov/product/act/act_prod_table.html} as Data Release Products for South 280 GHz, we also include the beam profiles and transforms.} which are also calibrated to the CMB. Details about their calibration are included in the Appendix~\ref{sec:calib_280}. {In summary, power spectrum calibration uncertainties are 2.0\%, 2.4\%, 3.6\%, 9.2\%, 3.6\%, 4.5\%, 9.1\% and 7.8\% for f150-2008, f220-2008, f150-2009, f220-2009, f150-2010, f220-2010, f280-2008 and f280-2010 respectively}

\begin{figure*}
\centering
    \includegraphics[width=17cm]{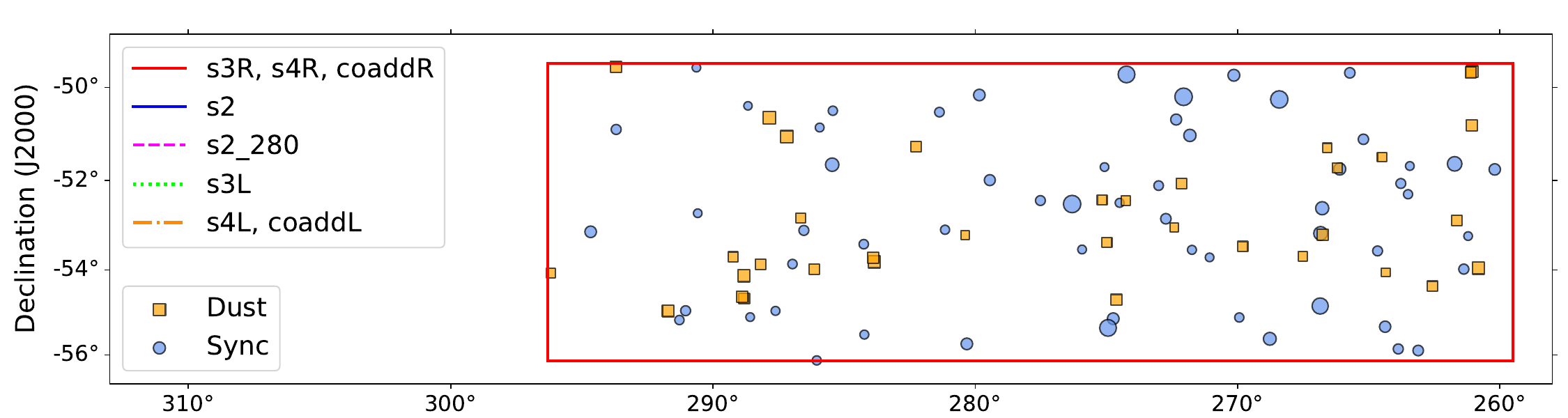}
    \includegraphics[width=17cm]{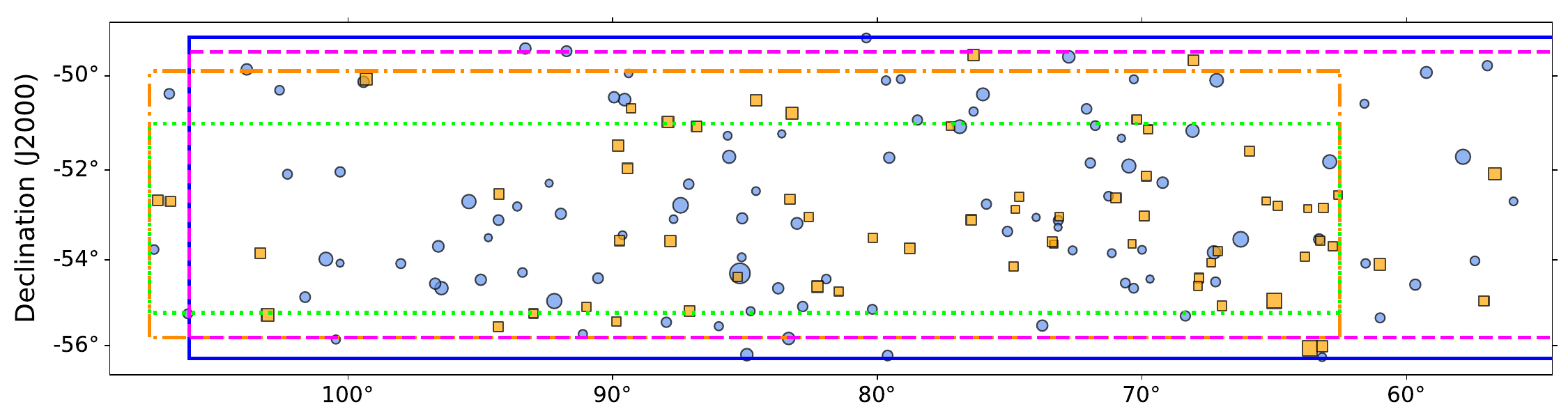}
    \includegraphics[width=17cm]{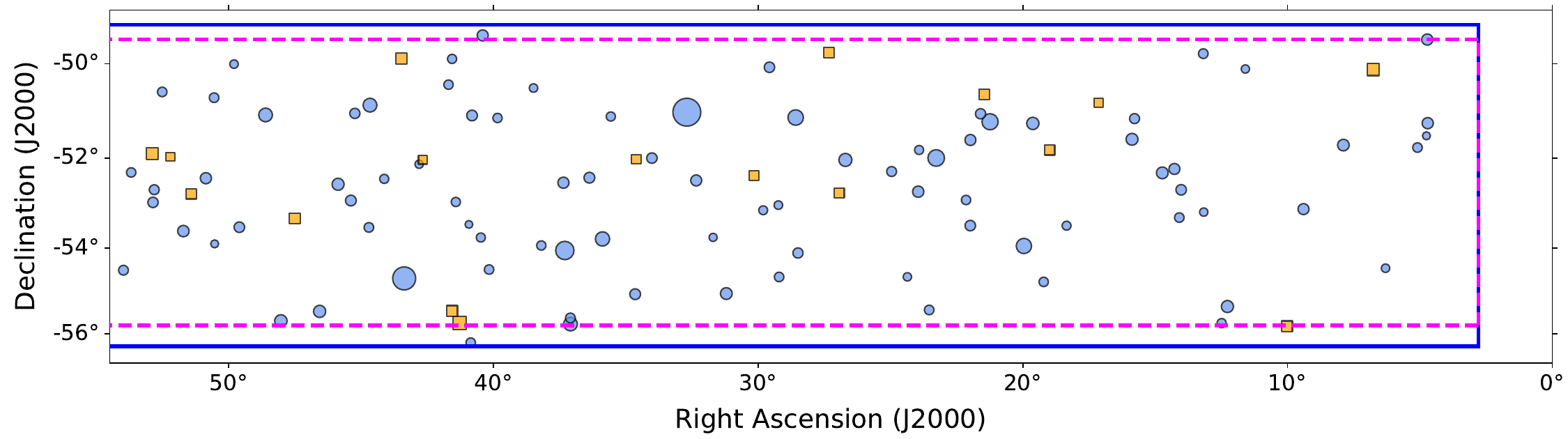}
    \caption{Fields considered in this work along with sources found in this study. Blue circles and orange squares mark the location of sources with a size proportional to the square root of the associated flux density at 220\,GHz ($S_{220}>8$\,mJy). Blue circles denote sources designated as synchrotron dominated, whereas orange squares denote dusty sources. {The two bottom panels are connected at right ascension 54.5\degr, which means that season 2 is a continuous strip running from approximately 3\degr to 106\degr in right ascension.} {The lower density of sources in the bottom panel is due to the exclusive presence of data from season 2 in that area, without any overlap from other seasons.}}
    \label{fig:footprints}
\end{figure*}

In a map, the number of observations (or "hits") is typically higher in the central part and falls toward the edges. The distribution of hits is similar for the same season at different frequencies, but the number of observations is usually lower at higher frequencies due to stricter data selection criteria, because the atmospheric contamination is worse at higher frequencies.
Motivated by this, and considering the implementation of our source extraction method (Sect.~\ref{sec:matchedfilter}), we chose fields consisting of rectangular patches for the different frequency band maps in each season. 
These areas represent portions of the full map, and they contain an approximately uniform amount of cross-linked scanning,\footnote{Ground based experiments favor constant elevation scans to avoid changing the atmosphere's column depth. ACT achieves “cross linking” by combining scans made at different times {and central azimuths}; as the Earth rotates, constant-elevation scans trace across the sky at different angles.} which improves the noise characteristics. The defining coordinates, area and relative extension of the selected patches for each season or frequency are presented in Table~\ref{tab:Fields} and drawn in Fig.~\ref{fig:footprints} respectively.

In addition, overlapping maps of the same frequency but different observing seasons were combined to form coadded maps with improved noise characteristics, at the expense of averaging out possible time variability of the sources. 

To obtain these coadded maps, each map is weighted by the white noise inverse variance per pixel, which we take to be proportional to the number of hit counts (see \citealt{Naess2020}). The coadded map {for frequency band $\nu$ and pixel $\mathbf{x}$ is then given by}

\begin{equation}
    M_\nu^c (\mathbf{x}) = \frac{\sum_i M_{\nu,i}(\mathbf{x}) W_{\nu,i}(\mathbf{x})}{\sum_i W_{\nu,i}(\mathbf{x})},
\label{eq:coadd}
\end{equation}

\noindent where $M_{\nu,i}(\mathbf{x})$ and  $W_{\nu,i}(\mathbf{x})$ are the individual map and its corresponding hit count map for season $i$ and frequency band $\nu$, respectively. {We make two coadded fields for L and R, called coaddL and coaddR respectively. The coaddL field contains data from all three seasons, and the rectangular area coincides with the s4 field; In coaddL case, we also include a portion of data from s3 outside this field weighted using Eq.~\ref{eq:coadd} (the part above and below $-55.22\degr$ and $-51.00\degr$ declination coincident with s4, see Fig.~\ref{fig:footprints}). The coaddR field combines data from s3 and s4 and uses the exact dimensions of these fields.}
As explained below in Sect.~\ref{sec:eff_beams}, beams at a given frequency are similar year-to-year at the percent level; for this reason we do not take into consideration differences in beams between seasons for these pixel-by-pixel coadded maps. Moreover due to the lower noise, we use the s4 beams to analyze the coadded maps.

 \begin{table*}
  \centering
  \caption{Description of the fields considered in this work. For each field we give the right ascension and declination defined by the two vertices (RA0, DEC0) and (RA1, DEC1) from lower left to top right. We also give an estimation of the effective area in square degrees. The s2, 280 GHz patch is smaller than its 150 and 220 GHz analogs because it has lower hit counts on the edges. The remaining fields are the same for the different available frequencies. {Coadded} fields combine the different seasons available for each frequency, i.e. coaddL combines season 2008, 2009 and 2010 at 150 and 220~GHz, but only seasons 2008 and 2010 for 280\,GHz}
  \begin{tabular}{rrrrrrrc}
    \hline
    Field & Season & Band & RA0~($^{\circ}$) & RA1~($^{\circ}$) & DEC0~($^{\circ}$) & DEC1~($^{\circ}$) & Eff area~($\mathrm{deg}^2$) \\
    \hline
    s2 & 2008 & 150, 220 & 2.8 & $106.0$ & $-56.30$ & $-49.20$ & $443.4$ \\
       &      & 280      & 2.8 & $106.0$ & $-55.80$ & $-49.50$ & $394.2$ \\
    \hline
    s3L & 2009 & 150, 220 & $62.52$ & $107.5$ & $-55.22$ & $-51.00$ & $113.7$ \\
    s3R & 2009 & 150, 220 & $259.5$ & $296.3$ & $-56.15$ & $-49.50$ & $147.7$ \\
    \hline
    s4L & 2010 & 150, 220, 280 & $62.52$ & $107.5$ & $-55.80$ & $-49.90$ & $160.1$ \\
    s4R & 2010 & 150, 220, 280 & $259.5$ & $296.3$ & $-56.15$ & $-49.50$ & $147.7$ \\
    \hline
    coaddL & 2008--2010 & 150, 220 & $62.52$ & $107.5$ & $-55.80$ & $-49.90$ & $160.1$\\
             & 2008, 2010 & 280 & $62.52$ & $107.5$ & $-55.80$ & $-49.90$ & $160.1$\\
    coaddR & 2009, 2010 & 150, 220 & $259.5$ & $296.3$ & $-56.15$ & $-49.50$ & $147.7$ \\
    \hline
  \end{tabular}
  \label{tab:Fields}
\end{table*}

\subsection{Effective source beams}\label{sec:eff_beams}

The instantaneous instrument beams are derived from planet observations as described in \citealt{Hasselfield2013}. From these observations beams are approximated as azimuthally symmetric; therefore, we worked with radial profiles in real and harmonic space. 
Point sources in the maps are the combination of measurements made in several passes of the telescope over that same region, and thus are affected by scatter from random pointing errors of the system, effectively broadening the point spread function. The pointing uncertainty of the telescope is roughly $5\arcsec$ \citep{Hasselfield2013}. 
Thus, assuming that the errors distribute evenly in all directions, we model them as a Gaussian with $\sigma_B = 5\arcsec$. The effective beam can be approximated by the convolution of the instantaneous beam and this 2D Gaussian. 
When transformed to harmonic space, we can express the effective beam as the multiplication of the instantaneous beam and a Gaussian with variance $\sigma_B^2$ in square radians,

\begin{equation}
    B_\ell^{\text{eff}} = B_\ell \exp\left({-\ell^2\sigma_B^2/2}\right),
\label{eq:beam_jitter}
\end{equation}

\noindent where $\ell$ is the spherical harmonic multipole number and $B_{\ell}$ is the instantaneous beam. Moreover, the empirical point spread function measured by the telescope corresponds to the weighted average of the monochromatic beam over the passband transmission function and the source spectra. 
The MBAC passbands are nearly 20\,GHz wide, meaning that the effective beam depends on the spectral properties of each source \citep[see][]{Page2003}.
Given this consideration, beams obtained from observations of planets are modeled assuming a Rayleigh-Jeans distribution, while other sources will be modeled by power laws of different spectral indices. 
To a good approximation, the effective beam can be modeled as monochromatic {at an} effective frequency for a given source spectrum \citep{Hasselfield2013}. 
In Table~\ref{tab:eff_freq} we show effective frequencies for point sources having different spectral indices, updated from \citealt{Swetz2011} (Hasselfield et al in prep.). 
{In this work, we use an average beam, which we call \emph{dustsync}. This beam is calculated assuming a frequency which is the mean of the effective frequencies of a synchrotron and a dust source.} 
Given the frequency $\nu_B$ at which the beam was measured, we obtain the effective dustsync beam as
\begin{table}
    \centering
    \caption{Frequency band information and effective frequency for different point sources, all given in gigahertz. For the matched filter we use the dustsync effective frequency. The uncertainty on each source effective frequency is roughly the uncertainty at the central frequency (3.5\,GHz), the uncertainties in the band centers are larger than the differences between dust and synchrotron effective frequencies.}
    \begin{tabular}{l|c|c|c}
        \hline
          Band          &   f150    &  f220     & f280       \\
%                       & 150 (GHz) & 220 (GHz) & 280 (GHz) \\
        \hline
        Central freq $\nu_0$  & $148.0 \pm 3.5$ & $218.7 \pm 3.5$ & $276.5 \pm 3.5$ \\
        Bandwidth $\Delta \nu$ & 18.0 & 16.9 & 20.5 \\
         $\nu_B$          & 148.2 & 218.7 & 276.5   \\
        \hline
         %Source & \multicolumn{3}{c}{Effective Frequencies~(GHz)}\\
        \multicolumn{4}{c}{Effective Frequencies for point sources}\\
        \hline
         Synchrotron      & 148.0 & 218.4 & 276.2   \\ 
         Dust             & 149.5 & 219.3 & 277.2   \\
         Dustsync         & 148.75 & 218.85 & 276.7 \\
         \hline
    \end{tabular}
    \label{tab:eff_freq}
\end{table}
\begin{table}
    \centering
    \caption{{Properties of the beams used in this work. Beams depend only on frequency and season.} These include a correction to the dustsync effective frequency from the planet map and a convolution with a Gaussian which models the pointing uncertainty.}
% 
%    \caption{Solid angle of the beams used in this work.  These include a correction to the dustsync effective frequency from the planet map and a convolution with a Gaussian which accounts for the pointing variance.}
    \begin{tabular}{l|c|c|c}
    \hline
Band & Season & {FWHM($\arcmin$)} &  {Solid Angle ($10^{-9}$\,sr)}\\
\hline
f150 &  2008  & {1.367}           & $216.3 \pm 3.9$            \\
     &  2009  & {1.375}           & $216.2 \pm 3.9$            \\
     &  2010  & {1.371}           & $214.8 \pm 3.9$            \\
\hline
f220 &  2008  & {0.997}           & $116.6 \pm 2.3$            \\
     &  2009  & {1.028}           & $124.4 \pm 2.5$            \\
     &  2010  & {1.023}           & $124.9 \pm 2.5$            \\
\hline
f280 &  2008  & {0.887}           & $99.1 \pm 2.0$             \\
     &  2010  & {0.892}           & $100.0 \pm 2.0$            \\
\hline
    \end{tabular}
    \label{tab:solid}
\end{table}

\begin{equation}
B_{\ell} = B\left(\ell \times \frac{\nu_B}{\nu_\mathrm{dustsync}}\right)
\label{eq:beam_freq}.
\end{equation}

As a result, the beam grows or shrinks by a sub-percent amount depending on the source spectral index, thus changing the beam solid angle, with a direct implication for the final flux density. Beam {properties} for each frequency band and season are presented in Table~\ref{tab:solid}.

{Using an average beam introduces a photometric uncertainty in the measured flux from the fact that synchrotron have a lower effective frequency (and thus slightly broader beam), and dust has a higher effective frequency and thus narrower beam (see Table~\ref{tab:eff_freq}). 
This photometric bias is 1.5\% at 150 GHz, 1.1\% at 220 GHz and 1.1\% at 280 GHz, positive for dust ($\alpha = 3.4$) and negative for synchrotron ($\alpha = -0.7$).
Moreover, point sources in the map are subject of mapping uncertainties from variable calibration errors during the season, which is reported as 3\% in \citep{Marsden2014}, and their flux measurement uncertainties are also proportional to the beam solid angle uncertainty (Table~\ref{tab:solid}). 
The total calibration uncertainty is then obtained by quadratically adding all the previous errors to the power spectrum calibration uncertainties (Sect.~\ref{sec:maps}), leading to 4.0\%, 4.1\%, 5.1\%, 9.9\%, 5.0\%, 5.7\%, 9.8\%, 8.7\% for f150-2008, f220-2008, f150-2009, f220-2009, f150-2010, f220-2010, f280-2008 and f280-2010 respectively. 
We approximate the calibration uncertainty for the coadded maps as roughly the average of the single-season maps, resulting in 5\%, 6\% and 9.5\% source calibration uncertainty for coadded source maps from f150, f220 and f280, respectively.}

\begin{figure}
  \resizebox{\hsize}{!}{\includegraphics{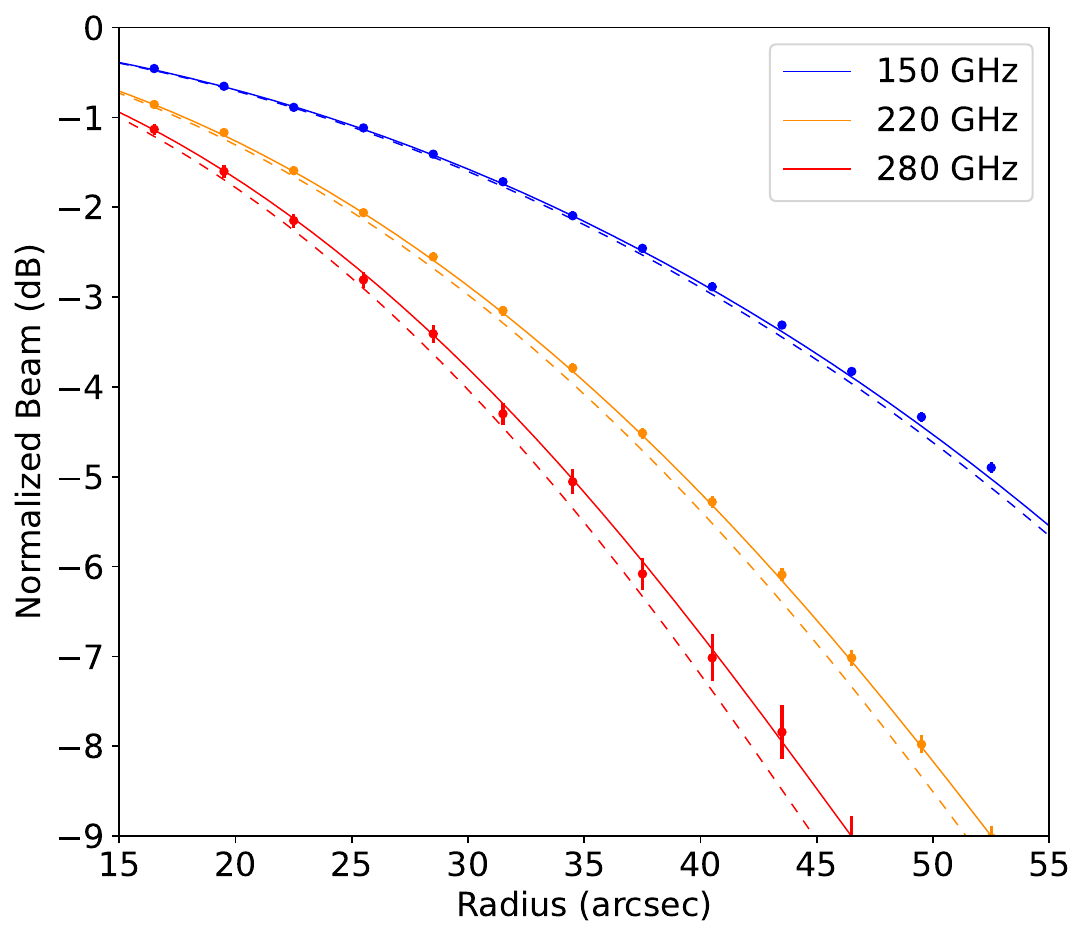}}
  \caption{Instantaneous beams (dashed lines) and effective beams (solid line). Data {with error bars} are obtained from a stacking of 16 bright sources with $S/N > 15$ taken from the 220\,GHz 2010 catalog. We see that data prefer the effective beam model over the instantaneous beam derived from planet observations. All the effective beam models use a broadening of $5\arcsec$; they are not fit to the data.}
  \label{fig:beam_stack}
\end{figure}

We test our beam model by comparing it to the empirical profile of stacked point sources in the maps. The stacking process is the following: at the position of each source we make sub-maps of {$20\arcmin \times 20\arcmin$},\footnote{Due to map projection, pixels are not squares in sky coordinates but rectangles; pixel scales are approximately $0.485\arcmin$ in right ascension and $0.505\arcmin$ in declination.} then we re-center the source in Fourier space (by the missing fraction of a pixel) and add it to a stack. 
Once the stack is complete, we take the inverse transform and reduce the pixel size of the map by a factor $2^4$ using Fourier interpolation.\footnote{Interpolation consists in adding zeros in the Fourier space, which over-sample the image, decreasing the pixel size to $0.030\arcmin$ by $0.032\arcmin$ in right ascension and declination, respectively.} 
We then subtract a component taken as the mean of measurements in a ring from 200 to 300 arcsec, and finally we calculate the profile in radial bins of 3 arcseconds each. Error bars are approximated as: $\sigma_{i}/\sqrt{N_{i}/2^4}$, where $\sigma_{i}$ is the standard deviation in the ring and $N_{i}$ is the number of pixels in the ring. 
The approximation assumes Poisson statistics, in which the error is $\sigma$ divided by the square root of the number of measurements. Since we interpolated the original data, we take the number of measurement as $N/2^4$. In Fig.~\ref{fig:beam_stack} we show the difference between instantaneous (planet) and effective beams, and how data from stacking prefer the effective beam.

\section{Source detection and flux density estimation}\label{sec:source_extraction}

Our main goal is to measure the flux density of point sources in our data. Since our maps include the CMB and noise, they require a matched filter method to optimally detect sources.

\subsection{Matched filter}
\label{sec:matchedfilter}

A matched filter maximizes the signal-to-noise of a known profile embedded in noise \citep[][]{Haehnelt1996}, which in our case is the {beam radial profile $B(\mathbf{x})$  }{(Eq.~\ref{eq:beam_freq} transform)}.
For our purposes, noise includes all the other components of the map, such as CMB, atmosphere contamination, and the detector noise, and we group them as

\begin{equation}
T_\text{other}(\mathbf{x}) = T_\mathrm{CMB}(\mathbf{x}) + T_\mathrm{atm,noise}(\mathbf{x}) + T_\mathrm{white,noise}(\mathbf{x}). 
\end{equation}

Thus, a temperature map $T$ with a source at the center is represented by

\begin{equation}
    T(\mathbf{x}) = T_0 B (\mathbf{x}) + T_\text{other}(\mathbf{x}).
\end{equation}

If we assume only two conditions, namely that the mean of $T_\mathrm{other}$ is zero and its statistical properties are independent of the position, we can adopt

\begin{equation}
T_\mathrm{est} = \int \Phi(\mathbf{x}) T (\mathbf{x}) d^2x
\end{equation}

\noindent as a linear estimator of the source temperature, $T_0$. The bias and the variance of this estimator are
\begin{align}
b &\equiv \langle T_\mathrm{est} - T \rangle, \\
\sigma^2 &\equiv \langle (T_\mathrm{est}-T)^2 \rangle.
\end{align}

The optimal unbiased ($b=0$) estimator is the one that minimizes the variance $\sigma$. In Fourier space, it becomes 
%We choose the \emph{best estimator} as the one that is unbiased ($b=0$), and minimizes the variance $\sigma$, implying:

\begin{equation}
    \Phi(\mathbf{k}) = \frac{{B}^\ast (\mathbf{k})}{P(\mathbf{k})} \left[ {\int \frac{\left | {B}(\mathbf{k}) \right |^2}{P(\mathbf{k})}d^2 \mathbf{k}} \right ]^{-1},
\end{equation}

\noindent where $\mathbf{k} = (k_x,k_y)$ is the angular wave-vector with $x$ and $y$ referring to Cartesian coordinates projected on the sphere, and $P(\mathbf{k}) = | {T}_\text{other}(\mathbf{k})|^2 $ is the power spectrum of the noise and other signals. This has the simple functional form $\Phi(\mathbf{k}) \propto {B}(\mathbf{k})/P(\mathbf{k})$; therefore, the matched filter prioritizes the Fourier modes where the beam profile stands out the most above the noise background. The normalization ensures that the peak value of the filtered sources corresponds to an unbiased value of the source temperature. (For a complete derivation see Appendix A of \citealt{Haehnelt1996}).
When applied in Fourier space the filtered map becomes

\begin{equation}
    T_{\mathrm{filt}}(\mathbf{x}) = \mathcal{F}^{-1} ({T}(\mathbf{k})\Phi(\mathbf{k})).
\label{eq:Tfilt}
\end{equation}

%\subsubsection{Background noise model}
%\label{sec:nm}

The noise model, $T_\mathrm{other}$, is obtained from the same maps by equalizing the noise and removing point source contamination. 
The result represents the spectral power density distribution of the unwanted components of the maps.
The noise distribution in the maps is inhomogeneous due to the different integration times across the maps. 
To account for this, we multiply the maps pixel by pixel by a weight function:

\begin{equation}
    w(\mathbf{x}) = \sqrt{H_\mathrm{pix}(\mathbf{x})/H_\mathrm{max}}.
\label{eq:weight}
\end{equation}

\noindent where $H_\mathrm{pix}(\mathbf{x})$ is the hit count of pixel $\mathbf{x}$, and $H_\mathrm{max}$ the maximum hit count in the map. 
This results in a map with approximately homogeneous noise level.

Point sources are removed in a two-pass process, using the cleaned map to estimate $T_\mathrm{other}$. 
On the first pass, we mask the strongest point sources ($S/N>50$) using $3'$ radius circular patches, filled with the average temperature of a ring around the source of diameter ranging $5'$ to $8'$. We typically mask around 10 sources per map. The second pass includes dimmer sources, down to $|S/N|>5$, typically involving around 50 sources per map.
Note that both positive and negative sources are removed to include 
the Sunyaev-Zel'dovich signal from galaxy clusters, which produces a decrement at 150 GHz, a null at 220 GHz and an increment at 280 GHz.
The masks represent less than 1\% of the map.
We stopped at this point because the purity of the detections significantly dropped below this $S/N$ (see Sect.~\ref{sec:purity}). Thus, masking more would affect the noise properties of $T_\mathrm{other}$.

The matched filter requires computing the power spectrum of $T_\text{other}$. 
This was done by smoothly tapering the patch from unity to zero with a cosine over a region {$10\arcmin$ from its edge.}
To better represent the noise power spectral density as a smooth function in harmonic space, we smoothed the power spectrum of $T_\mathrm{other}$ {by a Gaussian of $\sigma_s=10$ multipoles. }
This smoothing scale is important: If the value is too big, different angular scales mix, and the result is a noisier filtered map. 
On the other hand, if the value is too small, the power spectrum contains too much information from the particular realizations of the noise, introducing correlations in the map. We simulated different values and settled on {ten multipoles.}

Although the filter suppresses large scales, they are undersampled and can leak into other scales. For this reason, we applied a low-pass filter, removing modes below $\ell = 1000$. {Additionally,} we removed modes with $|k_x <100|$ {(vertical strip)}, which are contaminated by scan synchronous pickup. A similar treatment was used to process source maps in Fourier space.

\subsection{Source detection and flux measurement}
\label{sec:flux}

Maps where sources have not been masked are processed following the same procedure previously described, including noise equalization, tapering and filtering. We then divide the filtered map by its standard deviation, $\sigma$, effectively making a signal-to-noise map. From this map we detect peaks down to a certain $S/N$ threshold.

We detect the sources through an iterative process, progressively increasing the detection threshold and masking the newly detected sources in each iteration. This allowed us to avoid artifacts produced by the residual source content. We perform three iterations, using $S/N$ thresholds of $50$, $5.0$ and $3.5$.
Note that this iterative process is separate from the iterative process described above to obtain the noise model; we thus have two sets of iterations in our pipeline.

To measure flux densities, first we divide weighted filtered maps from Eq.~\ref{eq:Tfilt} by Eq.~\ref{eq:weight} (to undo the weighting), then we create small, over-sampled sub-maps around each candidate source. 
We first take sub-maps of {$4\arcmin \times 4\arcmin$} centered at each source, then increase the pixel resolution by a factor of $2^4$ using Fourier space interpolation. 
This allowed us to better constrain the location and amplitude of the peaks that we call $T_\mathrm{peak}$. We also de-convolve the map pixel window function in the higher resolution map.

The source flux densities are calculated as $T_\mathrm{peak}$ multiplied by the beam solid angle $\Omega_B$. 
And, to convert from CMB fluctuation temperature units ($\mu \mathrm{K}$) to flux (Jy), we multiply by the partial derivative of the Planck intensity function $B_\nu (T)$ with respect to temperature at the CMB mean temperature. Thus, the flux density becomes

\begin{equation}
    S_m = T_\mathrm{peak} \Omega_B \left. \frac{\partial B_\nu (T)}{\partial T} \right|_{T = T_\mathrm{CMB}}.
\label{eq:flux}
\end{equation}

\begin{figure*}
\centering
   \includegraphics[width=17 cm]{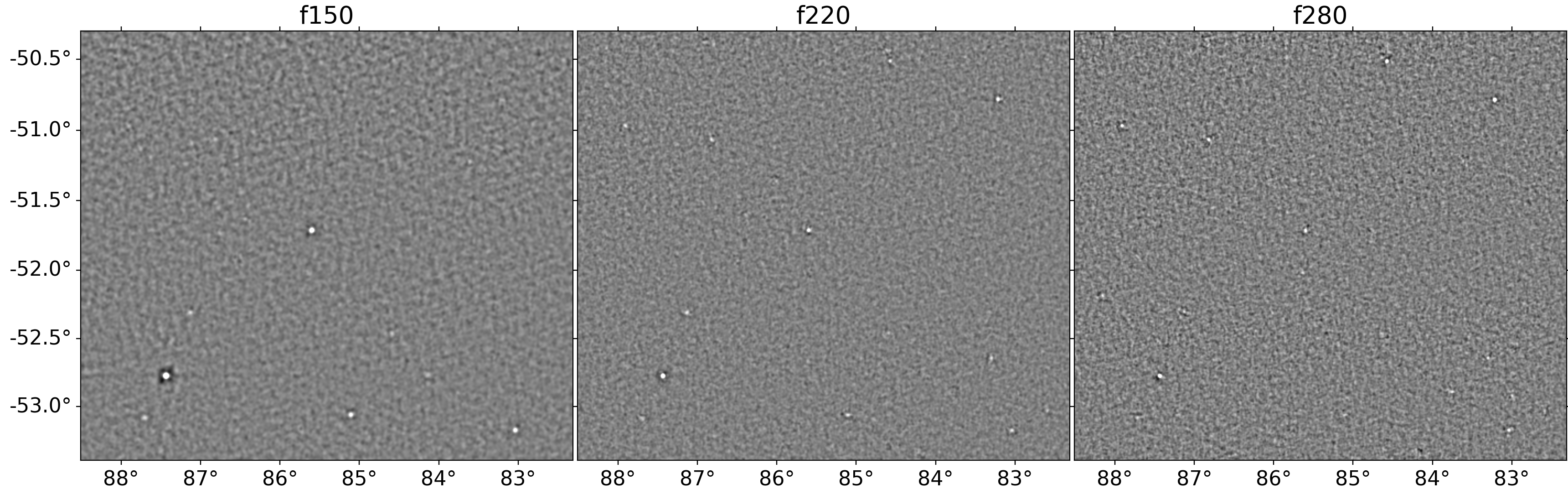}
     \caption{Filtered maps of a region observed in 2010 at the 150, 220, and 280\,GHz bands, from left to right, respectively. The gray scale is the same for the three maps going from $-$25\,mJy to +25\,mJy. The sources are the bright spots. Dusty sources are typically undetected or have low fluxes in 150\,GHz maps, but their flux rises at higher frequencies. The opposite is true for synchrotron sources: they have higher flux at the lowest frequency. Note that the angular scale of the noise fluctuations is related to the beam size in these filtered maps.}
     \label{fig:filtered_maps}
\end{figure*}

We refer to this as measured flux. We note that Eq.~\ref{eq:flux} requires the effective frequency $\nu$ of the sources, for which we use the intermediate frequency denoted as dustsync (see Table~\ref{tab:eff_freq}). 
{Fig.~\ref{fig:filtered_maps} shows unweighted filtered maps of flux density in the 150, 220, and 280 GHz bands. Visual inspection reveals point sources with varying flux density in all the frequency bands.}

The standard deviation of the flux density from Eq.~\ref{sec:flux}{, $\sigma_m$,} is determined by the noise in the filtered map at the location of the source. Under the assumption of a Gaussian distribution for the noise in the map,

\begin{equation}
    \sigma_m = \sigma_0/w(\mathbf{x}),
\label{eq:sigma}
\end{equation}

\noindent where $\sigma_0$ is the standard deviation of the weighted and filtered version of the noise map $T_\mathrm{other}$. Note that by definition $w(\mathbf{x}) \leq 1$, hence $\sigma_m \geq \sigma_0$.

Finally, we define the measured signal-to-noise:

\begin{equation}
     S/N = S_m/\sigma_m.
\label{eq:snr}
\end{equation}

This signal to noise is different from the one used in the source cleaning process, which was based on the noise of the current version of $T_\mathrm{other}$ and the peaks were measured without oversampling the pixels. From here on, $S/N$ refers to Eq.~\ref{eq:snr}.

\subsection{Statistical properties from simulations}\label{sec:sims}

\begin{figure}
  \resizebox{\hsize}{!}{\includegraphics{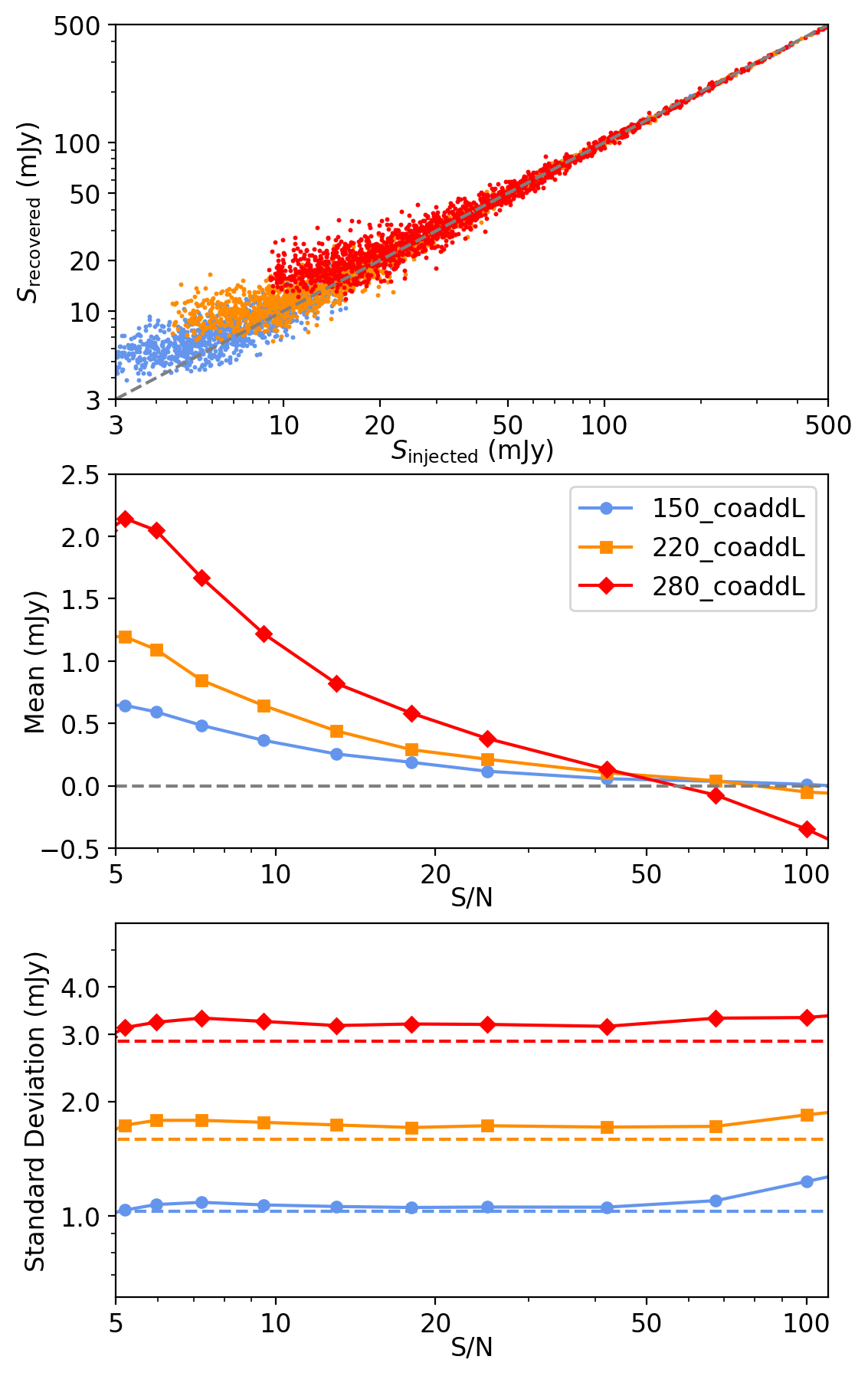}}
  \caption{From simulations, statistics of the difference between the flux densities of recovered and injected sources in bins of signal to noise. These results are from the coaddL patch at f150, f220 and f280; other patches show similar results. Top: Injected versus recovered fluxes at different frequencies. Middle: Mean difference between recovered and injected values across different signal to noise bins. From the mean we see that $S/N \lesssim 50$ sources {are boosted} in their flux measurements; this boosting is negligible at $S/N \gtrsim 50$. Bottom: Standard deviation of different bins compared to the measured value from the weighted filtered maps $\sigma_0$ shown as dotted lines. The $S/N \lesssim 50$ bins agree with a constant model, but the $S/N \gtrsim 50$ bins depart from it.}
  \label{fig:stats}
\end{figure}

\begin{figure*}
\sidecaption
\includegraphics[width=12 cm]{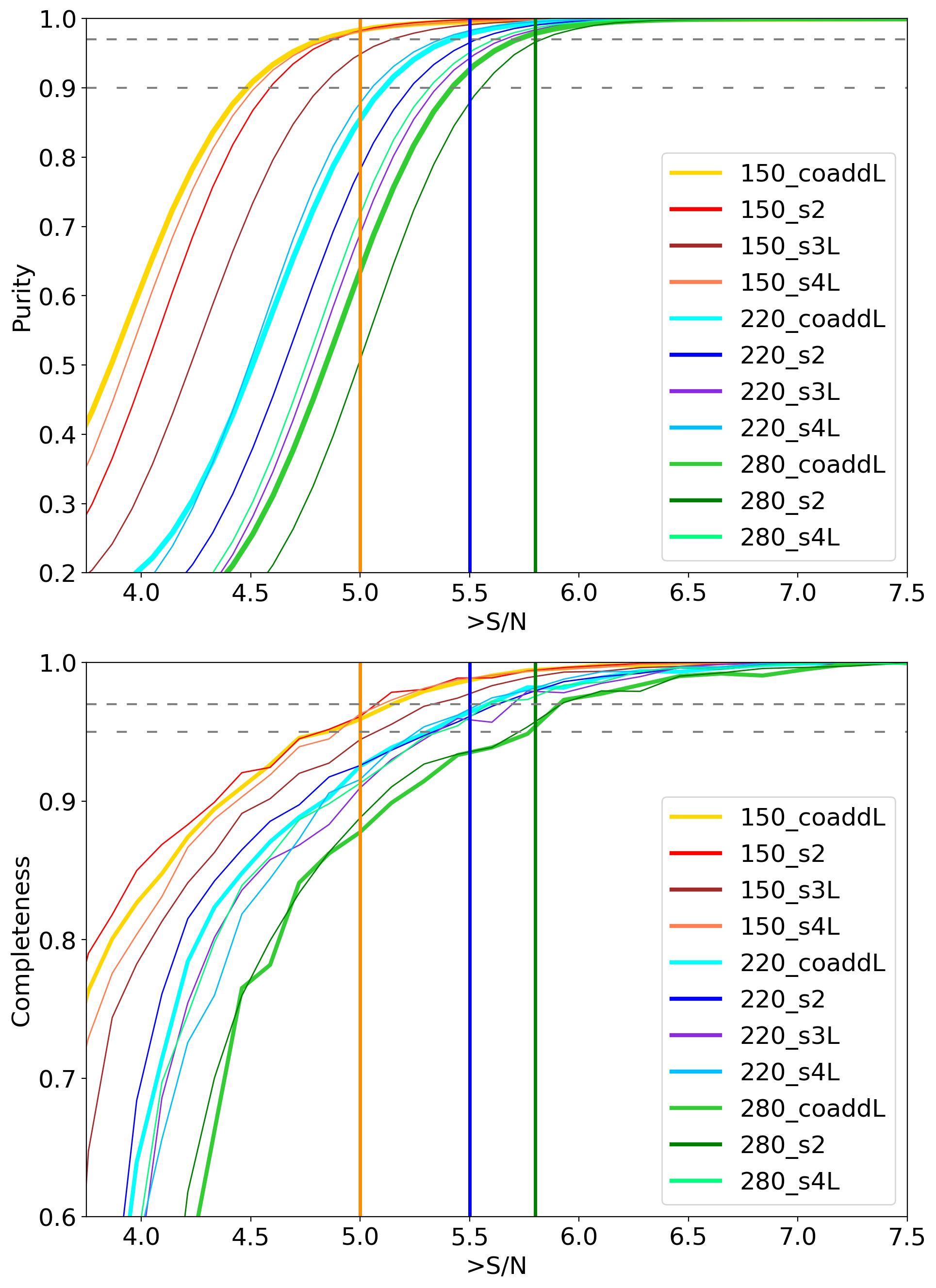}
  %\resizebox{\hsize}{!}{\includegraphics{Figures/pur_com.png}}
  \caption{Cumulative purity and completeness of the catalog at different $S/N$ cuts for different seasons. Top: Cumulative purity at different $S/N$ cuts for the 150, 220, and 280\,GHz frequency bands and different seasons, including the coadd maps. The criterion for inclusion of a source in the catalog for each map is taken as {$S/N_{150} \geq 5.0$, $S/N_{220} \geq 5.5$ and $S/N_{280} \geq 5.8$ at 150, 220 and 280\,GHz (orange, blue and green vertical lines, respectively)}. Bottom: Completeness of the catalog at different $S/N$ cuts for different seasons, including coadd maps. In this case, the same $S/N$ implies a lower completeness at higher frequencies. Note that different seasons behave similarly at a fixed frequency. The coadd maps generally have a greater purity compared to the single-frequency maps, which is expected.}
  \label{fig:pur_com}
\end{figure*}

To better understand the statistics noise and detection statistics, we simulated map realizations based on the background noise and the expected source population as follows. First, we used the noise model from Sect.~\ref{sec:matchedfilter} and calculated the 2D power spectrum. Then we randomly changed the phases of the power spectrum, multiplying it by Gaussian noise realizations in Fourier space with unitary amplitude. The result was a new realization of the same power spectrum. After that, we added sources following a \citet{Tucci2011} flux distribution. We normalized the distribution with a value of 0.0856 sources per square degree with fluxes between 50 and 5,000\,mJy, in agreement with current results (see Sect.~\ref{sec:number_counts}). {This resulted in around 220 sources injected on 160-square-degree patches (scaling linearly with patch size).\footnote{{The exact number depends on the lower flux cut. We used approximately three sigmas of the weighted filtered map noise as the lower flux limit. This cut is approximately 3, 4.5, and 9 mJy for the 150, 220, and 280 GHz coadd maps.}}} The exact value of the normalization is not important as long as the shape of the distribution follows the real distribution, that is, most sources lie in the low flux range, and only a few have high fluxes (over 1,000\,mJy). The sources are added in Fourier space, modeled as point spread functions of the desired flux density. The sources are placed at random locations inside the map.
    Finally, we run our source extraction pipeline on each of these simulated maps with thresholds of 50.0, 5.0 and 3.5 $S/N$. We repeat the process for each field at every frequency and season including coaddings, thus producing a set of 18 simulations.

For s2 maps, we make $10^3$ simulations, and for s3, s4 and coadded maps, we make $3 \times 10^3$ simulations on each patch, \texttt{L} and \texttt{R} (see Fig.~\ref{fig:footprints}). 
{This results} in roughly $10^6$ simulated source detections for the map areas in the study, allowing us to achieve good statistics.

For each simulation we produce a list of injected and recovered sources. 
We cross-match them with an association radius of $0.5'$, and we compute the difference in signal to noise between the injected and recovered sources; this is $\Delta_{\mathrm{S},i} = w(\mathbf{x}_i) (S_{\mathrm{R},i}-S_{\mathrm{I},i})$. 
We bin the results in $S/N$ and calculate the mean, variance, skewness and kurtosis of $\Delta_{\mathrm{S}}$ over each bin (see Fig.~\ref{fig:stats}). 
{From these,} we model the purity and completeness of the catalog, as well as other systematic effects present in our method.

\subsubsection{Purity and completeness analysis}\label{sec:purity}

{The cumulative purity is defined as $1-N_\mathrm{fp}/N_\mathrm{R}$, where $N_\mathrm{fp}$ is the number of false positives (recovered features that are not related to an injected source) and $N_\mathrm{R}$ is the total number of sources recovered up to a given $S/N$}.
In Fig.~\ref{fig:pur_com} we show purity curves for different frequencies and seasons. 
The curves reach unity at high signal-to-noise cuts ($S/N \gtrsim 6$), but decrease rapidly at lower values {($S/N \lesssim 5$)}.
We also see that the lowest frequency band f150 has the greatest purity, arguably due to the quality of the map and the larger beam. %; this implies fewer beams in the same area of the sky than for f220 or f280 and thus fewer possible false positives. 
{Purity is model-dependent. For a fixed number of false positives ($N_\mathrm{fp}$), a higher number of real sources ($N_\mathrm{R}$) results in greater purity. Therefore, the lower number of sources in higher-frequency maps results in a lower purity.}
We also calculated the purity using the noise map ($T_\mathrm{other}$) to measure $N_\mathrm{fp}$ and compare it to the number of sources of the real catalog $N_\mathrm{R}$, and the result is similar to a few percent level, $\lesssim 5\%$. 

We also calculate the cumulative completeness, defined as the fraction of injected ($N_I$) and recovered ($N_R$) sources up to a signal-to-noise $S/N$. 
The cumulative completenesses in Fig.~\ref{fig:pur_com} also shows better behavior at lower frequencies. 
Moreover, to correct number counts in Sect.~\ref{sec:number_counts}, we calculate the inverse of the differential completeness, i.e., $N_\mathrm{R} / N_\mathrm{I}$ binned by flux. 
We calculate this from each simulation and use the mean and the standard deviation in each bin as the correction factor and the uncertainty, respectively.

{Based on the completeness and purity of the maps depicted in Fig.~\ref{fig:pur_com}, we have selected conservative $S/N$ thresholds of 5.0, 5.5, and 5.8 for frequencies f150, f220, and f280, respectively. This selection ensures that the purity of the data remains consistently high, with an estimated purity level exceeding 97\% across all fields.}

\subsubsection{Boosting of the flux measurement}

An important systematic effect is the bias of the estimator. 
In Fig.~\ref{fig:stats}, we observe that fluxes tend to be overestimated in the lower $S/N$ bins, converging to the input values at higher bins. 
This is a consequence of the flux and centroid determination process since the algorithm searches for a positive peak while correcting its position.  
This favors regions of positive noise over negative, thus boosting the derived flux density. We refer to this as flux boosting. 
We estimate it as the mean of $\Delta_\mathrm{S}$ per $S/N$ bin, as shown in Fig.~\ref{fig:stats}.
%We use the results shown in Fig.~\ref{fig:stats}  to estimate the flux boosting for each source based on its signal-to-noise level.
As this correction is relevant for individual fluxes and also in the context of statistical studies such as number count measurements, we provide it as a separate parameter in our catalog. 
{We also compute how flux boosting depends on flux density, as opposed to $S/N$, which we use specifically in the number counts analysis presented in Sect.~\ref{sec:number_counts}.}

\subsubsection{Flux uncertainty from simulations}

We use the simulations is to estimate the uncertainty in the flux measurements, complementing the noise estimate presented in Eq.~\ref{eq:sigma}.
For this we use the standard deviation of $\Delta_\mathrm{S}$ per $S/N$ bin, $\sigma_\mathrm{db}$.
%This is done obtaining the standard deviation of the difference between recovered and injected flux, weighted and computed per $S/N$ bin, which we define as %$\sigma_\mathrm{db}^2 = \Sum (S_\mathrm{rec}-S_\mathbf{inj})^2 w(x)^2 / N$.
%\begin{equation}
%    \sigma_\mathrm{db} = \left[ \sum_{i = 1}^{N_\mathrm{bin}} (S_\mathrm{R,i}-S_\mathrm{I,i})^2 w(\mathbf{x}_i)^2 / N_\mathrm{bin} \right]^{1/2}.
%\end{equation}
Both of these statistics are useful and are included in the individual catalogs. However, we believe that the latter, which comes from simulations, is a more conservative representation of flux error. 
This is because the simulations incorporate all of the effects in the process, such as the Fourier interpolation of the source\footnote{Mathematically, the variance of the map should be preserved if we only add zeros outside the original Fourier Space, but finite map size and discontinuities can have an impact on the zero-padded map and make the variance higher. We check this with different sub-maps, and the result is always a higher standard deviation in the re-pixelated map} and effects of the filter itself such as ringing around bright sources (as explained in Sect.~\ref{sec:flux}). 
Since, typically, the first calculation does not include these effects, it underestimates the noise levels. 

In Fig.~\ref{fig:stats} we observe that the flux uncertainty from the simulations is typically above the pixel noise $\sigma_0$ {(the dashed line)}, and grows at higher $S/N$. 
This is indicative of errors dominated by {systematic effects}, which increase proportionally with the source flux. 
Also, the difference between $\sigma_0$ and $\sigma_\mathrm{db}$ is lower at 150\,GHz compared to 220\,GHz and 280\,GHz. 
We report values for both standard deviations for each detected source. 

We also calculate the kurtosis and skewness of each $S/N$ bin.  Non-zero values for either would indicate departures from the Gaussian nature of the errors. As their deviation from zero is $\lesssim 0.4$, we expect each bin distribution to be Gaussian. 
To test this assumption we perform \citealt{DAgostino1973} test of normality, which yields p-values$< 0.05$ for $S/N \gtrsim 5$. At $S/N \lesssim 5$, however, the normality tests begin to fail mostly due to negative skewness. Hence, our Gaussian assumption holds for most bins, but is an approximation as $S/N$ approaches 5.

\section{ACT southern catalog}\label{sec:act_catalog}

The main result from this work is a multi-frequency, multi-epoch catalog for the entire 600 square degrees of the survey area, containing {483 sources} detected at 150, 220, or 280\,GHz. The process for making this catalog is the following:

\begin{enumerate}
    \item Point sources are detected in each individual map. We select sources that lie above $S/N = 5.0, 5.5, 5.8$ for 150, 220 and 280 GHz respectively. We consider as valid all sources with at least one detection in one of the maps.
    \item We cross-match sources from all maps with an association radius of 0.5', making a data cube.
    \item We fill up the data cube by forcing a flux measurement at positions corresponding to valid detections in other maps (other seasons or bands), using the {$S/N$ weighted average} of the centroids from valid measurements of the same source.
    \item We make the distinction between valid and forced measures using a flag.
    \item We {assume} that forced measures are not susceptible to flux boosting because their centroids are fixed; thus we do not include boosting statistics for them.\footnote{{By definition, the filtered map is a Gaussian noise field with signals. When we already know the position of the source and measure it on a different filtered map, the result is an unbiased measurement of the source.}}
\end{enumerate}

We name each source using the International Astronomical Union (IAU) designation, based on the coordinates of the highest signal to noise detection. 
We include a single position corresponding to the average centroid (the same used for forced photometry). 
According to best practices for data publication in the astronomical literature \citep{Chen2022}, we round up to arcseconds for the naming of sources (which is the order of the positional accuracy, see Sect.~\ref{sec:astrometry}). 
And we retain the naming of sources previously found in \citet{Marsden2014}. We matched our catalog to the earlier one with an {association radius of $1\arcmin$}. 
There are no ambiguous matches, and the separation between sources matched in the two catalogs is roughly less than $20\arcsec$, except for the dusty source ACT-S\,J033134-515355 which has a separation of $32\arcsec$.  
We confirm the dusty nature of this galaxy at a high significance ($S/N = 14.5$) in the 280\,GHz-2008 map with a flux of $76.9 \pm 5.6$\,mJy. {We do not find a match for 11 of the 191 sources from \citealt{Marsden2014} (6\%); 8 of them are synchrotron type and 3 are dusty, all of them are close to their cut limit ($S/N \gtrsim 5$) at either 150 or 220 GHz.}

For each frequency, there are up to four measurements (2008, 2009, 2010, and the coadd), depending on whether the source falls inside the corresponding map. 
For example, sources in the \verb|coaddL| area have measurements in all seasons, whereas sources from \verb|coaddR| have measurements only in the 2009, 2010, and coadded maps.
{In Appendix~\ref{sec:cat_description}, we describe the entries in the catalog.}

 \begin{table}
 \centering
  \caption{Number of detected sources per frequency in each field and across all fields. Top table: ACT fields utilized in this study along with the corresponding counts of detected sources and the noise characteristics of the weighted filtered map (refer to Sect.~\ref{sec:flux}). {Bottom table: Number of sources identified at various significance levels, considering their presence in one, two, and three bands simultaneously.}}
  \begin{tabular}{lrcc}
    \hline
    Band  & Field   & Detections & $\sigma_0$\,(mJy)\\
    \hline
    f150   &  s2     & 172      & 2.04             \\
           &  s3L    & 36       & 1.53             \\
           &  s3R    & 48       & 1.91             \\
           &  s4L    & 72       & 1.47             \\
           &  s4R    & 49       & 1.82             \\
           &  coaddL & 95       & 1.03             \\
           &  coaddR & 65       & 1.41             \\
    \hline
    f220   &  s2     & 170      & 2.85             \\
           &  s3L    & 38       & 2.44             \\
           &  s3R    & 40       & 3.00             \\
           &  s4L    & 61       & 2.02             \\
           &  s4R    & 56       & 2.15             \\
           &  coaddL & 105      & 1.59             \\
           &  coaddR & 72       & 1.93             \\
    \hline
    f280   &  s2     & 142      & 4.98             \\
           &  s4L    & 60       & 3.15             \\
           &  s4R    & 38       & 4.00             \\
           &  coaddL & 91       & 2.88             \\
    \hline
    \hline
    Band          & $N_\mathrm{src} > 5 \sigma $ & $N_\mathrm{src} > 4 \sigma $ & $N_\mathrm{src} > 3 \sigma $ \\
    \hline
    f150          & 311                          & 327                          & 358                          \\
    f220          & 334                          & 372                          & 404                          \\
    f280          & 231                          & 281                          & 337                          \\
    f150 and f220 & 218                          & 264                          & 310                          \\
    f220 and f280 & 169                          & 228                          & 298                          \\
    f150 and f280 & 124                          & 168                          & 239                          \\
    Three bands   & 118                          & 163                          & 231                          \\
    \hline
  \end{tabular}
  \label{tab:Fields_results}
 \end{table}

\subsection{Spectral classification}\label{sec:specclass}
\label{sec:alpha}

By comparing fluxes from different frequency bands, we measure spectral indices for each source to help characterize their physical nature. 
We do this for all three frequency combinations, obtaining $\alpha_{150}^{220}$, $\alpha_{150}^{280}$, $\alpha_{220}^{280}$, respectively.
Since 280 GHz adds a new point, sources can display a falling (negative-negative), rising  (positive-positive), upturning (negative-positive), and peaking behavior (positive-negative). 
This can be seen in Fig. \ref{fig:color_color}.
Although the majority of sources are in the falling and rising quadrants, the objects in the peaking and upturning quadrants deserve a deeper explanation.
We found that, after removing known nearby galaxies from the sample (see Sect.~\ref{sec:external_associations}), most of the peaking objects go away, meaning that it is very likely that their excess flux at 220 GHz was driven by the CO(2-1) line.
Interestingly, another plausible explanation for this behavior could be very high redshift DSFGs ($z>5$) dominated by cold dust ($T \lesssim 30$ K).
Finally, it is also possible that they fell into that category due to their flux uncertainty. 
On the other hand, upturning spectra may imply a combination of synchrotron and thermal emission, spectral lines contaminating either the f150 or f280 bands, or misclassification due to flux uncertainties.
Yet, when the 150-280 slope is considered, we observed that most of the upturning sources maintain a decreasing spectral tendency toward higher frequencies, while the peaking objects follow a rising tendency. 

\subsection{Object classification}\label{sec:objclass}

To characterize and validate our sources, we first cross-checked with different catalogs. Then, we applied criteria on the spectral indices to classify the remaining galaxies.

\subsubsection{External counterparts}\label{sec:external_associations}

We spatially cross-matched the ACT catalog with the following nine public catalogs with spectral response ranging from radio to x-ray:

\begin{itemize}
\setlength\itemsep{0.2em}
    \item[--] The Sydney University Molonglo Sky Survey (SUMSS) at 843~MHz \citep{Mauch2003}

    \item[--] The Parkes-MIT-NRAO southern survey (PMN) at 4850\,MHz \citep{Wright1994}

    \item[--] The Australia Telescope 20\,GHz survey (AT20G) at 20\,GHz \citep{Murphy2010}

    \item[--] The South Pole Telescope Sunyaev-Zel'dovich (SPT-SZ) survey point source catalog at 95, 150 and 220~GHz \citep{Everett2020}

    \item[--] {The Second Planck Catalogue of Compact Sources (PCCS2) at 30, 44, 70, 100, 143, 217, 353, 545 and 857~GHz  (1\,cm to 350\,$\mu$m) \citep{Planck_Sources}}
    
    \item[--] The Planck Multi-frequency Catalog of Non-thermal Sources (PCNT) at 30, 44, 70, 100, 143, 217, 353, 545 and 857~GHz (1\,cm to 350\,$\mu$m) \citep{PCNT2018}

    \item[--] The Infrared Astronomical Satellite (IRAS) Faint Source Catalog at 60 and 100 $\mu$m \citep{Moshir1990}

    \item[--] The Wide Field Infrared Explorer (WISE) source catalog at 22 $\mu$m \citep{Wright2010}

    \item[--] The AKARI/FIS All-Sky Survey Bright Source catalog at 65, 90, 140, and 160\,$\mu$m \citep{Yamamura2010}

    \item[--] The AKARI/IRC All-Sky Survey Point Source catalog at 9 and 18\,$\mu$m \citep{Ishihara2010}

    \item[--] The Second ROSAT All-Sky Survey Point Source catalog (2RXS) at X-ray energies 0.1-2.4keV \citep{Boller2016}

\end{itemize}

We used association radii of 0.5, 1.0, 1.5, or 2.5 arcminutes in order to ensure a low probability of a random association ($P_{\mathrm{random}} \lesssim 1\%$).  The choice of association radius depended on the number of sources per square degree $\Sigma_{\mathrm{src}}$ and the area of association $A_{\mathrm{assoc}}$ as 

\begin{equation}
P_{\mathrm{random}} = 1 - \exp\left({-\Sigma_{\mathrm{src}} A_{\mathrm{assoc}}}\right).
\label{eq:P_random}
\end{equation}

We calculated the cross-matches and $\Sigma_{\mathrm{src}}$ for each catalog in the area where it overlaps with the ACT survey areas.
Table~\ref{tab:external_associations} shows a summary of the results from external associations. Most matches come from either low-frequency experiments such as SUMSS, PMN, AT20G, or catalogs constructed at frequencies similar to ACT, such as SPT. 
At higher frequencies, WISE has the most counterparts, mainly because of the high number of sources per unit area, as well as its good sensitivity.

{We also queried a catalog search on the \textit{Herschel} archive using the SPIRE 250, 350, and 500 and PACS 70, 100, and 160 micron catalogs \citep{SPIRE,PACS}, and we used the ALMA archive via ALminer \citep{ALminer2023} to find ALMA cross-matches. For Herschel, using a radius of $1\arcmin$, we found 40 cross-matches, of which 26 are targeted observations of SPT sources. Interestingly Herschel also has the AKARI Deep Field South (ADFS) survey coincident with our footprint, we found 15 Herschel sources within ADFS. For ALMA we found 51 matches, 32 are synchrotron sources which are usually used as calibrators, 19 are dusty sources of which 17 are SPT DSFGs. We also found 3 stars beta pictoris, V Tel, and R Hor, which are coincident to IRAS sources.} 

{Fig.~\ref{fig:color_color} shows the spectral indices $\alpha_{150}^{220}$, $\alpha_{220}^{280}$ and $\alpha_{150}^{280}$ along with a histogram of the spectral indices and Gaussian fits. We note that these fits differ slightly from those mentioned above due to the applied cut. We also overlay the cross-matches with SPT, SUMSS, IRAS, ROSAT, AT20G, \textit{Herschel} and ALMA. As expected, low-frequency experiments such as SUMSS and AT20G, and the X-ray catalog from ROSAT detect mostly synchrotron sources, whereas higher-frequency experiments such as IRAS find dusty ones. Of the total of 284 synchrotron sources, 251 have counterparts (88\%), and of the 184 dusty sources, 95 have counterparts (51\%) in any cross-matched catalog. This suggests that 90 dusty candidates are new detections.}

{From NED matches, we cataloged sources as nearby if they have a $z<0.15$. The known AGNs were classified as AGN, while the sample from SPT at high redshift was classified as DSFG. }

\subsubsection{Object classification using $\alpha$}

\begin{figure*}
\sidecaption
\includegraphics[width=12cm]{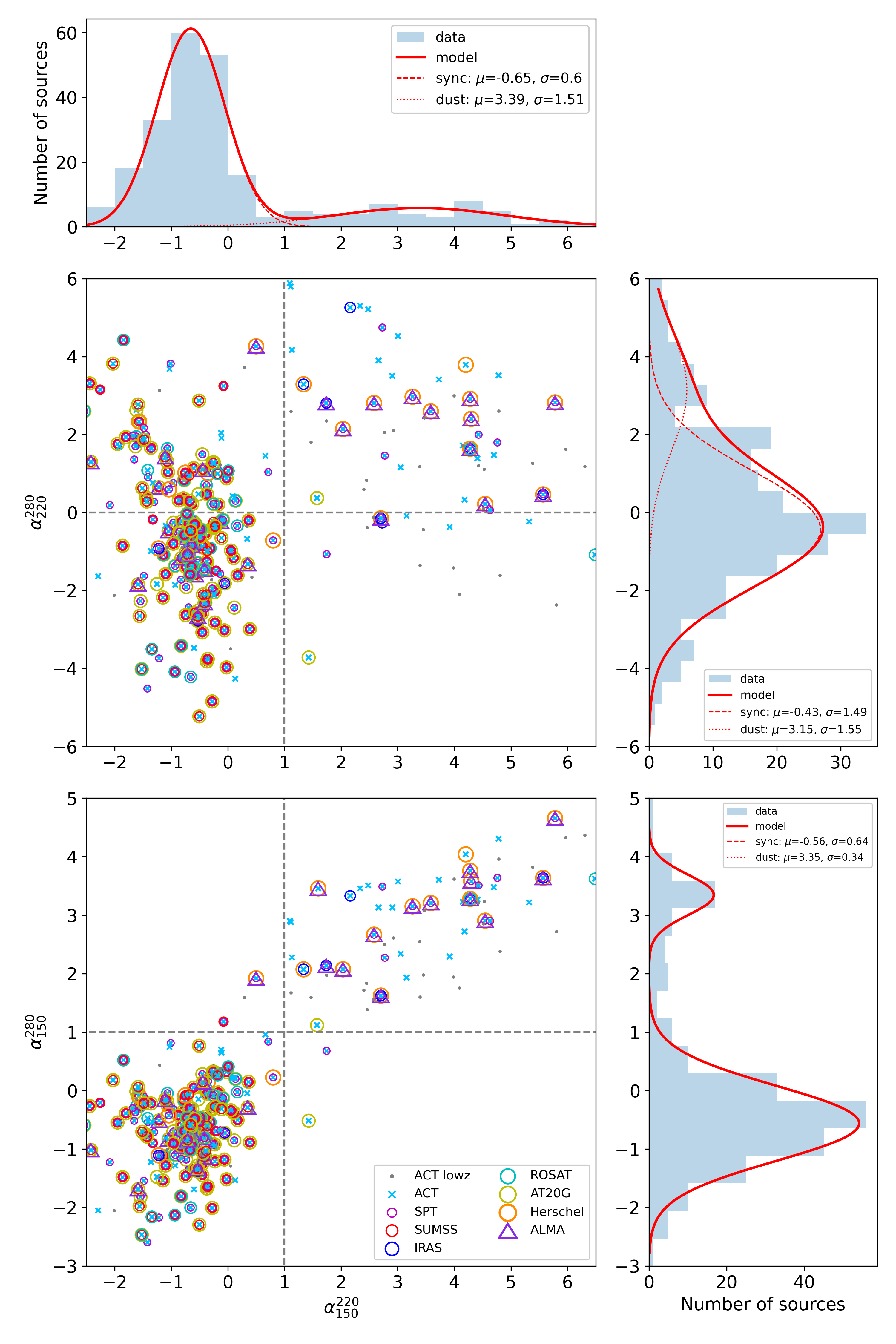}
     \caption{{Color-color diagram of sources, including cross-matches and spectral index histograms. {Here, we display sources with either ($S/N_{150}>5$, $S/N_{220}>3$ and $S/N_{280}>1$) or ($S/N_{150}>1$, $S/N_{220}>3$ and $S/N_{280}>5$)}. The populations of synchrotron (at left) and dusty (at right) sources are visibly separated in the 150--220\,GHz spectral index. A vertical line at $\alpha_{150}^{220} = 1.0$ shows our separation criterion. Counterparts from external catalogs are shown as circles. SPT has detected sources in both populations, and most synchrotron sources have counterparts in SUMSS, AT20G, and ROSAT; dusty sources are more likely to have only counterparts in SPT or IRAS. At top left and bottom right we show the histogram of $\alpha_{150}^{220}$ and $\alpha_{220}^{280}$, respectively. In the case of $\alpha_{150}^{220}$, both populations are clearly separated; we show the double Gaussian fit of separated components and the sum. In the case of $\alpha_{220}^{280}$, both populations are not clearly separated, and there is no clear local minima in this case.}}
     \label{fig:color_color}
\end{figure*}

{After removing nearby galaxies, a histogram of the spectral indices between 150 and 220 GHz shows a bimodal distribution, with two distinct populations of sources (See top histogram of Fig.~\ref{fig:color_color}).
We expect sources with flat or negative index: $\alpha_{150}^{220} \lesssim 1$, must be dominated by synchrotron radiation, most likely corresponding to AGNs.
On the other hand, sources with positive index ($\alpha_{150}^{220} \gtrsim 1$) are expected to be dominated by thermal emission from dust, so they may correspond to DSFG candidates.
We repeated the histogram using 150-280 and recovered a bimodal distribution with a clear separation between synchrotron and thermal-dominated sources. However, when we do this on 220-280, the populations tend to mix, making it harder to distinguish between them.}

{To distinguish between these populations, we fitted the distribution $\alpha_{150}^{220}$ with a double Gaussian function. 
This resulted in two components, one with a mean of $\mu = -0.65$ and a standard deviation of $\sigma = 0.60$, which we identify with synchrotron emission, and the other with a mean of $\mu = 3.39$ and a standard deviation of $\sigma = 1.51$, which we identify with dust emission. 
The two Gaussian curves intersect at approximately $\alpha = 1.0$, and the minimum of the double Gaussian model occurs also at roughly $\alpha=1.0$.
These parameters are in agreement with previous works, for instance \citealt{Gralla2020} found 
$\mu = -0.66$ and $\sigma = 1.2$ for synchrotron, and $\mu = 3.7$ and $\sigma = 1.8$ for dust, whereas \citealt{Everett2020} found $\mu= -0.6$ and $\sigma=0.6$ for synchrotron, and $\mu = 3.4$ and $\sigma = 0.8$ for dust.}

{Repeating this process for $\alpha_{150}^{280}$ results in a component with $\mu = -0.58$, $\sigma = 0.64$ and the other with $\mu = 3.35$, $\sigma = 0.34$. Both Gaussian curves are clearly separated and they do not intersect, the combined model reaches the minimum at a value between 1 and 2. The mean spectral index for synchrotron at 150--280 of -0.65 is slightly lower than 150--220 of -0.58, the uncertainty on the values are 0.03 on each case, thus the significance is around $2\sigma$. This flattening could be due to physical components in AGNs, such as dust.}

{The case of spectral indices between 220 and 280\,GHz is more complicated due to the higher uncertainties in fluxes, and the fact that both frequencies are closer to each other. In this case the fit results in a component at $\mu = -0.43$, $\sigma = 1.49$ and other component with $\mu = 3.15$ and $\sigma = 1.55$. The mean spectral index for synchrotron at 220--280 GHz is also lower than 150--220 GHz, but the difference is not statistically significant since the uncertainty in the fit is around 0.5}.

\begin{figure}
  \resizebox{\hsize}{!}{\includegraphics{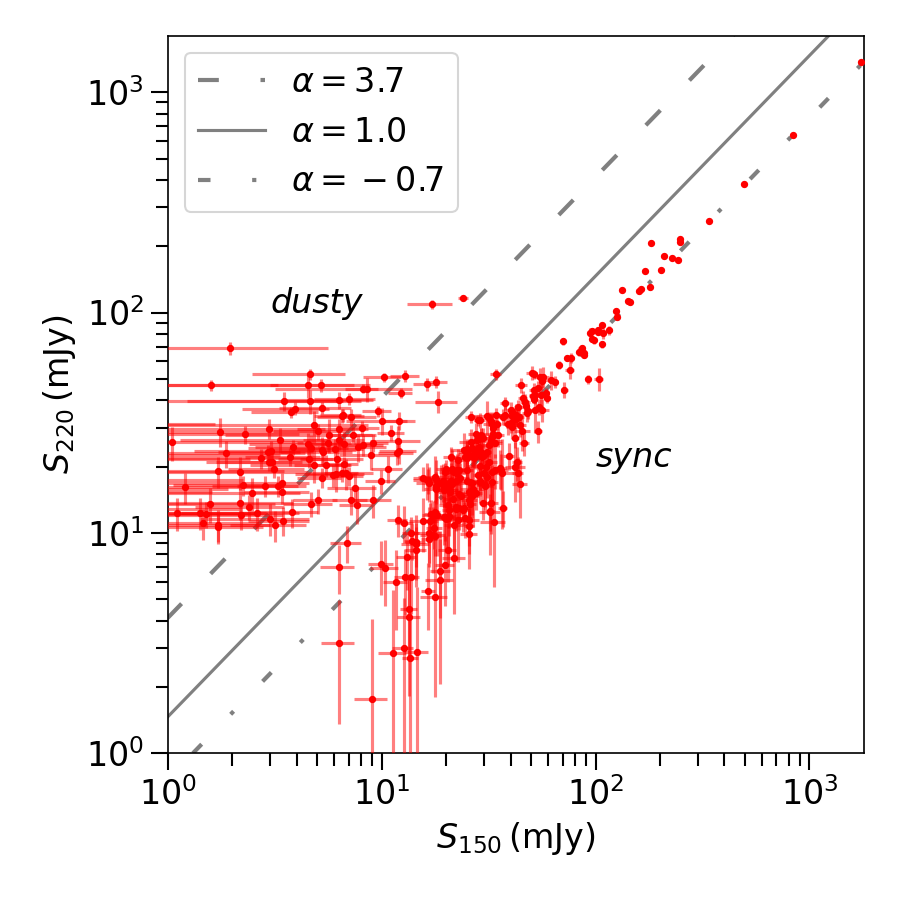}}
  \caption{Comparison of f150 flux density against f220 flux density. Sources with at least one detection in either of the two bands are shown as red circles. The population at the right that extends to high flux densities is composed of synchrotron dominated sources; the fainter population to the left, where f220 emission is stronger, is composed of dusty sources. The solid line is an spectral index of $1.3$ which is used to separate the populations. The dotted and dashed lines indicate the mean spectral indices of the whole sample for synchrotron and dusty sources of $-0.7$ and $3.8$ respectively.}
  \label{fig:f150_f220}
\end{figure}

\begin{figure}
  \resizebox{\hsize}{!}{\includegraphics{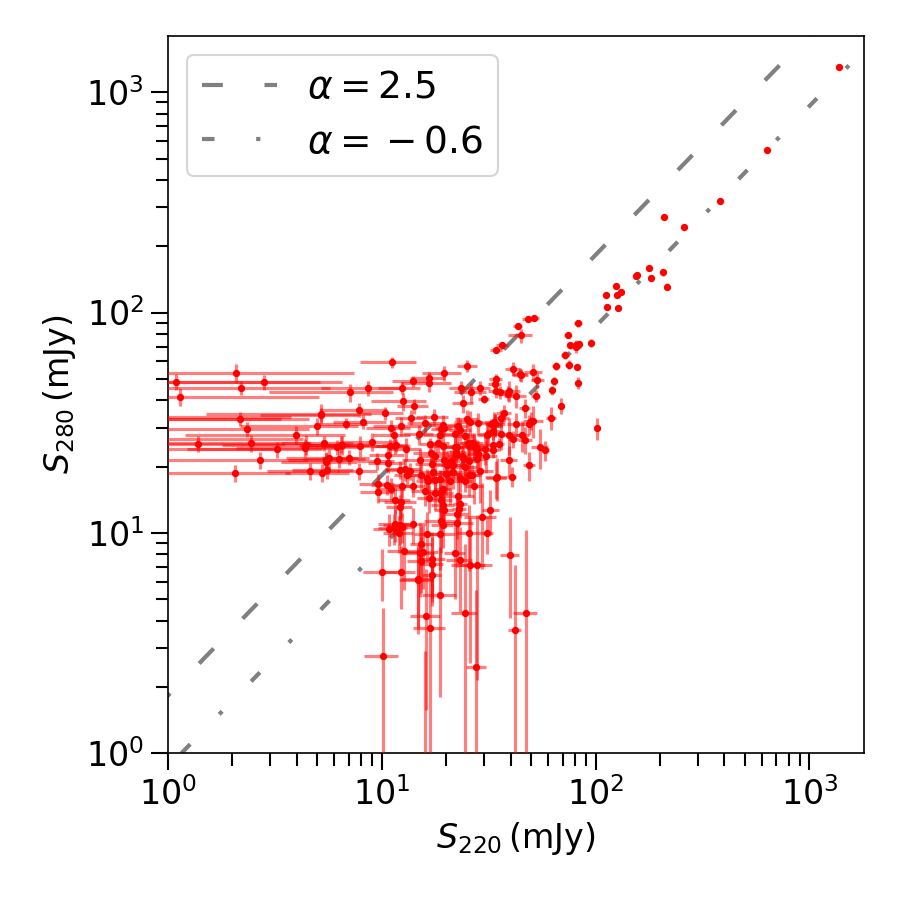}}
  \caption{Comparison of f220 flux density against f280 flux density. Sources with at least one detection in either of these two bands are shown as red circles. Here the populations are not clearly separated. The dotted and dashed lines indicate spectral indices of $-0.7$ and $2.5$, which are fits for bright synchrotron and dusty sources.  We note that the dusty source spectral index in this case, is lower compared to that between f150 and f220. This indicates a departure from a constant index, suggesting curvature in the spectral energy distribution.}
  \label{fig:f220_f280}
\end{figure}

Figures~\ref{fig:f150_f220} and \ref{fig:f220_f280} show individual fluxes in frequency pairs of 150--220 and 220--280\,GHz, respectively, with overlaid spectral indices from the Gaussian fits. The sources group around the spectral indices, with some scatter at the lower fluxes. The flux span of the full catalog ranges from tens to thousands of mJy. 
At 150--220\,GHz, a spectral index of $\alpha = 1.0$ separates the two populations. Below 1.0, the synchrotron-dominated population extends from the lowest fluxes to thousands of mJy. Above 1.0, the dusty population is only noticeable roughly below 20\,mJy at 150 GHz and below 70\,mJy at 220\,GHz. This is expected since the observations are in the Rayleigh-Jeans spectral regime of the observed dust emission, and highlights the importance of the 220\,GHz band for detecting dusty sources.
At 220--280\,GHz, there is no clear separation between the two populations, but {a visual inspection shows that the dusty population scatters around a spectral index of 2.5, and the highest dusty sources fluxes are around 100\,mJy at 280 GHz, which is higher compared to their highest flux on 220\,GHz, which are around 70\,mJy. The latter suggests that having higher frequency channels is also helpful for finding dusty sources.}

{To complete the classification of the remaining objects we use different spectral indices, if a source has both $\alpha_{150}^{220}>1$ and $\alpha_{150}^{280}>1$ is classified as DSFG, on the other hand if it has $\alpha_{150}^{220}<1$ and $\alpha_{150}^{280}<1$ is classified as AGN. The remaining sources that either do not have a consistent positive or negative pair tendency or are missing an index due to the lack of a measurement in any band are categorized using the index that has the highest pair of $S/N$ or the only available index with the convention $\alpha>1$ DSFG and $\alpha<1$ AGN. We verified that this classification is consistent when comparing the SPT DSFG sample; all of their sources were identified as DSFGs with this algorithm. On the other hand, all known radio sources, such as those from the PKS catalog, are classified as AGNs. From the classification algorithm, there are in total 284 AGNs, 140 DSFGs, 43 nearby galaxies, and 16 unclassified objects.}

%\begin{figure*}
%\sidecaption
%   \includegraphics[width=12cm]{Figures/alpha12_alpha23.png}
%     \caption{{Color-color diagram of sources, including cross-matches and spectral index histograms. Here, we select sources with either $S/N_{150}>3$ and $S/N_{220}>3$ or $S/N_{220}>3$ and $S/N_{280}>3$. The populations of synchrotron (at left) and dusty (at right) sources are visibly separated in the 150--220\,GHz spectral index. A vertical line at $\alpha_{150}^{220} = 1.3$ shows our separation criterion. Counterparts from external catalogs are shown as circles. SPT has detected sources in both populations, and most synchrotron sources have counterparts in SUMSS, AT20G, and ROSAT; dusty sources are more likely to have only counterparts in SPT or IRAS. At top left and bottom right we show the histogram of $\alpha_{150}^{220}$ and $\alpha_{220}^{280}$, respectively. In the case of $\alpha_{150}^{220}$, both populations are clearly separated; we show the double Gaussian fit of separated components and the sum. In the case of $\alpha_{220}^{280}$, both populations are not clearly separated, and there is no clear local minima in this case.}}
%     \label{fig:color_color}
%\end{figure*}

\begin{table*}
    \centering
    \caption{Summary of external catalogs and the cross-matching with ACT sources. The first column lists the catalog survey name, followed by the number of counterparts found, the band frequency or wavelength of the survey, the number of sources per square degree within the ACT fields, the radius of association used, and the probability that a given association is random.} 
    \begin{tabular}{lrrrccc}
        \hline
         Survey     & Matches & Band                      & Beam FWHM             & $\Sigma_\mathrm{src}$(src. per $\mathrm{deg}^2$) & $r_\mathrm{assoc}$ (arcmin) & $P_\mathrm{random}$(\%)\\
        \hline
        (This work)  &   {483}        &  150, 220, 280\,GHz        & 1.4, {1.0}, 0.9\,arcmin  &          {0.82}           &  {...}   &  {...}  \\
        %(this work)&           &  (2.0, 1.4, 1.1~mm)       &                        &                           &                    &                        \\
        SUMSS       &    156    &  843~MHz (36~cm)          & 45\,arcsec             &           30.37           &         0.5        &          0.66          \\
        PMN         &    214    &  4840~Mhz (6~cm)          & 4.2\,arcmin            &            2.14           &         2.5        &          1.16          \\
        AT20G       &    160    & 20~GHz (1.5~cm)           & 4.6\,arcsec            &            0.39           &         1.0        &          0.03          \\
        SPT-SZ      &    237    & 95,150,220~GHz            & 1.0-1.7\,arcmin        &            1.44           &         1.0        &          0.13          \\
        %           &           & (3.2, 2.0, 1.4~mm)        &                        &                           &                    &                        \\
        PCNT        &    30     & 30-857~GHz                & 4.3-32.2\,arcmin       &            0.66           &         1.5        &          0.13          \\
        IRAS        &    56     & 12, 25, 60, 100~$\mu$m    & 11-88\,arcsec          &            5.84           &         1.5        &          1.14          \\
        WISE        &   114     & 3.4, 4.6, 12, 22~$\mu$m   & 6.1-12\,arcsec         &           55.24           &         0.5        &          1.20          \\
        AKARI/FIS   &    54     & 65, 90, 140, 160~$\mu$m   & 24-59\,arcsec          &            1.43           &         1.5        &          0.28          \\
        AKARI/IRC   &    16     & 9, 18~$\mu$m              & 3.3-6.6\,arcsec        &            9.54           &         0.5        &          0.21          \\
        2RXS        &    62     & 0.1-2.4~keV               &  5\,arcsec             &            3.93           &         1.5        &          0.77          \\
        \hline
    \end{tabular}
    \label{tab:external_associations}
\end{table*}

\subsection{Astrometry}\label{sec:astrometry}

Due to its low source density and high positional accuracy, the AT20G catalog is an excellent tool to test the astrometry of our sources. 
We use the external associations from Sect.~\ref{sec:external_associations} to calculate the offset in right ascension and declination between the two catalogs. 
If we make matches with an association radius of $1.0'$ and ${S/N}>20$ (using the highest signal to noise available), we find 66 counterparts. 

The result is a mean offset of $0.63\arcsec$ and $0.31\arcsec$ in right ascension and declination, respectively, with standard deviations of $4.7\arcsec$ and $3.5\arcsec$.
This result is similar to results obtained in the equatorial catalogs, where positions were compared to those from the Faint Images of the Radio Sky at Twenty-cm survey (FIRST). \citet{Gralla2020} reported standard deviations of $2\arcsec$ in right ascension and declination for a sample of $S/N_{150} > 16$ and $5\arcsec$ for ${S/N}_{150} > 5$; furthermore \citet{Datta2019} reported standard deviations of $1.55\arcsec$ and $1.98\arcsec$ in right ascension and declination respectively for a sample with flux density higher than 50\,mJy.

\subsection{Comparison with SPT}\label{sec:comparison}

\begin{figure*}
\centering
   \includegraphics[width=17cm]{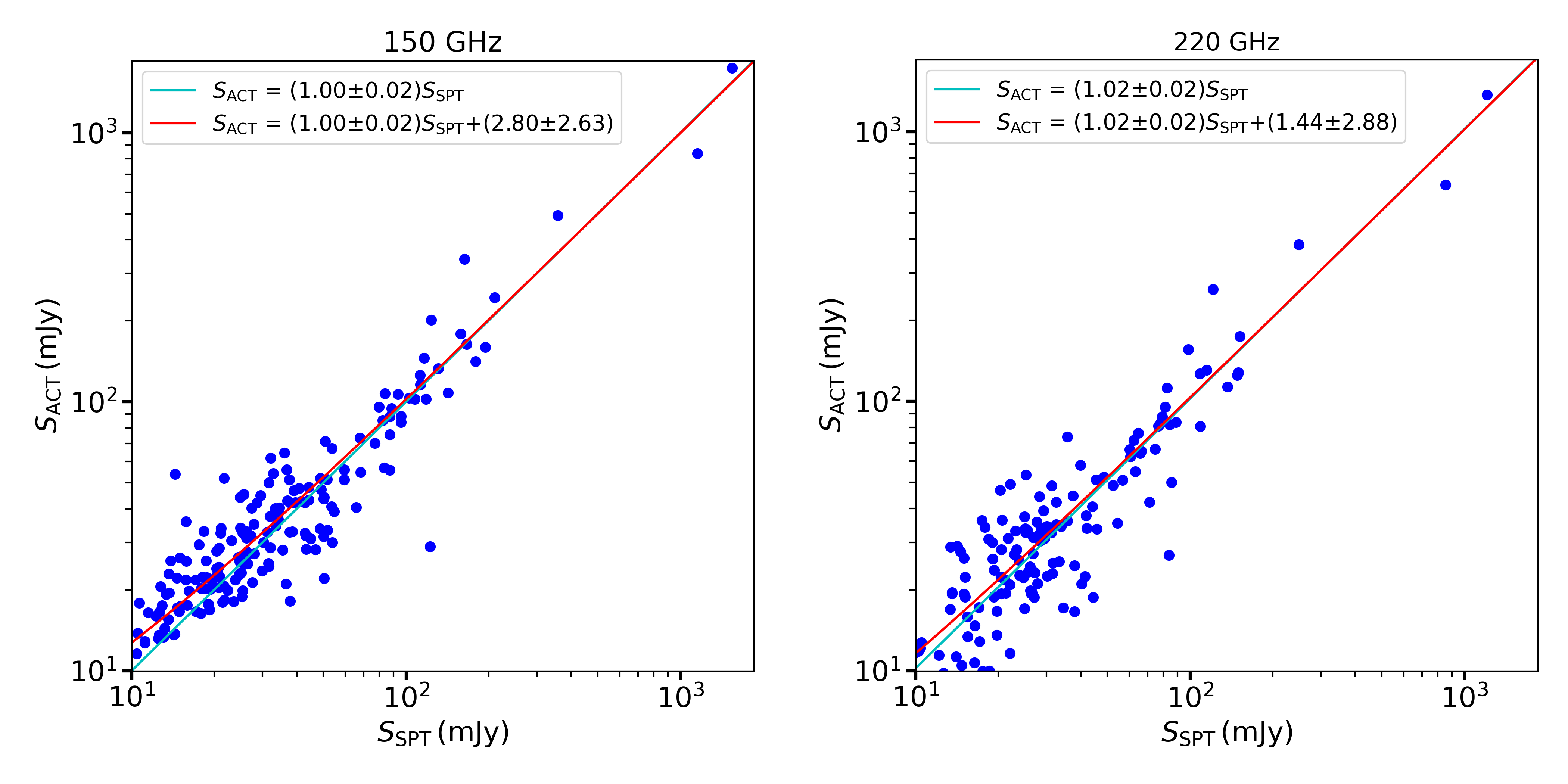}
     \caption{Comparison of ACT and SPT synchrotron source flux measurements at 150 and 220 GHz. Two types of linear fits are shown: a forced fit with a zero intercept (cyan) and a fit with a free intercept (red). The fits show agreement at the percent level within uncertainties, but there is significant scatter for individual sources. This scatter is likely due to flux variability, as it is similar to the level of scatter seen from year-to-year variations (Sect.~\ref{sec:variabilityanalysis}).}
     \label{fig:ACT_SPT}
\end{figure*}

{We can directly compare our fluxes with those measured by the SPT \citep{Everett2020}, which has significant overlap with ACT in both sky coverage and frequencies (150 GHz and 220 GHz bands).
We select synchrotron sources with a $S/N$ greater than or equal to 3 in each frequency, resulting in 196 and 140 matches for 150 and 220\,GHz, respectively. We then convert their fluxes from SPT effective frequencies of 154.5\,GHz and 220.4\,GHz to our effective frequencies of 148.75\,GHz and 218.85\,GHz using a spectral index of $-$0.7. This yields conversion factors of 1.026 and 1.005, respectively.
Figure~\ref{fig:ACT_SPT} shows a comparison of the ACT and SPT flux measurements. We perform linear fits to the data and find good correlation between the two experiments. At 150\,GHz, a forced fit with a zero intercept gives a slope of $1.00 \pm 0.02$, and a fit with a free intercept gives a slope of $1.00 \pm 0.02$ and an intercept of $2.68 \pm 2.44$. At 220\,GHz, the forced fit with a zero intercept gives a slope of $1.02 \pm 0.02$, and the fit with a free intercept gives a slope of $1.02 \pm 0.02$ and an intercept of $1.44 \pm 2.88$.
The individual flux densities for both frequencies analyzed show significant scatter (30\%). This scatter can be explained by source variability since the two experiments observed the sky in different years. We explore year-to-year variability in Sect.~\ref{sec:variabilityanalysis}.}

\section{Source analysis}\label{sec:source_analysis}

\subsection{Source variability}\label{sec:variabilityanalysis}

Our source catalog is slightly dominated by sources with synchrotron emission, presumably AGNs. In many cases, such sources are variable. 
Here we compare flux measurements from pairs of seasons, 2008--2009, 2009--2010 and 2008--2010 to identify sources with variability. Thus we determine the level of annual variability over a 3 year period. We used a measure of the variability defined as 

\begin{equation}
    \sigma_\textrm{var-12} = \frac{\left|S_\textrm{1}-S_\textrm{2}\right|}{\sigma_\textrm{tot}},
\end{equation}

\noindent where $\sigma_\textrm{tot} = \sqrt{\sigma_\textrm{1}^2 +\sigma_\textrm{2}^2}$ is the combined uncertainty.
This quantity measures the variation of fluxes normalized by the measurement uncertainty.

\begin{figure*}
\centering
   \includegraphics[width=17cm]{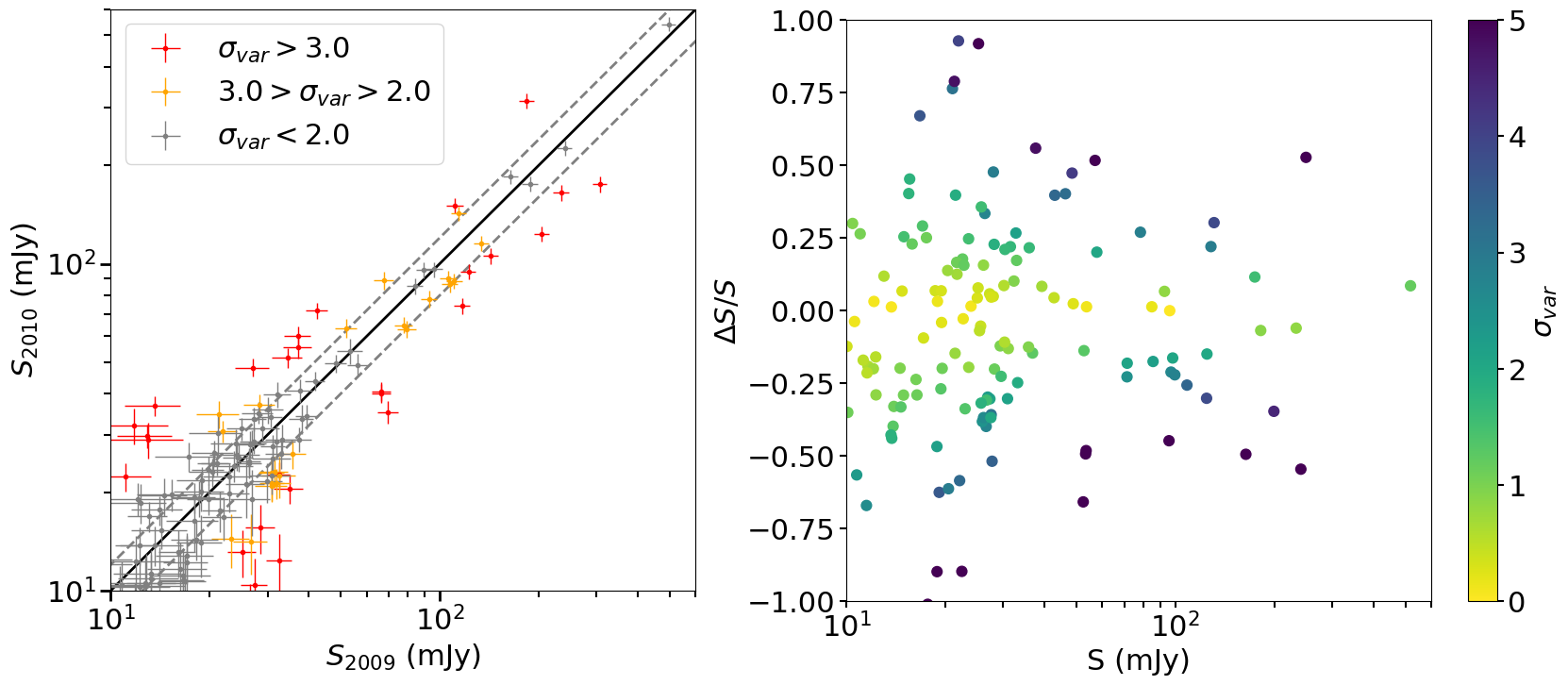}
   \caption{Variability between seasons 3 and 4 at 150~GHz. Top: Flux densities from 2009 versus 2010. The straight line marks the 1:1 relation while the dotted lines deviate 20\% from it.  Red crosses indicate highly variable sources with $\sigma_\textrm{var}>3.0$; orange crosses are mildly variable sources with $3.0 > \sigma_\textrm{var} > 2.0$; and gray crosses are sources that we consider non-variable. Bottom: Fractional variation of flux density versus the mean flux density; colors represent $\sigma_\textrm{var}$. {Other possible combinations, such as 2008-2010 or 2008-2009, yield similar results.}}
   \label{fig:variability}
\end{figure*}

We select sources as variable if they have, in at least one pair of seasons, $\sigma_\textrm{var} > 3.0$. Figure~\ref{fig:variability} shows the source variability between seasons 2009 and 2010. At the high flux density end, we find several sources with high significance variations. At the low flux density end, the variability, if any, is comparable to the uncertainty in the flux.

{The standard deviation of the fractional variability is $\mathrm{std}(\Delta S/ S) = 0.35 \pm 0.02$, implying that active galaxies vary their flux by about $35\pm2$\% year to year. Here we assume an uncertainty of $\mathrm{std}(\Delta S/ S) / \sqrt{2N-2}$, where $N$ is the number of available pairs of measurements. Results are similar for the 2008--2009 or 2009--2010 seasons, which yield $0.33 \pm 0.03$ and $0.39 \pm 0.03$, respectively.}

\begin{figure*}
\centering
   \includegraphics[width=17cm]{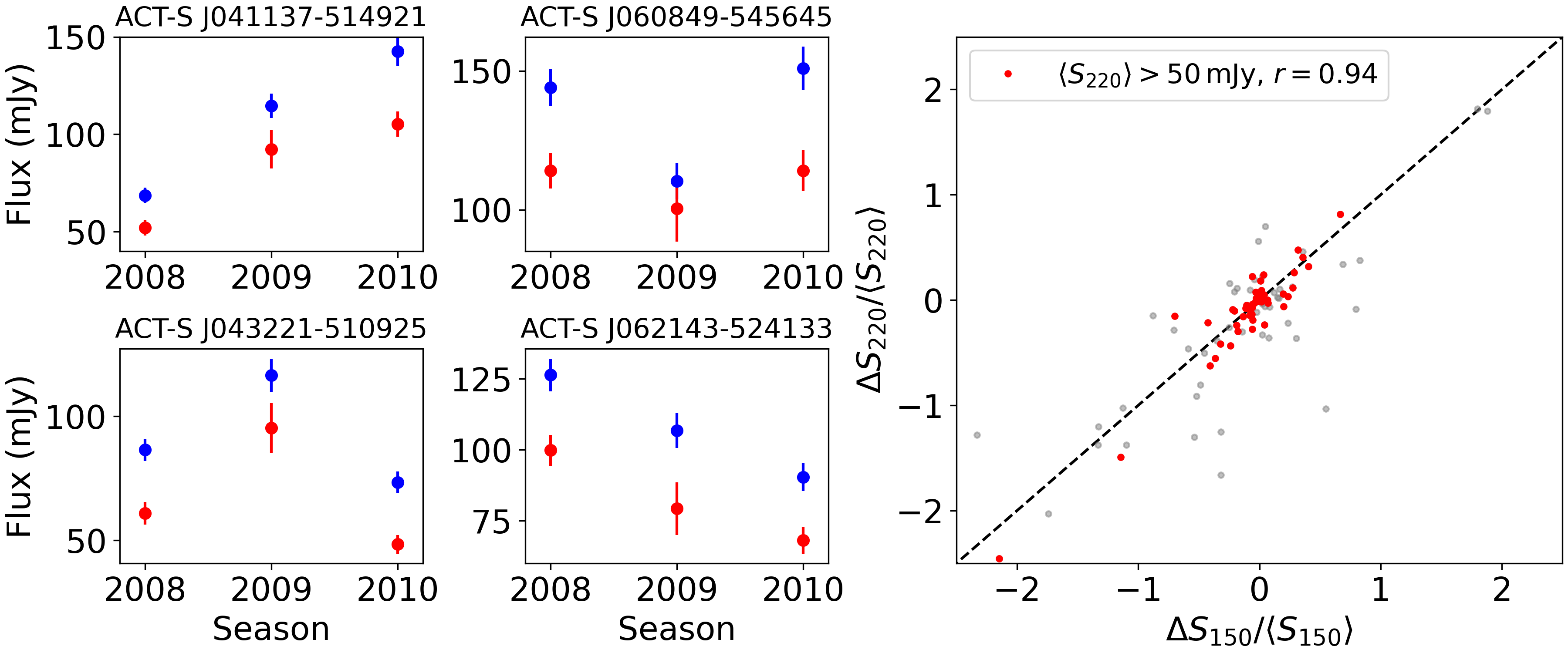}
   \caption{{Variability across all seasons at 150 and 220 GHz. Left: Flux densities of sources across three seasons, portraying data from f150 in blue and f220 in red. All sources shown are variable and classified as synchrotron emitters. We observe a correlation between frequencies, indicating that changes in flux at f150 correspond to analogous changes at f220. Right: Fractional variability for the entire sample from f150 and f220, encompassing all seasons. We highlight sources with a mean f220 flux exceeding 50 mJy in red. The Pearson correlation coefficient for these sources is $r = 0.94$, while for the entire sample $r=0.84$. These findings suggest a significant correlation between frequency bands.}}
   \label{fig:var_modes_corr}
\end{figure*}

{We investigated the evolution of the flux of individual sources throughout three seasons and in two frequency bands (f150 and f220). We observed multiple modes of variability, such as increment-increment, decrement-increment, increment-decrement, and decrement-decrement. Comparing the fractional variability of both bands, we found a Pearson correlation coefficient of $r=0.94$ for sources with mean f220 flux higher than 50\,mJy and $r=0.84$ for the whole sample, indicating a high correlation between frequencies (See Fig.~\ref{fig:var_modes_corr}).}

\subsection{Number counts}\label{sec:number_counts}

Here, we present an analysis of the differential number counts $dN/dS$ of synchrotron and dusty sources using our multi-wavelength observations at 150, 220 and 280\,GHz. 
%Our analysis covers a wide range of flux densities from roughly 10 to 1,000\,mJy. 
We compare with theoretical predictions for models of galaxy formation and evolution, and discuss the implications of our findings for the physics of accretion and star formation.

{To combine different sky patches we follow a method similar to \citealt{Carniani2015}.
%; for details we refer readers to that paper. 
For each frequency and population we combine three non overlapping fields: coaddL, coaddR and s2$^\star$ (which is s2 excluding the coaddL). These provide 160, 147 and 288 square degrees respectively for 150 and 220\,GHz. In the case of 280\,GHz the s2 region is 49 square degrees smaller due to a declination cut (see Sect.~\ref{sec:maps}).}

\begin{figure}
  \resizebox{\hsize}{!}{\includegraphics{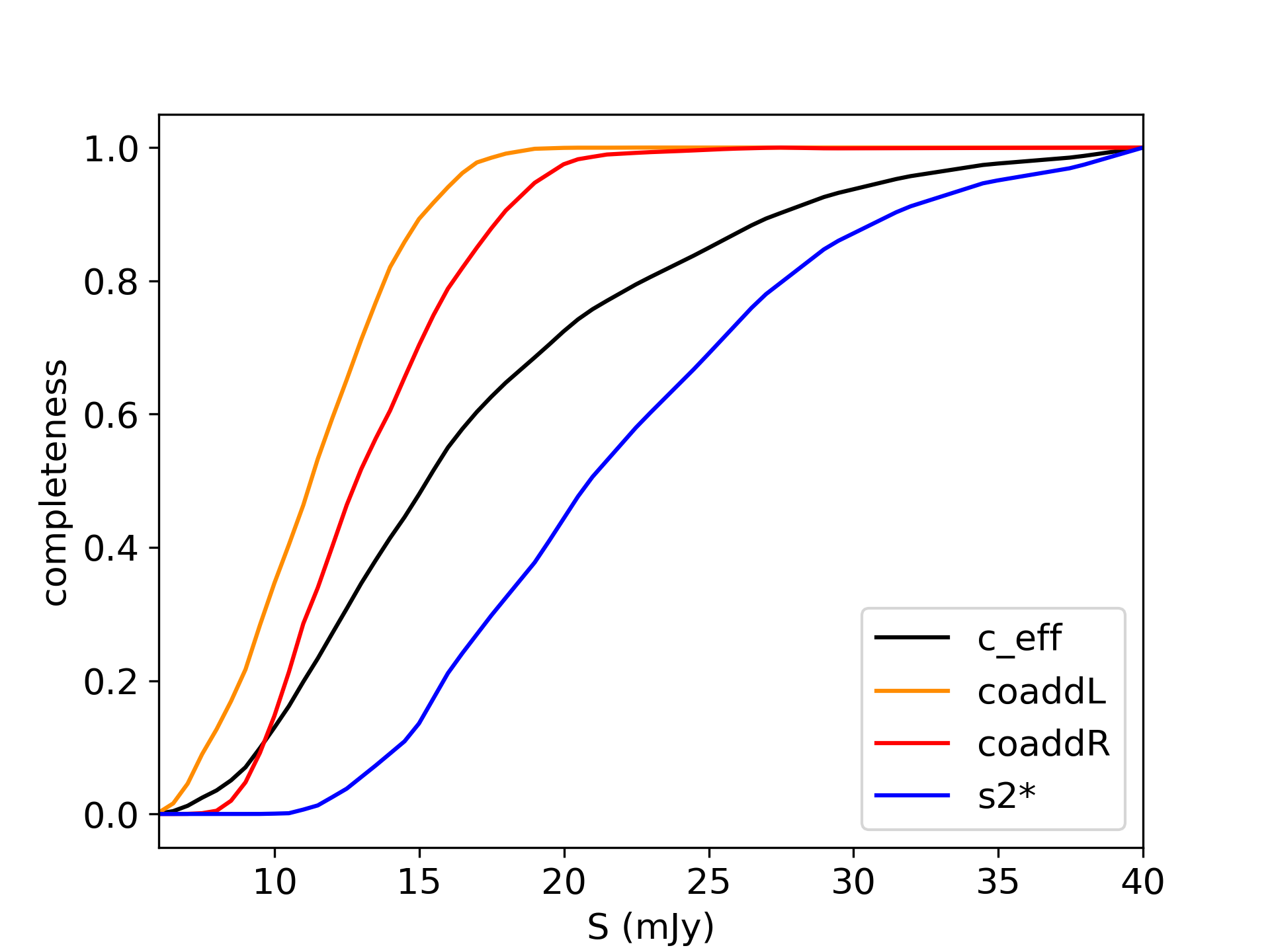}}
  \caption{Effective and individual completeness from different fields as a function of flux in 220 GHz, considered for number counts. {We use the black line to correct for number counts in Eq.~\ref{eq:number_counts}}}
  \label{fig:c_eff}
\end{figure}

{Each field has a set of simulations (Sect.~\ref{sec:sims}) from which we calculate the de-boosting and completeness by flux. 
The effective completeness $c_\mathrm{eff}$ of the full survey, as a function of flux $S$, is the weighted average (by area) of the completeness of each field}:

\begin{equation}
   c_\mathrm{eff}(S) = \frac{ A_1 \times c_1(S) + A_2 \times c_2(S) + A_3 \times c_3(S)}{A_1+A_2+A_3},
\end{equation}

\noindent where labels 1, 2 and 3 correspond to coaddL, coaddR and s2$^\star$, respectively. In Fig.~\ref{fig:c_eff}, we show the completeness curves for 220 GHz. It is clear that the completeness of s2$^\star$ is very low near the detection limit of $S \sim 15$ mJy. This means that below this flux density, the detected sources are drawn from a smaller effective area, which is approximately equal to the area of coaddL and coaddR combined.

{To count sources in a flux bin, we take account of flux-dependent completeness by weighting each source by the inverse of its effective completeness. This means that the corrected number of sources in the bin is}

\begin{equation}\label{eq:number_counts}
 dN = \sum_{i=0}^{N_\mathrm{src}} \frac{1}{c_\mathrm{eff}(\mathrm{S}_i)},
\end{equation}

\noindent {where $\mathrm{S}_i$ is the flux of the $i$th source in a bin with $N_\mathrm{src}$.
For example, in a high flux bin at 220 GHz (e.g., $S>50$\,mJy), each source contributes 1 to the corrected number of sources, while in a low flux bin (e.g., 15\,mJy), each source contributes roughly 2.
Thus, the completeness correction depends on the flux from individual sources in that bin. The $dS$ is simply the difference between the bin edges,} {and we need to divide $dN/dS$ by the total area of the survey.}

Differential number counts have steep slopes; so, to see features in graphical presentations, they are usually scaled by $S^{5/2}$. For this reason, we present source counts as $S^{5/2} dN/dS$ in units of $\mathrm{Jy}^{3/2} \mathrm{sr}^{-1}$. We take the flux $S$ to be the center of each flux density  bin, and we approximate the error bars as Poissonian. We also include a 10\% uncertainty from the completeness correction.

\begin{figure*}
\centering
\includegraphics[width=0.95\linewidth]{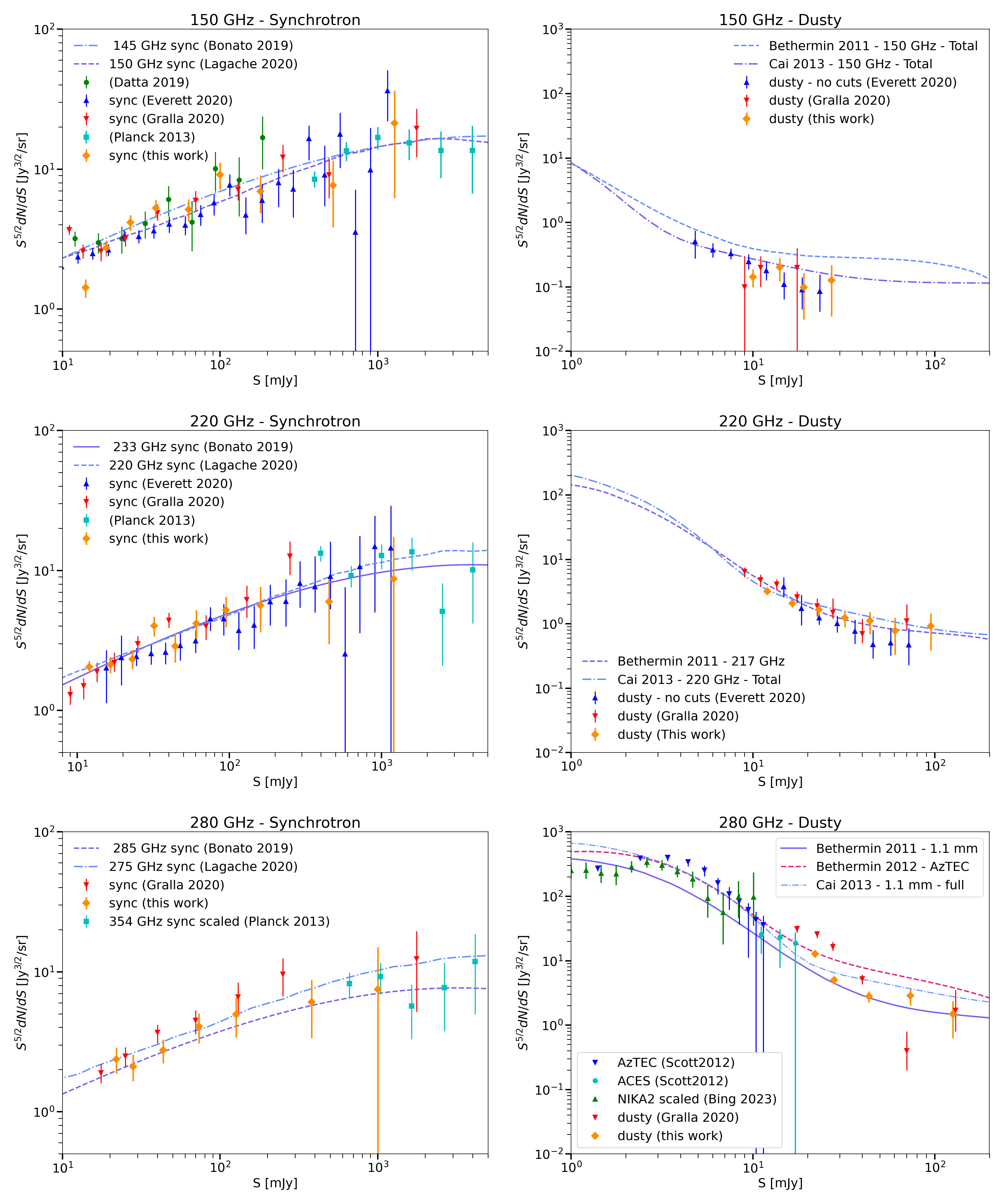}
\caption{Number counts at 150, 220 ,and 280\,GHz for synchrotron and dusty source populations. In the case of synchrotron sources we include available ACTPol \citep{Datta2019}, ACT equatorial \citep{Gralla2020}, Planck \citep{Planck2013}, and SPT \citep{Everett2020} number counts, and models from \citealt{Lagache2020} and \citealt{Bonato2019}. For dusty sources we include ACT equatorial and SPT results, and models from \citealt{Bethermin2011,Bethermin2012} and \citealt{Cai2013}. For 280\,GHz we also include results from \citealt{Scott2012} at 1.1 mm (273\,GHz) and \citealt{2023arXiv230507054B} at 1.2 mm (255\,GHz) scaled to 277\,GHz using $\alpha = 2.5$.}
\label{fig:number_counts}
\end{figure*}

Figure~\ref{fig:number_counts} shows number counts results for synchrotron and dusty sources at the three available frequencies.
Overlaid, we show previous results from ACT \citep{Gralla2020, Datta2019}, SPT \citep{Everett2020} and Planck \citep{Planck2013}, as well current models for AGNs \citep{Bonato2019,Lagache2020}, and DSFGs \citep{Bethermin2011, Bethermin2012, Cai2013}. 
We do not correct for slight differences in effective frequency between experiments (e.g, 149.5, 154.5, and 144.1\,GHz or 219.3, 220.4, and 223.7\,GHz for ACT, SPT, and Planck respectively, for a dust spectrum). We discuss the results of synchrotron counts in Sect.~\ref{sec:sync_nc} and dusty counts in Sect.~\ref{sec:dust_nc}. We also include a version of dusty counts without IRAS counterparts {which we compare with lensed DSFG models \citep{Negrello2007,Cai2013}} that we discuss in Sect.~\ref{sec:dustNI_nc}. Details about the number counts are available in Table~\ref{tab:number_counts}.

\begin{table*}
    \centering
    \caption{Differential number counts for different populations. The dusty-cut sample does not include sources with IRAS counterparts, effectively removing nearby luminous infrared galaxies. These number counts come from a combination of three fields: coaddL, coaddR and s2 excluding the coaddL patch. The fields encompass an area of 160, 147 and 288 sq deg, respectively; thus the full area is 595 sq deg (except for 280\,GHz where the s2 field is smaller by 49 sq deg). The effective area of each flux bin is proportional to the completeness value at the center of the bin. This entry is only informative since the calculation is done using individual fluxes from sources.}
    \begin{tabular}{ccccc}
\hline
\multicolumn{5}{c}{150\,GHz} \\
\hline
Flux Bin  &   $S^{2.5}dN/dS$, sync        &  $S^{2.5}dN/dS$, dusty       & $S^{2.5}dN/dS$, dusty-cut            & Effective Area       \\
 (mJy)    &  (Jy$^{1.5}$sr$^{-1}$)  & (Jy$^{1.5}$sr$^{-1}$) & (Jy$^{1.5}$sr$^{-1}$)     & (deg$^2$)  \\
\hline
$     8$ -- $    12$ & $0.30 \pm 0.10$ & $0.15 \pm 0.05$ & $0.08 \pm 0.03$ & $299$\\ 
$    12$ -- $    16$ & $1.50 \pm 0.20$ & $0.21 \pm 0.08$ & $0.14 \pm 0.07$ & $453$\\ 
$    16$ -- $    22$ & $2.90 \pm 0.40$ & $0.10 \pm 0.07$ & $0.10 \pm 0.07$ & $555$\\ 
$    22$ -- $    32$ & $4.30 \pm 0.50$ & $0.13 \pm 0.09$ &       & $593$\\ 
$    32$ -- $    46$ & $5.30 \pm 0.80$ &       &       & $595$\\ 
$    46$ -- $    80$ & $5.20 \pm 0.90$ &       &       & $595$\\ 
$    80$ -- $   120$ & $9.20 \pm 2.00$ &       &       & $595$\\ 
$   120$ -- $   240$ & $7.00 \pm 2.10$ &       &       & $595$\\ 
$   240$ -- $   800$ & $7.70 \pm 3.80$ &       &       & $595$\\ 
$   800$ -- $  1750$ & $21.30 \pm 15.10$ &       &       & $595$\\ 
\hline
\hline
\multicolumn{5}{c}{220\,GHz} \\
\hline
Flux Bin  &   $S^{2.5}dN/dS$, sync        &  $S^{2.5}dN/dS$, dusty       & $S^{2.5}dN/dS$, dusty-cut            & Effective Area       \\
 (mJy)    &  (Jy$^{1.5}$sr$^{-1}$)  & (Jy$^{1.5}$sr$^{-1}$) & (Jy$^{1.5}$sr$^{-1}$)     & (deg$^2$)  \\
\hline
$    10$ -- $    14$ & $2.1 \pm 0.2$ & $3.2 \pm 0.3$ & $2.8 \pm 0.2$ & $138$\\ 
$    14$ -- $    19$ & $2.1 \pm 0.3$ & $2.1 \pm 0.3$ & $1.7 \pm 0.3$ & $330$\\ 
$    19$ -- $    27$ & $2.3 \pm 0.4$ & $1.7 \pm 0.3$ & $0.5 \pm 0.2$ & $490$\\ 
$    27$ -- $    37$ & $4.0 \pm 0.6$ & $1.2 \pm 0.4$ & $0.6 \pm 0.3$ & $585$\\ 
$    37$ -- $    51$ & $2.9 \pm 0.7$ & $1.1 \pm 0.4$ & $0.6 \pm 0.3$ & $595$\\ 
$    51$ -- $    70$ & $4.2 \pm 1.1$ & $0.8 \pm 0.5$ & $0.5 \pm 0.4$ & $595$\\ 
$    70$ -- $   120$ & $5.2 \pm 1.3$ & $0.9 \pm 0.5$ &       & $595$\\ 
$   120$ -- $   200$ & $5.6 \pm 2.0$ &       &       & $595$\\ 
$   200$ -- $   700$ & $6.0 \pm 3.0$ &       &       & $595$\\ 
$   700$ -- $  1700$ & $8.7 \pm 8.7$ &       &       & $595$\\ 
\hline
\hline
\multicolumn{5}{c}{280\,GHz} \\
\hline
Flux Bin  &   $S^{2.5}dN/dS$, sync        &  $S^{2.5}dN/dS$, dusty       & $S^{2.5}dN/dS$, dusty-cut            & Effective Area       \\
 (mJy)    &  (Jy$^{1.5}$sr$^{-1}$)  & (Jy$^{1.5}$sr$^{-1}$) & (Jy$^{1.5}$sr$^{-1}$)     & (deg$^2$)  \\
\hline
$  20.0$ -- $  24.0$ & $2.4 \pm 0.5$ & $12.8 \pm 1.2$ & $11.2 \pm 1.1$ & $105$\\ 
$  24.0$ -- $  32.0$ & $2.1 \pm 0.5$ & $5.0 \pm 0.7$ & $4.1 \pm 0.6$ & $222$\\ 
$  32.0$ -- $  55.0$ & $2.8 \pm 0.5$ & $2.8 \pm 0.5$ & $1.5 \pm 0.4$ & $468$\\ 
$  55.0$ -- $  92.0$ & $4.1 \pm 1.0$ & $2.9 \pm 0.8$ & $1.7 \pm 0.6$ & $545$\\ 
$  92.0$ -- $ 160.0$ & $5.0 \pm 1.6$ & $1.5 \pm 0.9$ & $1.5 \pm 0.9$ & $546$\\ 
$ 160.0$ -- $ 600.0$ & $6.1 \pm 2.7$ &       &       & $546$\\ 
$ 600.0$ -- $1400.0$ & $7.5 \pm 7.5$ &       &       & $546$\\ 
\hline
    \end{tabular}
    \label{tab:number_counts}
\end{table*}

\begin{figure*}
\centering
   \includegraphics[width=17cm]{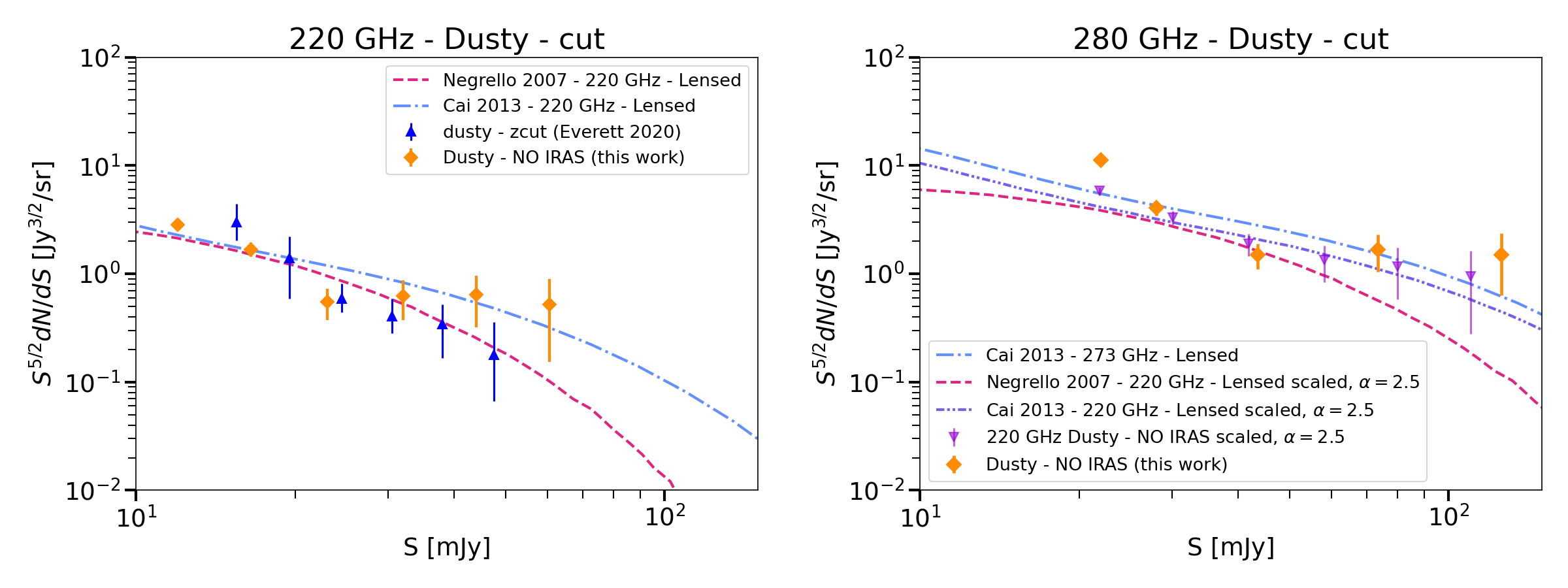}
    \caption{Differential number counts at 220 and 280\,GHz for dusty sources excluding IRAS counterparts, thus removing nearby luminous infrared galaxies.  Consequently, the population consists mainly of high redshift lensed galaxies. At 220\,GHz we include \citet{Everett2020} "dusty - zcut" counts, which also exclude IRAS sources. We note that the IRAS cut is more aggressive than just removing nearby sources, as almost all nearby sources are in IRAS.}
    \label{fig:number_counts_lensed}
\end{figure*}

\subsubsection{Synchrotron sources}\label{sec:sync_nc}

The AGNs dominate the counts at 150 GHz. Most of them are present in radio surveys, but the number counts change at millimeter wavelengths because the {population begins to include sources with spectra that differ from those common at lower radio frequencies} \citep{Tucci2011}. Current models include this correction; as shown in \citet{Marriage2011} we cannot simply extrapolate radio counts to mm wavelengths.
In this frequency range, there is agreement between experiments and models. ACT and SPT can sample the low-flux end reasonably well ($S<500$\,mJy), whereas Planck counts better constrain the high flux end ($S>500$\,mJy) as a result of its full sky coverage.

At 220\,GHz, we include ACT MBAC equatorial, SPT 220\,GHz, and Planck 217\,GHz results, which are directly comparable. As for 150 GHz, we see good agreement between both data and models and that ACT and SPT can constrain the low end reasonably well, which is, in this case, roughly $S<300$\,mJy. 

The 280\,GHz data set is unique; there are no similar observations at this frequency aside from the earlier ACT equatorial survey.  Hence, for comparison purposes, we must extrapolate measurements of AGN counts from other bands, such as 220 and 343\,GHz. We show a scaled version of \textit{Planck} 343\,GHz counts using the mean spectral index between 217 and 343 of $-0.45$ (measured by \emph{Planck}), as well as the ACT equatorial counts. \citet{Lagache2020} provides a model at 275\,GHz by scaling the 220\,GHz counts with a spectral index between 220 and 280\,GHz of $-0.6$, and \citet{Bonato2019} provides an empirical fit based on ALMA band 7 (285\,GHz) measurements of \textit{Planck} sources. Here we see agreement between ACT MBAC equatorial and southern surveys.  Our counts also favor the \citet{Lagache2020} model, which is expected since the scaling is $-0.6$ and we found a mean spectral index for synchrotron sources of $-0.55$ in Sect.~\ref{sec:alpha}.

\subsubsection{Dusty sources}\label{sec:dust_nc}

The number counts of dusty star-forming galaxies provide important information on the nature and evolution of galaxies at different epochs of the universe. By comparing the observed number counts with theoretical predictions, we can test models of galaxy formation and evolution and constrain the properties of the interstellar medium, such as its dust content and temperature \citep[For a number counts review see][]{Casey2014}. Such counts thus provide an important tool for understanding the processes that shape the evolution of galaxies over cosmic time.

Dusty sources at 150\,GHz are typically faint, with fluxes below 50 mJy. Thus, CMB experiments detect only the tip of the distribution. In this frequency band, ACT and SPT surveys agree. However, the \citet{Bethermin2011} and to a lesser extent the \citet{Cai2013} models tend to overestimate the number of sources at all flux levels.

{At 220\,GHz, the number of sources is higher at all fluxes due to the positive spectral index. The highest fluxes are around 100\,mJy.} In this case, both models \citet{Bethermin2011,Bethermin2012} and \citet{Cai2013}, are similar and agree with observations.  Our results constrain counts in  a similar range of flux densities to the equatorial survey, but the number of sources at fluxes higher than ${\sim}25$ mJy is more significant because of the larger area.

At 280 GHz, the source counts presented here agree with the equatorial source counts \citep{Gralla2020} above $\sim$30\,mJy, but there is a discrepancy at lower flux densities, likely due to unresolved systematic effects that predominate at low signal to noise flux bins. 
The current analysis was significantly simplified from the one presented in \citet{Gralla2020}. The most relevant difference is that in this analysis, we restrict the source detection and flux density measurements to single-band maps. 
This simplified approach may have reduced the number of detected sources and altered the error bars in the source counts.
However, it makes our analysis of the counts at the low flux density regime more straightforward, making it a more favorable choice for comparison with models such as those presented in \citet{Bethermin2011} and \citet{Cai2013}.

Additionally, the case of 280\,GHz is special because all the existing data available at comparable frequencies correspond to fainter fluxes ($\lesssim 15$\,mJy).
There have been multiple campaigns from JCMT and ASTE to map a few regions of roughly one square degree at 1.1\,mm (273\,GHz).
\citealt{Scott2012} presented an analysis of six fields, with combined counts going roughly from 1 to 11\,mJy in a 1.6 square degree area.
This work also presents the ACES field counts going from 10 to 16\,mJy, which is close to the faint end of our catalog at 280\,GHz. 
Although their uncertainties are high owing to the limited area, we can see the continuity of the number counts function going from \citet{Scott2012} 1--16\,mJy counts to ours at 14--100\,mJy. 
Also recently \citealt{2023arXiv230507054B} observed the GOODS-N and COSMOS fields with NIKA at 2\,mm (150\,GHz) and 1.2\,mm (255\,GHz). We scale the latter to 277\,GHz using an spectral index of 2.5, and we find that the results of this survey are similar to those of \citealt{Scott2012}.

The predictions of \citet{Bethermin2012} and \citet{Cai2013} agree with \citealt{Scott2012} {at fluxes below 10\,mJy} but the \citealt{Bethermin2011} model does not.    
Our data prefer the \citealt{Bethermin2011} model at high fluxes.

\subsubsection{Dusty-cut}\label{sec:dustNI_nc}

We also present in Fig.~\ref{fig:number_counts_lensed} dusty counts excluding IRAS counterparts, effectively removing nearby luminous infrared galaxies. {We label these counts as "dusty-cut"}. Since the number of sources is very small at 150\,GHz (5 sources) we only show the 220 and 280\,GHz cases. At 220\,GHz {it} is possible to compare with SPT results from \citet{Everett2020}. The SPT paper shows different data cuts, such as one excluding local IRAS galaxies from the counts (labeled “dusty-zcut”). Our counts generally agree with the dusty-zcut counts, but our counts favor the \citet{Cai2013} lensed galaxies model{, which is a subset of the model shown in Sect.~\ref{sec:dust_nc}, considering only the contribution of lensed galaxies. On the other hand} SPT counts favor the \citet{Negrello2007} lensed galaxies model. We note that the counts deviate from the models below 15\,mJy; we  suggest that this is because at this flux level we begin to have a contribution from un-lensed galaxies. At 280\,GHz there are no other measurements, so we include a scaled version of our 220\,GHz counts assuming a spectral index of $2.5$ motivated by the mean spectral $\alpha_{220}^{280} \approx 2.5$ found in Sect.~\ref{sec:alpha}. A \citet{Cai2013} {lensed} model is available for 273\,GHz (1.1\,mm), which is roughly the 220\,GHz {lensed} model scaled by $\alpha_{220}^{280} \approx 3.0$. As with the \citet{Cai2013} 220\,GHz lensed model, there is agreement at 280\,GHz between our counts and this model. We note this is not the case for the full model as shown in Fig.~\ref{fig:number_counts}, which overpredicts the number of sources at 280\,GHz. Our counts at $S \gtrsim 25$\,mJy prefer a scaling between $2.5$ and $3.0$.

\subsubsection{{Sources contribution to the power spectrum}}\label{sec:power_spectrum}

The high range of multipoles on the power spectrum of the sky at millimeter wavelengths are dominated by synchrotron and dusty galaxies as well as thermal SZ from clusters. In particular, the power spectrum at the highest multipoles constrains the Poisson distributed components of the source population. The contribution to the power spectrum by a Poisson-distributed population of sources is a function of the number counts as 

\begin{equation}
C_\nu^{\mathrm{PS}} (S_\mathrm{c}) =  \left(\frac{\partial B_\nu}{\partial T}  \right)^{-2}  \int_0^{S_\mathrm{c}} S^2 \frac{dN}{dS} dS,
\end{equation}

\noindent where $S_\mathrm{c}$ is the flux density cut in the data. This constant is usually expressed as the contribution to the CMB power spectrum at $\ell = 3000$, defined as $A_\nu^\mathrm{PS}(S_\mathrm{c}) = C_\nu^\mathrm{PS} \ell (\ell + 1)/(2\pi) $. Fitting our data to \citet{Lagache2020} model and using a cut of 15\,mJy and 100\,mJy gives $A_{150}^\mathrm{sync}(15\,\mathrm{mJy}) = 3.58 \pm 0.24\,\mu\mathrm{K}^2$ and $A_{150}^\mathrm{sync}(100\,\mathrm{mJy}) = 20.53 \pm 1.40\,\mu\mathrm{K}^2$. We can directly compare these results with ACT DR4 power spectrum results, \citealt{Choi2020} found $A_{150}^\mathrm{ACT-sync}(15\,\mathrm{mJy}) = 3.74 \pm 0.24\,\mu\mathrm{K}^2$ and $A_{150}^\mathrm{ACT-sync}(100\,\mathrm{mJy}) = 22.56 \pm 0.33\,\mu\mathrm{K}^2$. Hence, the results are consistent. 

We can also fit the dusty model, which represents the Poisson component of the CIB (not the clustered term). The result for \citet{Cai2013} model is $A_{150}^\mathrm{CIB}(15\,\mathrm{mJy}) = 7.95 \pm 1.60\,\mu\mathrm{K}^2$ and $A_{150}^\mathrm{CIB}(100\,\mathrm{mJy}) = 8.20 \pm 1.65\,\mu\mathrm{K}^2$, here we see that sources with flux densities higher than 15 mJy do not contribute significantly to the power spectrum. Results from ACT DR4 \citet{Choi2020} give a contribution of the CIB Poisson term of $A_{150}^\mathrm{ACT-CIB} = 6.58 \pm 0.37\,\mu\mathrm{K}^2$ slightly lower than our result but consistent, whereas SPT-SZ + SPTpol Surveys \citet{Reichardt2021} reports  $A_{150}^\mathrm{SPT-CIB} = 7.24 \pm 0.63\,\mu\mathrm{K}^2$ also consistent. %At higher frequencies, the CIB contribution is more relevant

%%%%%%%%%%%%%%%%%%%%%%%%%%%%%%%%%%%%%%%%%%%%%%%%%%%%%%%%%%%%%%%%%%

\section{Summary}\label{sec:summary}

This work presents a multi-frequency multi-epoch catalog of extragalactic point sources from ACT MBAC, including new flux measurements at 280\,GHz.
{From the combination of 11 frequency--season images, we constructed a catalog. We found 284 synchrotron and 183 dusty source candidates with $-0.65$ and $3.39$, $-0.43$ and $3.15$,  $-0.56$, and $3.35$ spectral indices between 150--220, 150--280, and 220--280 GHz, respectively.
We cross-matched the catalog with other surveys from radio to X-ray and found hundreds of counterparts, with most synchrotron sources being present in radio catalogs. 
In contrast, dusty source counterparts were found in infrared catalogs. 
These cross-matches helped us validate the catalog via direct comparison in the case of SPT and assess the pointing accuracy of our positions using AT20G.}

{After excluding known nearby galaxies, we separated the populations into those dominated by synchrotron versus thermal emission based on their measured spectral indexes.
We showed how the f280 band serves an important role in this regard.}

The multi-epoch nature of the catalog allowed us to study the variability of the sources. We found significant variability in AGNs {of $0.35 \pm 0.02$\%} and no significant variability for DSFGs.

We agree with the literature regarding source counts except in the less observationally constrained 280\,GHz frequency channel, where dust models tend to overpredict our measurements.
These observational counts will help constrain the models and contribute to greater understanding of galaxy evolution in the universe.
We also presented an example of SED fitting of the high redshift $z=4.1$ galaxy ACT-S J065207-551605 using ACT, APEX, and \textit{Herschel} data, giving insights into the system's physical properties.

Overall, the release of the calibrated 280\,GHz maps and catalogs provide a valuable tool for cross-correlation and cross-matching with the forthcoming Simons Observatory \citep{SO2019} and CMB-S4 \citep{S42019} high-frequency maps and catalogs at 280\,GHz as well as with future Fred Young Submillimeter Telescope \citep{CCAT2023} 280\,GHz surveys. 
Moreover, we have presented hundreds of new bright DSFG candidates using a different selection function compared to 220\,GHz catalogs. These are excellent candidates for follow-up programs on facilities such as the ALMA, APEX, and the Atacama Large Aperture Submm Telescope (\citealt{Klaassen2020, Ramasawmy2022, Mroczkowski2023}).

\begin{acknowledgements}
Support for ACT was through the U.S.~National Science Foundation through awards AST-0408698, AST-0965625, and AST-1440226 for the ACT project, as well as awards PHY-0355328, PHY-0855887 and PHY-1214379. Funding was also provided by Princeton University, the University of Pennsylvania, and a Canada Foundation for Innovation (CFI) award to UBC. ACT operated in the Parque Astron\'omico Atacama in northern Chile under the auspices of the Agencia Nacional de Investigaci\'on y Desarrollo (ANID). The development of multichroic detectors and lenses was supported by NASA grants NNX13AE56G and NNX14AB58G. Detector research at NIST was supported by the NIST Innovations in Measurement Science program. Computing for ACT was performed using the Princeton Research Computing resources at Princeton University, the National Energy Research Scientific Computing Center (NERSC), and the Niagara supercomputer at the SciNet HPC Consortium. SciNet is funded by the CFI under the auspices of Compute Canada, the Government of Ontario, the Ontario Research Fund–Research Excellence, and the University of Toronto. We thank the Republic of Chile for hosting ACT in the northern Atacama, and the local indigenous Licanantay communities whom we follow in observing and learning from the night sky.
We thank G. Morales and F. Rivas for their contributions to the source extraction pipeline.
CV received partial support from Center of Excellence in Astrophysics and Associated Technologies (PFB 06).
CHLC acknowledges financial support by the Spanish Ministry of Science and Innovation under the projects AYA2017-84185-P and PID2020-120514GB-I00, and the ACIISI, Consejeria de Economia, Conocimiento y Empleo del Gobierno de Canarias and the European Regional Development Fund (ERDF) under grant with reference ProID2020010108.
RD thanks ANID for grants BASAL CATA FB210003, FONDECYT 1141113 and Anillo ACT-1417.
MH acknowledges financial support from the National Research Foundation of South Africa.
ADH acknowledges support from the Sutton Family Chair in Science, Christianity and Cultures, from the Faculty of Arts and Science, University of Toronto, and from the Natural Sciences and Engineering Research Council of Canada (NSERC) [RGPIN-2023-05014, DGECR-2023-00180].
FR acknowledges support from the European Union’s Horizon 2020 research and innovation program under the Marie Sklodowska-Curie grant agreement No. 847523 ‘INTERACTIONS’ and from the Cosmic Dawn Center (DAWN), funded by the Danish National Research Foundation under grant No. 140.
CS acknowledges support from ANID through FONDECYT grant no.\ 11191125 and BASAL project FB210003.
This research has made use of the NASA/IPAC Extragalactic Database (NED), which is funded by the National Aeronautics and Space Administration and operated by the California Institute of Technology.
We acknowledge the use of the Legacy Archive for Microwave Background Data Analysis (LAMBDA), part of the High Energy Astrophysics Science Archive Center (HEASARC). HEASARC/LAMBDA is a service of the Astrophysics Science Division at the NASA Goddard Space Flight Center.
Some of the results in this paper have been produced using the following software: 
\textsc{Astropy} \citep{Astropy2018}; \textsc{Astroquery} \citep{Astroquery2019};\textsc{DS9} \citep{Joye_Mandel2003};\textsc{emcee} \citep{2013PASP..125..306F};\textsc{healpy} \citep{Zonca2019};\textsc{Matplotlib} \citep{Hunter2007};\textsc{NumPy} \citep{Harris2020};\textsc{SciPy} \citep{Virtanen2020}.
\end{acknowledgements}

\bibliographystyle{aa}

\bibliography{references}

\begin{appendix}

\section{Calibration of 280~GHz maps}\label{sec:calib_280}

Calibration of high-frequency channels is a challenge for ground-based astronomy due to a combination of factors such as higher atmospheric noise and variability and a reduction in the cosmic microwave background flux in the sub-millimeter.

As mentioned in Sect.~\ref{sec:data}, 150 GHz season 2 maps were calibrated using the WMAP CMB power spectrum \citep{Jarosik2011}, and then the rest of the maps were calibrated using the former \citep{Das2014}. This process is difficult for 280 GHz season 2 since the calibration factor changes across multipoles (it has a tilt). To resolve this issue, we first performed a calibration of 280\,GHz season\,4, which does not show a scale-dependent change in calibration in the multipole range $500<\ell<2000$, and then we calibrated 280\,GHz season 2 using 280 GHz season\,4 at high multipoles ($\ell > 4000$)

We follow \citealt{Hajian2011} method of calibration using cross-correlations (for details we refer to that publications and \citealt{Das2011}; we use the notation of the latter). Seasonal data is split evenly into four parts, each one called a four-way split; thus, the 2D power spectrum of a patch is the average of six power spectra:

\begin{equation}
C_{\bm{\ell}} = \frac{1}{6} \sum_{i,j;i<j}^{1 \leq j \leq 4} C_{\bm{\ell}}^{iA\times jB},
\end{equation}

\noindent where $ C_{\bm{\ell}}^{iA\times jB} = \mathrm{Re} \left[ {T}^{*iA}({\mathbf{k}}) {T}^{jB}({\mathbf{k}}) \right]$ are cross-power spectra of season maps for the patch and $iA$ and $jB$ are the 4-way splits of that patch. The final power spectrum is given by the average of all patches considered.

The cross power spectrum of different maps, for example 150 GHz season 4 and 280 GHz season 4, is computed in a similar way but in this case $iA$ and $jB$ correspond to each respective map, and there are 12 cross-spectra for each patch:

\begin{equation}\label{ec:cross_spectra}
C_{\bm{\ell}} = \frac{1}{12} \sum_{i,j;i<j}^{1 \leq j \leq 4} \left(  C_{\bm{\ell}}^{iA\times jB}+C_{\bm{\ell}}^{jB\times iA}  \right).
\end{equation}

We estimate the 2D variance of this spectral estimator as 

\begin{equation}
\sigma^2 \left( C_{\bm{\ell}}  \right) = \left(C_{\bm{\ell}}^{\mathrm{th}} {b_\ell^{A}}^2+N_{\bm{\ell}}^{AA}/4   \right)\left(C_{\bm{\ell}}^{\mathrm{th}} {b_\ell^{B}}^2+N_{\bm{\ell}}^{BB}/4   \right),
\end{equation}

\noindent where $C_\ell^\mathrm{th}$ is the theoretical CMB power spectrum, for which we use the theoretical spectrum of the best-fit model from Planck PR3 \citep{Planck2018CMB}. 
Also, $b_\ell$ is the beam window function\footnote{As explained in Sect.~\ref{sec:eff_beams} the beams depend on the frequency and to a lesser extent on the season; thus we approximate the beam for each 4-way split to be the same as for the full season}. 
Finally, $N_\ell$ is the noise power spectrum, calculated as the difference between the auto and cross power spectra for each frequency and season.

To estimate the power spectrum from our maps, we first apply a low-$\ell$ filter to remove under-sampled modes from each set of 4-way splits (8 maps in total). 
The filter goes from amplitude 0 to 1 as a squared sine from $\ell$ = 0 to 400. We apply it in the region $63<\mathrm{ra}<108$, $-55.5<\mathrm{dec}<50.5$ (wider than the patches that we use). 
Before calculating the Fourier transform, 
a Gaussian apodization of 10 pixels is performed, and we apply a pre-whitening to each set of 4-way splits (see \citealt{Das2011}). 
Then we used two patches, one at $65<\mathrm{RA}<85$, $-55.22<\mathrm{dec}<51.15$ and the other $85<\mathrm{RA}<105$, $-55.22<\mathrm{dec}<51.15$. From the set of 12 cross-spectra, we calculated the power spectrum per patch (Eq.~\ref{ec:cross_spectra}), and then we averaged both patches.  We performed this process for $150s4 \times 150s4$ and ${150s4 \times 280s4}$. Since the former has already been calibrated, the calibration factor for $280s4$ is

\begin{equation}
\alpha_{280s4} = C_\ell^{150s4\times 150s4}/C_\ell^{150s2\times 280s4}.
\end{equation}

\begin{figure}
\resizebox{\hsize}{!}{\includegraphics{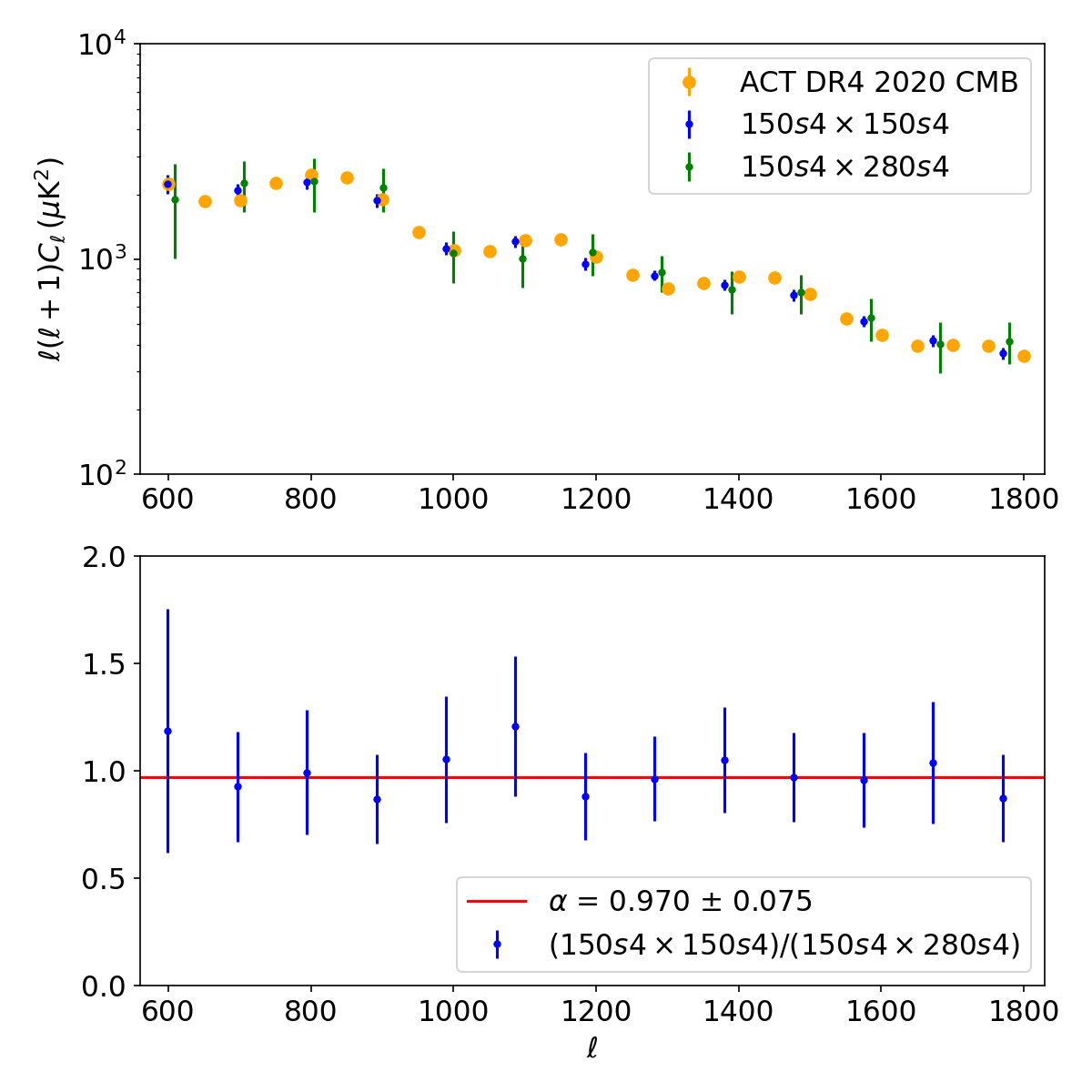}}
\caption{Power spectrum calibration of the 280 GHz 2010 map. Top: In blue, $150s4$ power spectrum and, in green, $150s4 \times 280s4$ power spectrum before the calibration (the latter has been displaced by $\ell=10$ for visibility).  For comparison we show the results from ACT Data Release 4 \citep{Choi2020}. Bottom: Ratio of the two spectra. The calibration factor derived from the weighted mean is shown as a red line.}
\label{fig:calib_280s4}
\end{figure}

Taking the ratio of the two spectra and applying a weighted mean results in a calibration factor of $\alpha_{280s4} = 0.970 \pm 0.075$ (Fig.~\ref{fig:calib_280s4}), a calibration uncertainty of roughly 8.1\%, roughly half the uncertainty expected from planet measurements (15\%). 
{Here we propagated the uncertainty from $150s4$, which is 3\%.}
This demonstrates that the map is well calibrated in the multipole regime used for CMB studies and can be used for large scale studies.
We remark that this is the first time that a cross-spectrum using 280\,GHz has been shown. We also note here that we are assuming that the CMB is dominant in this multipole regime for both frequencies. Typically, high frequencies are susceptible to dust contamination, which poses a challenge in power spectrum measurements. However, we have carefully selected a region that exhibits remarkably low dust contamination ($D_\ell^{\mathrm{Dust}} \lesssim 10\,\mu\mathrm{K}^2$) at 280\,GHz. This estimation is obtained through a scaling procedure using the 353\,GHz Planck dust emission data. Consequently, based on this, we assert that it is justifiable to disregard the contribution of dust emission in the cross maps within this particular region.

As in the previous calculation, we use $280s4 \times 280s4$ and $280s4 \times 280s2$ to calibrate 280s2 maps. Here the calibration factors are

\begin{equation}
\alpha_{280s2} \alpha_{280 s4} = C_\ell^{280 s4\times 280s4}/C_\ell^{280s4\times 280s2}.
\end{equation}

In this case, however, we used only high multipoles where the CMB is negligible ($D_\ell^{\mathrm{CMB}} \lesssim 1\,\mu\mathrm{K}^2$), that is, the range $4200<\ell<12000$. Since we mask bright sources, the power spectrum consists mainly of faint sources, Galactic dust, CIB, tSZ and kSZ. We assume this ensemble is constant through time. In this case we get a factor of $\alpha_{280s2} \alpha_{280s4} = 2.816 \pm 0.140 $.  Applying the calibration of 280s4, the final calibration of 280s2 is $\alpha_{280s2} = 2.733 \pm 0.251 $.  Hence the map has a calibration uncertainty of 9.2\%. 

\begin{figure}
\resizebox{\hsize}{!}{\includegraphics{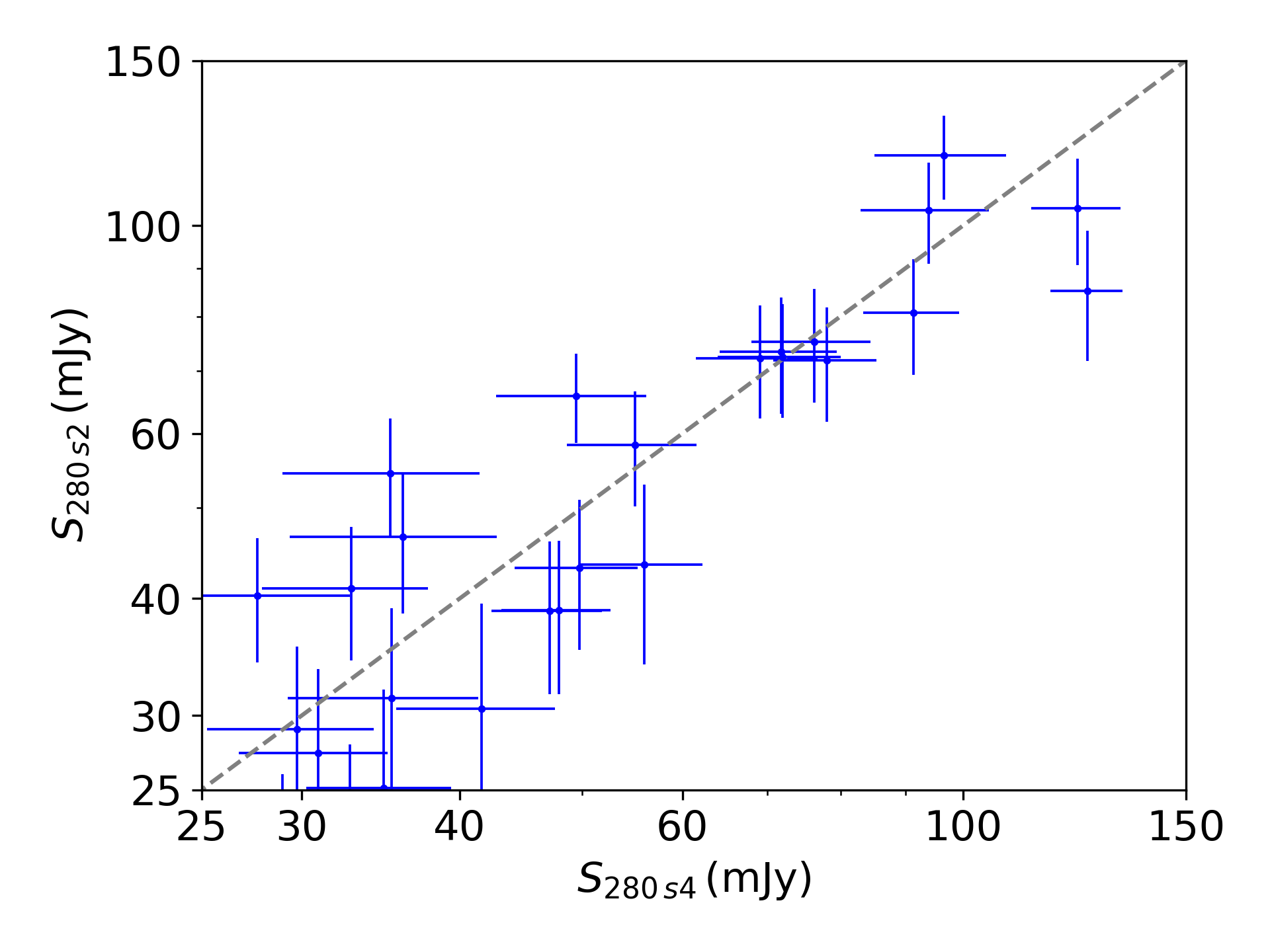}}
\caption{Comparing 280\,GHz point source fluxes from calibrated maps. }
\label{fig:S280s4s2}
\end{figure}

In Fig.~\ref{fig:S280s4s2} we show point source fluxes from the calibrated maps, excluding highly variable sources. 
When fitting the data with an intercept of zero, we obtained a slope of $0.98 \pm 0.04$, suggesting that the calibrated fluxes are consistent. 
Also, when we allowed the intercept to vary, the slope became $1.03 \pm 0.09$, and an intercept of $-3.08 \pm 4.47$\,{mJy} was determined. {The mutual consistency of the two independent observables, namely the power spectrum and the fluxes of the bright sources, provides strong evidence for the reliability of the calibration. }

\section{Description of the catalog}\label{sec:cat_description}

The catalog is available online in LAMBDA\footnote{\url{https://lambda.gsfc.nasa.gov/product/act/act_prod_table.html} under "Millimeter-wave source and SZ cluster catalog" as "Extragalactic Southern Sources Catalog"} as a Flexible Image Transport System (FITS) file. It has following entries: 

\begin{itemize}
\setlength\itemsep{0.1em}
    \item[--] \verb|ID|: Source IAU designation. 

    \item[--] \verb|ra|: Right ascension (J2000) in degrees.

    \item[--] \verb|dec|: Declination (J2000) in degrees.

    \item[--] \verb|S_xxx_yyy|: Flux density $S_m$ measured for a frequency array and season in mJy, where \verb|xxx| specifies the frequency (150, 220 or 280) and \verb|yyy| the season (2008, 2009, 2010 or coadd), up to 11 measurements per source.

    \item[--] \verb|snr_xxx_yyy|: Signal-to-noise ratio of the detection $S/N$. 

    \item[--] \verb|sigma_xxx_yyy|: Uncertainty of the flux density $\sigma_m$ in mJy. 

    \item[--] \verb|mudb_xxx_yyy|: Flux density boosting $\mu_\mathrm{db}$ in mJy inferred from the simulations. Only available for detected sources.

    \item[--] \verb|sigmadb_xxx_yyy|: Standard deviation in the flux density $\sigma_\mathrm{db}$ in mJy inferred from the simulations. Only available for detected sources.

    \item[--] \verb|sigmacal_xxx_yyy|: Uncertainty in the flux density $\sigma_\mathrm{cal}$ in mJy related to the calibration of each map.

    \item[--] \verb|alpha_150-220|: Spectral index between 150 and 220\,GHz $\alpha_{150}^{220}$ calculated using both fluxes from the season with the highest signal to noise. 

    \item[--] \verb|alpha_220-280|: Spectral index between 220 and 280\,GHz $\alpha_{220}^{280}$ calculated using both fluxes from the season with the highest signal to noise.

    \item[--] \verb|alpha_150-280|: Spectral index between 150 and 280\,GHz $\alpha_{150}^{280}$ calculated using both fluxes from the season with the highest signal to noise.

    \item[--] \texttt{behavior}: Spectral behavior of the source as falling, rising, upturning and peaking (see Sect.~\ref{sec:specclass}).

    \item[--] \verb|object classification|: {Classification of the galaxy as AGN, DSFG, nearby or star depending on external counterparts and $\alpha_{150}^{220}$, $\alpha_{150}^{280}$, $\alpha_{220}^{280}$ (see Sect.~\ref{sec:objclass}).}

    \item[--] \verb|external counterparts|: External catalogs where a source has an spatial cross-match with an offset smaller than the association radius. These are as described in Sect.~\ref{sec:external_associations}. (1) SUMSS, (2) PMN, (3) AT20 (4) IRAS, (5) WISE, (6) AKARI, (7) PCNT, (8) SPT, (9) ROSAT, (10) PCCS2, (11) Herschel, (12) ALMA.
\end{itemize}

%\section{Catalog sample}\label{sec:cat_sample}

In Table~\ref{tab:cat} we present a sample of the catalog of sources detected in the coadd maps at 150, 220, and 280 GHz in the region $65<\mathrm{RA}<90$, $-55.2<\mathrm{dec}<51.0$. We only include sources with at least one significant detection of $S/N>7$ in any band, resulting in a sample of 85 of the brighter sources. We note that most sources without any counterparts are classified as dust, and in most cases, they have robust detections in 220 and 280 GHz

\section{Example of spectral energy distribution characterization}
\label{sec:characterization_seds}

Our catalog provides multi-frequency and multi-epoch flux densities,  which are helpful in the analysis of high-redshift candidates and AGN variability.
The 280\,GHz maps allowed us to improve the characterization of the spectral energy distribution (SED) of dusty sources, in particular high-redshift DSFGs (some of which have been reported in previous ACT, SPT, and \textit{Herschel} works).
However, new candidates need new observations to confirm their high redshift nature. 
We have followed up some of these candidates using the Atacama Pathfinder EXperiment (APEX) and other astronomical facilities in Chile; this is the main aim of future work.
In addition, our catalog provides an opportunity to study some variability properties of AGNs on time scales of a year. We note that the characterization of variability can be useful for constraining the physical mechanisms at work in AGN emission.
To illustrate the potential of this data for such analyses, we carry out SED modeling of one high-redshift dusty galaxy. 

\begin{figure}
  \resizebox{\hsize}{!}{\includegraphics{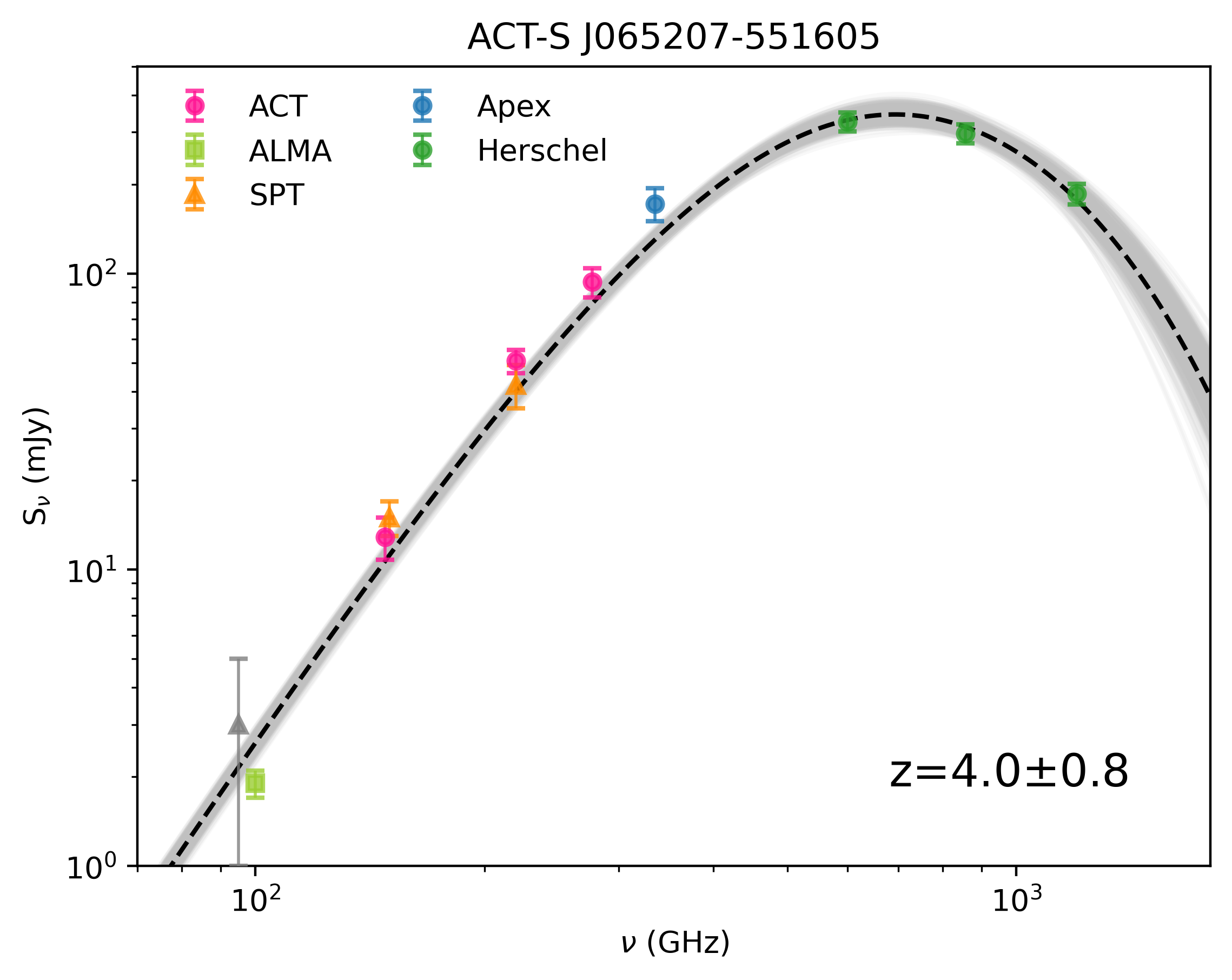}}
  \caption{Spectral energy distribution of the high-redshift dusty galaxy ACT-S\,J065207-551605.
The colored measurements identify the ACT (red), SPT (orange), APEX (blue) and \textit{Herschel} (green) data from Table~\ref{tab:fluxes2sed_ancillary}. The ACT measurements (red dots) correspond to flux densities from the {\sc coadd} analysis. Grey points identify flux densities with signal to noise ratio lower than 3.
The black dashed line corresponds to the modified blackbody function for a single-temperature, while the gray "shadow" is the uncertainty at $1~\sigma$. The obtained photometric redshift $z=4.0\pm0.8$ is presented in the bottom-right corner.}
  \label{fig:sed_dusty_ACT-S-J065207-551607}
\end{figure}

We model ACT-S J065207-551605, which has also been observed by SPT, APEX, ALMA and \textit{Herschel} \citep[][and references therein]{Everett2020}.
These measurements, listed in Table~\ref{tab:fluxes2sed_ancillary}, and our ACT coadd flux densities (labeled $ACT$), allowed us to display its SED (Fig.~\ref{fig:sed_dusty_ACT-S-J065207-551607}).
As a reference for the SED fitting, we follow the procedure presented in \cite{2017MNRAS.464..968S}.
The observed flux density can be modeled as a modified blackbody function for a single temperature as

\begin{equation}
S^{\rm obs}_{\nu}(z,T_{\rm d},\beta_{\rm d},M_{\rm d},d) \ = \Omega \left( 1 - {e}^{-\tau_{\nu}(z,\beta_{\rm d},M_{\rm d},d)} \right) B_{\nu} (T_{\rm d}/(1+z)),
\label{eq:sed_model_td}
\end{equation}

\noindent where the $T_{\rm d}$ is the dust temperature while the term $(1+z)$ accounts for the relationship between the observed ($\nu_{\rm obs}$) and rest-frame ($\nu_{\rm r}$) frequency, $\nu_{\rm obs} = \nu_r / (z+1)$.
$B_{\nu}(T)$ is the Planck function of spectral radiance.
The solid angle is $\Omega =\pi  \left(\sqrt{\mu}\,d/2\right)^2 / D_{A}^2$, where $D_A$ is the angular diameter distance to the source and $\mu$ corresponds to the possible gravitational lens magnification.
The optical depth $\tau_{\nu}(z,\beta_{\rm d},M_{\rm d},d)$ can be written as

%\frac{\mu\,M_{\rm d}}{\pi (\sqrt{\mu}\,d/2)^2}
\begin{equation}
\tau_{\nu}(z,\beta_{\rm d},M_{\rm d},d) = k_{0} \ \left( \frac{\nu}{\nu_{0}}\right)^{\beta_{\rm d}} \ \left( 1+z\right)^{\beta_{\rm d}} \  \frac{M_{\rm d}}{\pi (d/2)^2},
\label{eq:tau_dust_sed}
\end{equation}

\noindent which depends on the mass opacity coefficient $k_{0}$ (referred to the frequency $\nu_0$), the spectral emissivity $\beta_{\rm d}$, the redshift $z$, the total dust mass $M_{\rm d}$ and the diameter of the dust emission region ($d$).
There is a high possibility of a magnification $\mu$ due to gravitational lensing, but this does not affect the dust surface mass density in Eq.~\ref{eq:tau_dust_sed}. The estimation of the magnification $\mu$ is beyond the scope of this analysis.
Therefore, we solved the SED model using $\beta_{\rm d}=2$ and $k_{850\mu \rm{m}}=0.15$\,m$^2$\,kg$^{-1}$ at 850\,$\mu$m \cite[from ][]{2017MNRAS.464..968S}, and we allowed $z$, $T_{\rm d}$ (K), $\log( \mu M/M_\odot)$ and $\sqrt{u}d$ (kpc) to be free parameters in the fit. %(in our parameter space $\theta$).

\begin{table}
\centering
\caption{Ancillary data used to characterize the spectral energy distribution of ACT-S\,J065207-551605, which include {ALMA, SPT, APEX and \textit{Herschel} measurements where the absolute calibration uncertainties have been considered \citep{2020ApJ...902...78R}}.}
\begin{tabular}{ccc}
\hline
$\nu$ & $S_{\nu}$ & Facility\\
(GHz) & (mJy) & \\
\hline
95&$3\pm2$&SPT\\
100&$1.9\pm0.2$ &ALMA\\
150&$15\pm2$&SPT\\
220&$42\pm7$&SPT\\
345 &172$\pm22$ & APEX\\
600&$325\pm24$& \textit{Herschel}\\
857&$297\pm22$& \textit{Herschel}\\
1200&$186\pm15$& \textit{Herschel}\\
\hline
\end{tabular}
\label{tab:fluxes2sed_ancillary}
\end{table}
%
%95&$3\pm2$&SPT\\
%150&$15\pm1$&SPT\\
%220&$42\pm6$&SPT\\
%345 &172$\pm$7 & APEX\\
%600&$325\pm8$& \textit{Herschel}\\
%857&$297\pm8$& \textit{Herschel}\\
%1200&$186\pm8$& \textit{Herschel}\\

We used the Markov chain Monte Carlo (MCMC) method to determine the parameter space $\theta$, which allowed us to compute the marginalized posterior probability function for each parameter.
We used the publicly available affine-invariant MCMC sampler algorithm called {\sc emcee}\footnote{\url{https://emcee.readthedocs.io/en/stable/\#}} \citep{2013PASP..125..306F}, where the proposed model is fitted to the data by maximizing the corresponding logarithmic posterior distribution

\begin{equation}
\label{eq:likelihood}
\ln\left( \mathcal{L}(\theta) \ \prod_j P(\theta_j) \right)  \propto -\dfrac{1}{2} \sum_{i}^{n} \dfrac{\left( {\rm d}_i  - S^{\rm obs}_i(\theta) \right)^2}{\sigma_i^2} 
+ \sum_{j} \ln\left(P(\theta_j)\right),
\end{equation}

\noindent where $\mathcal{L}(\theta)$ is the likelihood and $P(\theta_j)$ corresponds to the prior distribution of the $j$-th parameter.
The variables d$_i$ and $\sigma_i$ are the flux densities and uncertainties in frequency band $i$ (observed frequency), respectively.
In Eq.~\ref{eq:likelihood}, we have assumed that measurements are independent of each other, and thus that the noise is uncorrelated.  Consequently, the covariance noise matrix is diagonal and each element of its inverse is given by $(N^{-1})_{i,j} = \delta_{i,j}/\sigma_i^2$.
As noted in \cite{2017MNRAS.464..968S}, there is a high correlation between parameters appearing in Eq.~\ref{eq:sed_model_td}, which requires the use of priors.
Firstly, we establish strict positive priors on all parameters.
For the dust temperature and total mass ($ \log(\mu M_{\rm d}/M_\odot)$), we adopt normal Gaussian distributions $\mathcal{N}(\overline{x},\sigma_x)$ with means and standard deviations used by \cite{2017MNRAS.464..968S} and based on results from \cite{Weiss2013}: $P(\scalebox{0.8}{$T_{\rm d}$}) = \mathcal{N} (40,12)$ and $P( \scalebox{0.8}{$\log(\mu M_{\rm d}/M_\odot)$})= \mathcal{N}(9.4,0.25)$.
Then the characterization of the posterior distributions is performed running the MCMC with 32 chains and 15000 iteration steps, where the first 2500 steps of each chain are excluded (these form the {\it burn-in} sample).
In order to recover the parameters from the SED modeling, we marginalize the posterior distributions over our parameter set, $\theta: \{z, T_{\rm d},  \log(\mu M_{\rm d}/M_\odot), \sqrt{\mu}d \}$, 
computing the median (the 50th percentile), and estimating the 16th and 84th percentiles to provide their uncertainties.

The parameter estimation provides {$z=4.0\pm0.8$, $T_{\rm d}=52.0^{+7.8}_{-7.2}$\,K, $\log(\mu M_{\rm d}/M_\odot) = 9.57\pm0.19$ and $d=6.9^{+1.1}_{-0.8}$\,kpc} for the high-redshift dusty Galaxy ACT-S\,J065207-551605 (dashed line in Fig.~\ref{fig:sed_dusty_ACT-S-J065207-551607}).
Our dust temperature is larger than the temperature value of $40\pm4$\,K reported by \cite{2020ApJ...902...78R}, although the difference can be understood in terms of the models and assumptions considered.
For instance, \cite{2017MNRAS.464..968S} obtained a dust temperature of $43.2^{+8.2}_{-7.2}$\,K and a redshift of $z=4.1^{+1.1}_{+1.0}$ (median and one-$\sigma$ standard deviation for their nine sources sample), from the modified blackbody model with a power law dust temperature distribution and without the assumption of optically thin emission. In contrast, they provided a higher dust temperature ($52.1^{+12.5}_{-9.6}$\,K) and a lower redshift ($3.4^{+1.0}_{-0.8}$) for their full sample, when the modified blackbody model with a single-temperature model without the assumption of optically thin dust is considered. Thus our dust temperature value falls in the range obtained by \citet{2017MNRAS.464..968S} using the same model.
Our photometric redshift is in agreement with the $z=4.2\pm0.5$ from \cite{2020ApJ...902...78R}, which is also compatible at $1~\sigma$ with respect to the spectroscopic redshift of $z_{\rm spec}=3.347$ reported by them.

\clearpage
\onecolumn
\begin{landscape}
%\begin{small}
\begin{center}
\begin{longtable}{|l|l|l|c|c|c|c|c|c|l|l|l|}
\caption{Sample from the catalog showing the coadded map fluxes {(from f150 coaddL, f220 coaddL, f280 coaddL)} in the area $65<\mathrm{RA}<90$, $-55.2<\mathrm{dec}<-51.0$. We include sources with at least one significant detection of $S/N>7$ in any band (85 of the brighter sources). Counterparts are found in: 1:SUMSS, 2:PMN, 3:AT20G, 4:IRAS, 5:WISE, 6:AKARI, 7:PCNT, 8:SPT, 9:ROSAT, 10:PCCS2, 11:Herschel, 12:ALMA.} \label{tab:cat} \\

\hline \multicolumn{1}{|c|}{ID (ACT-S)} & \multicolumn{1}{|c|}{ra\,(deg)} & \multicolumn{1}{|c|}{dec\,(deg)} & \multicolumn{1}{|c|}{$S_{150}$ (mJy)} & \multicolumn{1}{|c|}{$S_{220}$\,(mJy)} & \multicolumn{1}{|c|}{$S_{280}$\,(mJy)} & \multicolumn{1}{|c|}{$\alpha_{150}^{220}$} & \multicolumn{1}{|c|}{$\alpha_{220}^{280}$} & \multicolumn{1}{|c|}{$\alpha_{150}^{280}$} & \multicolumn{1}{|c|}{behavior} & \multicolumn{1}{|c|}{objclass} & \multicolumn{1}{|c|}{counterparts}\\ \hline 
\endfirsthead

\multicolumn{12}{c}%
{{\bfseries \tablename\ \thetable{} -- continued from previous page}} \\
\hline \multicolumn{1}{|c|}{ID (ACT-S)} & \multicolumn{1}{|c|}{ra\,(deg)} & \multicolumn{1}{|c|}{dec\,(deg)} & \multicolumn{1}{|c|}{$S_{150}$ (mJy)} & \multicolumn{1}{|c|}{$S_{220}$\,(mJy)} & \multicolumn{1}{|c|}{$S_{280}$\,(mJy)} & \multicolumn{1}{|c|}{$\alpha_{150}^{220}$} & \multicolumn{1}{|c|}{$\alpha_{220}^{280}$} & \multicolumn{1}{|c|}{$\alpha_{150}^{280}$} & \multicolumn{1}{|c|}{behavior} & \multicolumn{1}{|c|}{objclass} & \multicolumn{1}{|c|}{counterparts}\\ \hline 
\endhead

\hline \multicolumn{12}{|r|}{{Continued on next page}} \\ \hline
\endfoot

\hline \hline
\endlastfoot
J042000-545622 & 65.0032 & -54.9382 & 24.0 $\pm$ 1.2 & 116.2 $\pm$ 2.3 & 71.1 $\pm$ 4.3 & 4.1 & -2.1 & 1.8 & peaking & Nearby & 2,4,5,6,8,9,10,11,12 \\
J042347-513558 & 65.9443 & -51.5977 & 4.6 $\pm$ 1.3 & 21.7 $\pm$ 1.9 & 8.2 $\pm$ 3.3 & 4.0 & -4.1 & 0.9 & peaking & Nearby & 4,5,6,10 \\
J042501-525453 & 66.2546 & -52.9148 & ...  & 11.6 $\pm$ 1.6 & 25.0 $\pm$ 3.1 & ... & 3.3 & ... & ... & DSFG & ... \\
J042504-533159 & 66.2676 & -53.5327 & 133.1 $\pm$ 1.1 & 126.5 $\pm$ 1.7 & 119.8 $\pm$ 3.5 & -0.1 & -0.2 & -0.2 & falling & AGN & 1,2,3,5,8,9,10,12 \\
J042618-522855 & 66.5788 & -52.4820 & 4.9 $\pm$ 1.1 & 8.9 $\pm$ 1.7 & 21.0 $\pm$ 3.5 & 1.5 & 3.6 & 4.0 & rising & DSFG & ... \\
J042832-534819 & 67.1297 & -53.8017 & 3.4 $\pm$ 1.1 & 16.9 $\pm$ 1.8 & 25.2 $\pm$ 3.5 & 4.1 & 1.7 & 3.2 & rising & DSFG & ... \\
J042851-543005 & 67.2176 & -54.5025 & 20.0 $\pm$ 1.3 & 19.5 $\pm$ 1.9 & 12.7 $\pm$ 3.0 & -0.1 & -1.8 & -0.7 & falling & AGN & 2,3,4,5,8 \\
J042852-514612 & 67.2168 & -51.7701 & 19.7 $\pm$ 1.3 & 18.8 $\pm$ 1.9 & 29.5 $\pm$ 4.2 & -0.1 & 1.9 & 0.7 & upturning & AGN & ... \\
J042907-534943 & 67.2787 & -53.8277 & 85.3 $\pm$ 1.1 & 66.3 $\pm$ 1.8 & 54.0 $\pm$ 3.5 & -0.7 & -0.9 & -0.7 & falling & AGN & 1,2,3,7,8,9 \\
J043020-514743 & 67.5852 & -51.7954 & 0.2 $\pm$ 1.2 & 7.0 $\pm$ 1.7 & 22.0 $\pm$ 3.0 & 9.1 & 4.9 & 7.5 & rising & DSFG & ... \\
J043125-542451 & 67.8530 & -54.4140 & 2.2 $\pm$ 1.2 & 16.4 $\pm$ 1.8 & 12.9 $\pm$ 4.2 & 5.2 & -1.0 & 2.8 & peaking & Nearby & 4,5,11 \\
J043221-510927 & 68.0885 & -51.1567 & 88.3 $\pm$ 1.5 & 64.2 $\pm$ 1.9 & 48.8 $\pm$ 3.5 & -0.8 & -1.2 & -1.0 & falling & AGN & 1,2,3,5,8,12 \\
J043639-524338 & 69.1646 & -52.7275 & 3.5 $\pm$ 1.1 & 11.3 $\pm$ 1.7 & 10.5 $\pm$ 3.2 & 3.1 & -0.3 & 1.8 & peaking & DSFG & ... \\
J043651-521639 & 69.2164 & -52.2770 & 40.6 $\pm$ 1.2 & 35.4 $\pm$ 1.7 & 25.2 $\pm$ 3.2 & -0.4 & -1.4 & -0.8 & falling & AGN & 3,8 \\
J043920-520754 & 69.8330 & -52.1329 & 6.5 $\pm$ 1.2 & 18.7 $\pm$ 1.7 & 17.6 $\pm$ 3.0 & 2.7 & -0.3 & 1.6 & peaking & DSFG & 4,5,6,10 \\
J043936-530054 & 69.9029 & -53.0140 & 6.6 $\pm$ 1.1 & 20.4 $\pm$ 1.7 & 33.5 $\pm$ 3.1 & 2.9 & 2.1 & 2.6 & rising & Nearby & 4,5,6,8,10 \\
J043937-530927 & 69.9066 & -53.1575 & 1.5 $\pm$ 1.1 & 12.2 $\pm$ 1.7 & 13.1 $\pm$ 3.0 & 5.4 & 0.3 & 3.5 & rising & DSFG & 11 \\
J044116-543849 & 70.3130 & -54.6493 & 12.7 $\pm$ 1.3 & 11.1 $\pm$ 2.1 & 14.1 $\pm$ 4.1 & -0.3 & 1.0 & 0.2 & upturning & AGN & 2,8,11 \\
J044153-540357 & 70.4705 & -54.0672 & 3.7 $\pm$ 1.1 & 6.8 $\pm$ 1.8 & 31.3 $\pm$ 3.4 & 1.6 & 6.5 & 3.5 & rising & DSFG & 11,12 \\
J044158-515454 & 70.4934 & -51.9150 & 102.4 $\pm$ 1.2 & 82.0 $\pm$ 1.7 & 56.3 $\pm$ 3.1 & -0.6 & -1.6 & -1.0 & falling & AGN & 1,2,3,7,8,10,12 \\
J044230-543146 & 70.6275 & -54.5286 & 20.3 $\pm$ 1.2 & 15.1 $\pm$ 2.0 & 9.1 $\pm$ 4.2 & -0.8 & -2.1 & -1.1 & falling & AGN & 4,8,11 \\
J044306-511859 & 70.7761 & -51.3164 & 13.7 $\pm$ 1.3 & 10.0 $\pm$ 1.8 & 6.7 $\pm$ 3.6 & -0.8 & -1.7 & -1.2 & falling & AGN & 2,8 \\
J044502-523425 & 71.2605 & -52.5748 & 22.2 $\pm$ 1.1 & 17.2 $\pm$ 1.8 & 6.4 $\pm$ 3.3 & -0.7 & -4.2 & -2.0 & falling & AGN & 8,9 \\
J044702-510257 & 71.7627 & -51.0468 & 20.6 $\pm$ 1.6 & 12.0 $\pm$ 2.3 & 14.8 $\pm$ 3.6 & -1.4 & 0.9 & -0.5 & upturning & AGN & 2,8 \\
J044747-515058 & 71.9502 & -51.8501 & 27.2 $\pm$ 1.2 & 23.3 $\pm$ 1.8 & 13.5 $\pm$ 2.9 & -0.4 & -2.3 & -1.1 & falling & AGN & 1,2,3,8,9 \\
J045029-534657 & 72.6190 & -53.7830 & 17.7 $\pm$ 1.1 & 12.4 $\pm$ 1.7 & 10.7 $\pm$ 3.4 & -0.9 & -0.6 & -0.8 & falling & AGN & 2,8 \\
J045139-524723 & 72.9138 & -52.7898 & 1.6 $\pm$ 1.1 & 7.8 $\pm$ 1.7 & 19.1 $\pm$ 3.0 & 4.2 & 3.8 & 4.0 & rising & DSFG & 11 \\
J045239-530637 & 73.1622 & -53.1094 & 23.0 $\pm$ 1.1 & 19.4 $\pm$ 1.8 & 15.8 $\pm$ 2.9 & -0.5 & -0.9 & -0.6 & falling & AGN & 1,2,3,8,11 \\
J045240-531548 & 73.1709 & -53.2636 & 13.8 $\pm$ 1.1 & 9.2 $\pm$ 1.7 & 10.6 $\pm$ 3.0 & -1.0 & 0.6 & -0.4 & upturning & AGN & 8,11 \\
J045319-533814 & 73.3316 & -53.6383 & 3.5 $\pm$ 1.1 & 6.1 $\pm$ 1.7 & 20.6 $\pm$ 3.4 & 1.4 & 5.2 & 3.7 & rising & DSFG & 11 \\
J045333-533505 & 73.3897 & -53.5847 & 1.7 $\pm$ 1.1 & 10.6 $\pm$ 1.7 & 16.5 $\pm$ 3.4 & 4.7 & 1.9 & 3.7 & rising & DSFG & 11 \\
J045336-513015 & 73.3996 & -51.5065 & 13.4 $\pm$ 1.3 & 4.5 $\pm$ 1.8 & 4.8 $\pm$ 3.4 & -2.8 & 0.3 & -1.7 & upturning & AGN & 1,2,3,5,8,9 \\
J045457-540552 & 73.7381 & -54.0978 & 2.4 $\pm$ 1.1 & 13.2 $\pm$ 1.8 & 18.2 $\pm$ 3.3 & 4.4 & 1.4 & 3.3 & rising & DSFG & ... \\
J045559-530237 & 73.9981 & -53.0429 & 17.7 $\pm$ 1.2 & 9.6 $\pm$ 1.8 & 16.7 $\pm$ 3.0 & -1.6 & 2.3 & -0.1 & upturning & AGN & 1,2,3,8,11 \\
J045832-523518 & 74.6353 & -52.5883 & 3.2 $\pm$ 1.1 & 16.3 $\pm$ 1.8 & 17.6 $\pm$ 3.3 & 4.2 & 0.3 & 2.7 & rising & DSFG & ... \\
J050019-532125 & 75.0808 & -53.3567 & 28.2 $\pm$ 1.1 & 23.0 $\pm$ 1.6 & 12.8 $\pm$ 3.3 & -0.5 & -2.5 & -1.3 & falling & AGN & 3,8 \\
J050220-523943 & 75.5863 & -52.6622 & 3.8 $\pm$ 1.2 & 12.4 $\pm$ 1.8 & 16.3 $\pm$ 3.1 & 3.1 & 1.2 & 2.3 & rising & DSFG & ... \\
J050328-515539 & 75.8689 & -51.9275 & 3.2 $\pm$ 1.2 & 10.8 $\pm$ 1.7 & 10.4 $\pm$ 3.1 & 3.2 & -0.2 & 1.9 & peaking & DSFG & ... \\
J050329-524455 & 75.8766 & -52.7491 & 13.3 $\pm$ 1.2 & 10.4 $\pm$ 1.8 & 17.8 $\pm$ 3.0 & -0.6 & 2.3 & -0.2 & falling & AGN & 2,8 \\
J050731-510414 & 76.8848 & -51.0710 & 31.0 $\pm$ 1.5 & 42.2 $\pm$ 2.1 & 42.5 $\pm$ 3.5 & 0.8 & 0.0 & 0.2 & upturning & AGN & 3,5,7,8,9,10 \\
J050746-515602 & 76.9466 & -51.9351 & 13.6 $\pm$ 1.2 & 2.7 $\pm$ 1.7 & 4.2 $\pm$ 3.1 & -4.2 & 1.9 & -1.9 & upturning & AGN & 1,2,3,8 \\
J050758-512553 & 76.9932 & -51.4314 & 2.2 $\pm$ 1.3 & 12.1 $\pm$ 1.8 & 19.3 $\pm$ 3.5 & 4.4 & 2.0 & 3.5 & rising & DSFG & 8 \\
J050813-525243 & 77.0556 & -52.8787 & ...  & 5.3 $\pm$ 1.7 & 18.8 $\pm$ 3.0 & ... & 5.4 & ... & ... & DSFG & ... \\
J051333-525553 & 78.3897 & -52.9315 & 2.5 $\pm$ 1.1 & 2.0 $\pm$ 1.7 & 18.8 $\pm$ 3.0 & -0.5 & 9.4 & 3.2 & upturning & DSFG & ... \\
J051424-521923 & 78.6003 & -52.3232 & 1.7 $\pm$ 1.3 & 10.9 $\pm$ 1.8 & 24.9 $\pm$ 3.3 & 4.8 & 3.5 & 4.3 & rising & DSFG & ... \\
J051506-534420 & 78.7772 & -53.7395 & 2.9 $\pm$ 1.2 & 23.3 $\pm$ 1.9 & 31.3 $\pm$ 3.9 & 5.4 & 1.3 & 3.8 & rising & Nearby & 4,5,6,8,10 \\
J051507-511920 & 78.7832 & -51.3223 & ...  & 3.2 $\pm$ 2.0 & 23.9 $\pm$ 3.6 & ... & 8.5 & ... & ... & DSFG & ... \\
J051812-514358 & 79.5490 & -51.7345 & 28.3 $\pm$ 1.3 & 32.5 $\pm$ 1.9 & 31.1 $\pm$ 3.1 & 0.4 & -0.2 & 0.2 & falling & AGN & 1,2,3,7,8 \\
J052040-533013 & 80.1676 & -53.5037 & 2.8 $\pm$ 1.2 & 16.4 $\pm$ 1.8 & 17.2 $\pm$ 3.4 & 4.5 & 0.2 & 2.9 & rising & DSFG & 8,11,12 \\
J052046-550823 & 80.1875 & -55.1392 & 21.8 $\pm$ 1.5 & 18.9 $\pm$ 2.4 & 14.1 $\pm$ 4.1 & -0.4 & -1.2 & -0.7 & falling & AGN & 1,2,3,8 \\
J052237-512554 & 80.6548 & -51.4319 & 4.0 $\pm$ 1.4 & 6.2 $\pm$ 2.0 & 24.5 $\pm$ 3.6 & 1.1 & 5.9 & 2.9 & rising & DSFG & ... \\
J052317-530833 & 80.8254 & -53.1423 & 16.5 $\pm$ 1.2 & 5.5 $\pm$ 1.8 & 21.5 $\pm$ 3.0 & -2.9 & 5.8 & 0.4 & upturning & AGN & 1,2,3,8 \\
J052459-513510 & 81.2460 & -51.5861 & 12.9 $\pm$ 1.4 & 6.3 $\pm$ 2.1 & 17.8 $\pm$ 3.3 & -1.8 & 4.4 & 0.5 & upturning & AGN & 1,2,8,9 \\
J052516-520816 & 81.3188 & -52.1379 & 2.3 $\pm$ 1.3 & 9.5 $\pm$ 1.9 & 21.2 $\pm$ 3.0 & 3.7 & 3.4 & 3.6 & rising & DSFG & ... \\
J052524-545808 & 81.3509 & -54.9689 & 0.5 $\pm$ 1.4 & 15.9 $\pm$ 2.2 & 15.4 $\pm$ 4.1 & 8.8 & -0.1 & 5.4 & peaking & DSFG & ... \\
J052743-542609 & 81.9273 & -54.4370 & 20.1 $\pm$ 1.4 & 16.7 $\pm$ 2.1 & 14.4 $\pm$ 4.0 & -0.5 & -0.6 & -0.5 & falling & AGN & 1,2,3,8 \\
J052833-543353 & 82.1397 & -54.5649 & 13.1 $\pm$ 1.4 & 7.8 $\pm$ 2.2 & 11.4 $\pm$ 4.2 & -1.4 & 1.6 & -0.2 & upturning & AGN & 8 \\
J052903-543650 & 82.2647 & -54.6128 & 6.6 $\pm$ 1.4 & 34.2 $\pm$ 2.3 & 67.8 $\pm$ 4.2 & 4.3 & 2.9 & 3.8 & rising & DSFG & 8,10,11,12 \\
J053021-530205 & 82.5896 & -53.0350 & 5.2 $\pm$ 1.2 & 17.8 $\pm$ 1.9 & 17.4 $\pm$ 3.0 & 3.2 & -0.1 & 1.9 & peaking & DSFG & ... \\
J053117-550425 & 82.8200 & -55.0734 & 31.0 $\pm$ 1.5 & 23.2 $\pm$ 2.5 & 17.8 $\pm$ 4.0 & -0.8 & -1.1 & -0.9 & falling & AGN & 5,8,9 \\
J053159-525518 & 82.9962 & -52.9219 & 1.6 $\pm$ 1.2 & ...  & 27.8 $\pm$ 3.0 & ... & ... & 4.6 & ... & DSFG & ... \\
J053208-535803 & 83.0334 & -53.9678 & 51.4 $\pm$ 1.3 & 40.7 $\pm$ 1.9 & 18.0 $\pm$ 3.4 & -0.6 & -3.5 & -1.7 & falling & AGN & ... \\
J053208-531033 & 83.0355 & -53.1768 & 51.5 $\pm$ 1.3 & 40.7 $\pm$ 1.9 & 44.0 $\pm$ 3.0 & -0.6 & 0.3 & -0.3 & upturning & AGN & 1,2,3,5,8 \\
J053311-523827 & 83.2956 & -52.6425 & 3.8 $\pm$ 1.3 & 24.5 $\pm$ 1.9 & 16.8 $\pm$ 3.3 & 4.8 & -1.6 & 2.4 & peaking & Nearby & 4,5,6,8,10 \\
J053426-511323 & 83.6099 & -51.2233 & 14.4 $\pm$ 1.6 & 9.1 $\pm$ 2.1 & 18.9 $\pm$ 3.7 & -1.2 & 3.1 & 0.4 & upturning & AGN & 1,2,3,5,8 \\
J053458-543906 & 83.7443 & -54.6513 & 31.8 $\pm$ 1.5 & 33.0 $\pm$ 2.4 & 25.1 $\pm$ 4.2 & 0.1 & -1.2 & -0.4 & falling & AGN & 1,2,3,5,8 \\
J053647-532053 & 84.1976 & -53.3482 & 1.5 $\pm$ 1.3 & 11.1 $\pm$ 1.8 & 15.8 $\pm$ 3.4 & 5.2 & 1.5 & 3.8 & rising & DSFG & ... \\
J053820-522744 & 84.5834 & -52.4624 & 11.9 $\pm$ 1.3 & 11.5 $\pm$ 1.9 & 11.0 $\pm$ 3.6 & -0.1 & -0.2 & -0.1 & falling & AGN & 3,8,9 \\
J053909-551055 & 84.7867 & -55.1827 & 24.0 $\pm$ 1.6 & 12.8 $\pm$ 2.7 & 19.0 $\pm$ 4.1 & -1.6 & 1.7 & -0.4 & upturning & AGN & 2,3,8 \\
J054025-530348 & 85.1048 & -53.0630 & 34.6 $\pm$ 1.3 & 31.1 $\pm$ 1.9 & 10.0 $\pm$ 3.0 & -0.3 & -4.8 & -2.0 & falling & AGN & 1,2,3,8 \\
J054030-535626 & 85.1234 & -53.9400 & 16.9 $\pm$ 1.2 & 12.1 $\pm$ 2.0 & 10.8 $\pm$ 3.6 & -0.9 & -0.5 & -0.7 & falling & AGN & 2,3,8 \\
J054045-541821 & 85.1910 & -54.3063 & 492.2 $\pm$ 1.3 & 382.1 $\pm$ 2.1 & 321.9 $\pm$ 3.7 & -0.7 & -0.7 & -0.7 & falling & AGN & 1,2,3,5,7,8,9,10,12 \\
J054104-542311 & 85.2682 & -54.3865 & 0.2 $\pm$ 1.4 & 15.3 $\pm$ 2.1 & 7.5 $\pm$ 4.0 & 11.0 & -3.1 & 5.7 & peaking & DSFG & ... \\
J054223-514256 & 85.5978 & -51.7157 & 73.3 $\pm$ 1.4 & 62.5 $\pm$ 2.0 & 44.4 $\pm$ 3.1 & -0.4 & -1.5 & -0.8 & falling & AGN & 3,8,9,12 \\
J054544-513802 & 86.4341 & -51.6340 & 12.6 $\pm$ 1.4 & 3.0 $\pm$ 2.1 & 8.0 $\pm$ 3.3 & -3.7 & 4.2 & -0.7 & upturning & AGN & 8,9 \\
J054717-510405 & 86.8196 & -51.0682 & 9.1 $\pm$ 1.7 & 25.7 $\pm$ 2.4 & 24.8 $\pm$ 3.8 & 2.7 & -0.1 & 1.6 & peaking & DSFG & 4,5,6,8,10,11,12 \\
J054830-521836 & 87.1260 & -52.3105 & 26.4 $\pm$ 1.4 & 24.4 $\pm$ 1.9 & 17.2 $\pm$ 3.4 & -0.2 & -1.5 & -0.7 & falling & AGN & 1,2,8 \\
J054930-523142 & 87.3781 & -52.5284 & 9.9 $\pm$ 1.3 & 7.2 $\pm$ 2.0 & 3.6 $\pm$ 3.6 & -0.8 & -2.9 & -1.6 & falling & AGN & ... \\
J054944-524626 & 87.4327 & -52.7742 & 163.8 $\pm$ 1.3 & 127.7 $\pm$ 1.9 & 104.9 $\pm$ 3.2 & -0.6 & -0.8 & -0.7 & falling & AGN & 1,2,3,7,8,9,10,12 \\
J055047-530454 & 87.6973 & -53.0832 & 25.6 $\pm$ 1.4 & 10.7 $\pm$ 2.0 & 22.5 $\pm$ 3.1 & -2.3 & 3.2 & -0.2 & upturning & AGN & 1,2,8 \\
J055115-533435 & 87.8143 & -53.5759 & 3.7 $\pm$ 1.3 & 35.2 $\pm$ 2.0 & 20.2 $\pm$ 3.6 & 5.8 & -2.4 & 2.7 & peaking & Nearby & 4,5,6,8 \\
J055235-534926 & 88.1437 & -53.8238 & 13.7 $\pm$ 1.3 & 6.3 $\pm$ 2.0 & 3.8 $\pm$ 3.6 & -2.0 & -2.1 & -2.1 & falling & AGN & 1,2,8 \\
J055423-530447 & 88.5986 & -53.0799 & 7.2 $\pm$ 1.3 & 14.0 $\pm$ 2.0 & 10.9 $\pm$ 3.1 & 1.7 & -1.1 & 0.7 & peaking & DSFG & 8 \\
J055829-532626 & 89.6241 & -53.4415 & 17.5 $\pm$ 1.3 & 12.1 $\pm$ 1.9 & 10.0 $\pm$ 3.7 & -1.0 & -0.8 & -0.9 & falling & AGN & 1,2,3,8 \\
J055857-533350 & 89.7377 & -53.5641 & 3.0 $\pm$ 1.3 & 21.0 $\pm$ 2.0 & 16.8 $\pm$ 3.8 & 5.1 & -1.0 & 2.8 & peaking & Nearby & 4,5,6,8 \\
\end{longtable}
\end{center}
%\end{small}
\end{landscape}

\end{appendix}

\end{document}